\documentclass[a4paper,11pt]{article}

\pdfoutput=1

\usepackage[a4paper,left=2.73cm,right=2.7cm,top=3cm,bottom=3.5cm]{geometry}

\usepackage[T1]{fontenc} 
\usepackage{graphicx}
\usepackage{epsfig}
\usepackage{amssymb}
\usepackage{amsfonts}
\usepackage{dsfont}
\usepackage{amsmath,euscript,array,mathrsfs,nccmath}
\usepackage{multirow}
\usepackage{bbold,bm,bbm}
\usepackage{epsf}
\usepackage{slashed}
\usepackage{comment}
\usepackage{colortbl}
\usepackage{lmodern}
\usepackage[utf8]{inputenc}
\usepackage[noadjust]{cite}

\usepackage{jheppub}

\usepackage{caption,subcaption,wrapfig}
\usepackage[colorlinks=true,linktocpage=true,linkcolor=blue,citecolor=blue]{hyperref}

\usepackage{tikz}
\usetikzlibrary{decorations.markings,decorations.pathmorphing}

\newcommand{\be}{\begin{equation}}
\newcommand{\ee}{\end{equation}}
\newcommand{\beqs}{\begin{eqnarray}}
\newcommand{\eeqs}{\end{eqnarray}}

\newcommand{\dd}{\mathrm{d}}

\newcommand{\AAA}{\mathcal{A}}
\newcommand{\FF}{\mathcal{F}}
\newcommand{\OO}{\mathcal{O}}
\newcommand{\GG}{\mathcal{G}}
\newcommand{\MM}{\mathcal{M}}
\newcommand{\NN}{\mathcal{N}}

\newcommand{\BB}{\mathcal{B}}

\newcommand{\CP}{\mathds{C}\mathds{P}}

\newcommand{\at}{\tilde{a}_1}
\newcommand{\ah}{\hat{a}_1}

\newcommand{\AJ}{\mathcal{A}_J}
\newcommand{\BJ}{\mathcal{B}_J}
\newcommand{\BX}{\mathcal{B}_X}
\newcommand{\AO}{\mathcal{A}_t}
\newcommand{\AH}{\widehat{\mathcal{A}}_t}
\newcommand{\AT}{\widetilde{\mathcal{A}}_t}
\newcommand{\BET}{\overline{b}}

\newcommand{\cfh}{f_h}
\newcommand{\cgh}{g_h}
\newcommand{\clh}{\lambda_h}
\newcommand{\chh}{h_h}
\newcommand{\cbh}{\mathsf{b}_h}
\newcommand{\cbjh}{\textsc{J}_h}
\newcommand{\cbxh}{\textsc{X}_h}
\newcommand{\cajh}{\textsc{A}_h}
\newcommand{\caoneh}{\alpha_{h}}
\newcommand{\cahh}{\hat{\alpha}_{h}}
\newcommand{\cath}{\tilde{\alpha}_{h}}

\newcommand{\cfi}{f_{\text{\tiny IR}}}
\newcommand{\cbi}{\mathsf{b}_{\text{\tiny IR}}}
\newcommand{\cli}{\lambda_{\text{\tiny IR}}}
\newcommand{\icli}{{\lambda_{\text{\tiny IR}}}^{-1}}
\newcommand{\chir}{\mathbf{h}_{\text{\tiny IR}}}
\newcommand{\cbji}{\chi_J}
\newcommand{\cbxi}{\chi_X}
\newcommand{\caji}{\alpha_J}
\newcommand{\caone}{\alpha_{\text{\tiny IR}}}
\newcommand{\cah}{\hat{\alpha}_{\text{\tiny IR}}}

\newcommand{\R}{\mathds{R}}
\newcommand{\cfUV}[2]{\mathbf{f}^{\text{\tiny UV}}_{\, #1,#2}}
\newcommand{\cgUV}[2]{\mathbf{g}^{\text{\tiny UV}}_{\, #1,#2}}
\newcommand{\clUV}[2]{\mathbf{l}^{\text{\tiny UV}}_{\, #1,#2}}
\newcommand{\cbUV}[2]{\mathbf{b}^{\text{\tiny UV}}_{\, #1,#2}}
\newcommand{\cbJUV}[2]{\mathbf{B}^{\text{\tiny UV}}_{J;\, #1,#2}}
\newcommand{\cbXUV}[2]{\mathbf{B}^{\text{\tiny UV}}_{X;\, #1,#2}}
\newcommand{\caJUV}[2]{\mathbf{A}^{\text{\tiny UV}}_{J;\, #1,#2}}
\newcommand{\chUV}[2]{\mathbf{h}^{\text{\tiny UV}}_{\, #1,#2}}
\newcommand{\caoneUV}[2]{\mathbf{A}^{\text{\tiny UV}}_{#1,#2}}
\newcommand{\cahUV}[2]{\widehat{\mathbf{A}}^{\text{\tiny UV}}_{#1,#2}}
\newcommand{\catUV}[2]{\widetilde{\mathbf{A}}^{\text{\tiny UV}}_{#1,#2}}

\newcommand{\gs}{g_s}
\newcommand{\ls}{\ell_s}

\newcommand{\QD}{Q_{\text{\tiny D2}}}
\newcommand{\LQCD}{\Lambda_{\text{\tiny QCD}}}
\newcommand{\UM}{{\rm U}_{_{\cal M}}(1)}
\newcommand{\UU}{{\rm U}}

\newcommand{\Bphys}{\mathsf{B}}
\newcommand{\Mphys}{\mathsf{M}}

\newcommand{\Iren}{I_{\text{\tiny ren}}}

\DeclareMathOperator{\arccsc}{arccsc}

\newcommand{\vev}[1]{\left\langle #1 \right\rangle}

\begin{document}

 \begin{titlepage}

\thispagestyle{empty}

\begin{flushright}
\hfill{NORDITA 2022-090}
\end{flushright}

\vspace{40pt}  
	 
\begin{center}

{\huge \textbf{Monopoles and confinement in three dimensions from holography}}

\vspace{30pt}
		
{\large \bf Ant\'on F. Faedo,$^{1,\,2}$   Carlos Hoyos,$^{1,\,2}$   \\ [1mm]

and Javier G. Subils$^{3}$
}

\vspace{25pt}

{\normalsize  $^{1}$ Departamento de F\'{i}sica, Universidad de Oviedo, \\ Federico Garc\'ia Lorca 18, ES-33007, Oviedo, Spain.}\\
\vspace{15pt}
{ $^{2}$Instituto Universitario de Ciencias y Tecnolog\'{\i}as Espaciales de Asturias (ICTEA), \\ Calle de la Independencia 13, ES-33004, Oviedo, Spain.}\\
\vspace{15pt}
{ $^{3}$ Nordita, Stockholm University and KTH Royal Institute of Technology,\\
Hannes Alfvéns väg 12, SE-106 91 Stockholm, Sweden.}\\
\vspace{15pt}

\vspace{60pt}
			
\textbf{Abstract}
\end{center} 
We study the phase diagram of a confining three-dimensional ${\cal N}=1$ supersymmetric $\UU(N)\times \UU(N+M)$ theory with holographic dual corresponding to a known string theory solution. The theory possesses a global $\UU(1)$ symmetry under which magnetic monopoles are charged. We introduce both temperature and an external magnetic field for monopoles and find that there are deconfinement phase transitions as any of the two is increased, supporting monopole condensation as the possible mechanism for confinement. We find that the transition as the magnetic field is increased is second order, providing the first example in holographic duals of a deconfinement transition which is not first order. We also uncover a rich structure in the phase diagram, with a triple point and a critical point where a line of first order transitions end. 
\end{titlepage}

\tableofcontents

\hrulefill
\vspace{10pt}

\section{Introduction}\label{sec:intro}

One of the main unresolved issues of QCD is understanding its phase diagram. In vacuum, QCD is a confining theory but, due to asymptotic freedom, for large enough temperatures or chemical potentials it should be well described by a deconfined plasma composed of weakly interacting quarks and gluons. At zero chemical potential lattice QCD can be employed to show a crossover between the confined and deconfined phases as the temperature increases \cite{Aoki:2006we,Bhattacharya:2014ara}. Unfortunately, the sign problem prevents from applying lattice QCD to regions where the quark chemical potential is comparable or larger than the temperature \cite{deForcrand:2009zkb}, and we lack a first principles approach that can determine the properties of QCD in the intermediate sector of the phase diagram that lies between the region where lattice QCD is applicable and the asymptotic region where QCD becomes weakly coupled.

This is not just of academic interest, but some parts of the intermediate strongly-coupled region are accessible through experiments of heavy ion collisions and through astrophysical observations of neutron stars and binary mergers (see e.g. \cite{Brambilla:2014jmp} for a review). Collision experiments may be able to reach a critical point marking the end of a line of first order chiral symmetry breaking phase transitions in the temperature-baryon chemical potential plane, conjectured to exist from phenomenological models. Neutron stars observations are sensitive to the equation of state in regions of low temperature and high baryon density.

A possible way to learn about QCD deconfinement transitions is to study similar strongly coupled gauge theories for which we have a known gauge/gravity dual pair. Even though we do not expect that there is a perfect equivalence between the phase diagrams, they may serve to understand the dynamics behind the transitions at strong coupling. There is a handful of confining theories with a gravity dual that have a realization in string theory. For instance, in four dimensions the best known are the Witten QCD (WQCD) \cite{Witten:1998zw} and Klebanov--Strassler (KS) \cite{Klebanov:2000hb} models. However, the WQCD model is really dual to a compactification of a higher dimensional theory and there is no separation between the confinement and Kaluza--Klein scales in the regime where the theory is under control. The KS model is more realistic, but it is technically quite challenging, among other things because of its exotic UV behaviour. One can avoid the issues of the WQCD and KS models if one is willing to move a bit further from QCD by going to three dimensions. Studying confining theories in lower dimensions has a venerable history, and it was in fact in this context where confinement was first proved by Polyakov for QED${}_3$ \cite{Polyakov:1976fu}, and confirmed later by lattice simulations \cite{DeGrand:1981yq}.

In \cite{Faedo:2017fbv} the gauge/gravity duals of  three-dimensional theories with a mass gap were studied in detail as a family of solutions of type IIA supergravity originally constructed in \cite{Cvetic:2001pga,Cvetic:2001ma}. The solutions are similar to the gravitational duals to ABJM \cite{Aharony:2008ug} and ABJ \cite{Aharony:2008gk} theories - indeed connected to them by RG flows - so they are expected to be dual to a quiver gauge theory with a rank and Chern--Simons level determined by the supergravity fluxes. The uplift to M-theory further revealed that despite having a mass gap, most of the solutions were actually not dual to a confining theory, just those for which the Chern--Simons level in the dual field theory vanishes. These last will be the focus of the present work. 

In the absence of temperature and magnetic field the aforementioned theories have ${\cal N}=1$ supersymmetry (i.e. 2 real supercharges). Contrary to higher supersymmetric models, there are no non-renormalization theorems based on holomorphicity. Furthermore, localization techniques cannot be applied, so ${\cal N}=1$ theories are almost on par with non-supersymmetric models concerning the analysis of their non-perturbative properties. Although there has been some work studying the vacuum structure of ${\cal N}=1$ theories \cite{Bashmakov:2018wts,Bashmakov:2021rci}, as far as we are aware non-Abelian quiver theories with vanishing Chern--Simons levels like the ones considered here have not been studied before. In this regard, our analysis might also shed further light on the phase structure of ${\cal N}=1$ theories.

The phase diagram as a function of the temperature was studied in \cite{Elander:2020rgv}. In addition to the solution dual to the confined phase, there are black hole solutions that are dual to the deconfined phase of the same theory. The dominant phase is the one with lowest free energy, that is determined from the supergravity action of the dual solution. It was found that when the temperature is increased, there is a first order deconfinement transition, as it is commonly the case in large-$N$ gauge theories. In this work we will explore further the phase diagram by turning on an external magnetic field for a global $\UU(1)$ symmetry. The properties of the supergravity solution, that will be discussed in more detail in the main text, suggest that the gauge group is $\UU(N)\times \UU(N+M)$ in which case the global symmetry would correspond to a magnetic $\UM$ symmetry, as discussed in detail in e.g.~\cite{Aharony:2008ug,Bergman:2020ifi} for ABJM, with the conserved current equal to the Hodge dual to the Abelian flux of the diagonal $\UU(1)$ gauge group.

The external gauge field couples to color magnetic vortices that have a dual description in terms of D0- and wrapped D2-branes. These can in principle be created by the insertion of a local magnetic monopole operator, with the brane attached to the location of the monopole at the asymptotic boundary. It is worth recalling at this point that the Abelian confinement in three dimensions described by Polyakov is produced precisely by a gas of Abelian monopoles akin to the ones we are describing. Similarly, in the Seiberg--Witten solution of ${\cal N}=2$ super Yang-Mills in four dimensions, confinement at a generic point of the moduli space is produced by the condensation of Abelian monopoles\footnote{In this case monopoles are line rather than local operators.}, after the non-Abelian part has been Higgsed \cite{Seiberg:1994rs}. In our setup the $\UM$ external magnetic field could have the effect of introducing a gap for the monopoles, thus suppressing their contribution to the path integral. Therefore our analysis can serve as a check of whether Polyakov's form of confinement still plays a role in a less supersymmetric strongly coupled non-Abelian theory, beyond lattice calculations.  

\begin{figure}[t!]
	\begin{center}
		\includegraphics[width=.85\textwidth]{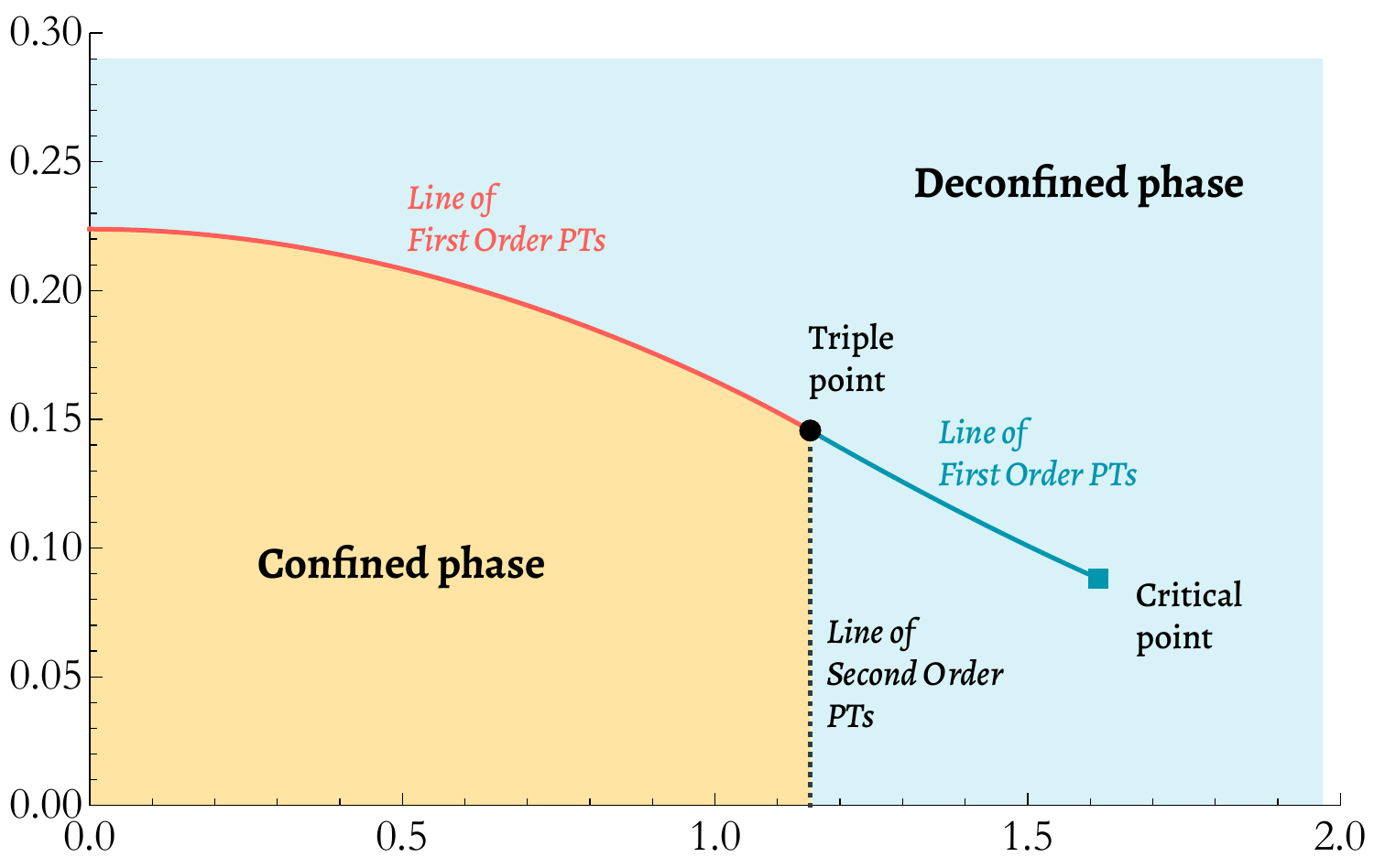} 
		\put(-365,240){$T/\LQCD$}
		\put(-40,-15){$\Bphys/\LQCD^2$}
		\caption{\small Phase diagram in the temperature, monopole magnetic field plane. The confined phase lies in the small temperature and magnetic field orange region, while the rest of the phase diagram corresponds to deconfined phases. The red, solid curve indicates a line of first order deconfinement phase transitions. The blue solid curve, however, stands for first order phase transitions between different deconfined phases. It ends at a critical point, represented by the blue square, where the phase transition becomes second order. Finally, the black, dotted line corresponds to second order deconfinement phase transitions. These three lines meet at a triple point, where coexistence of the three phases can occur.} \label{fig:PhaseDiagram}
	\end{center}
\end{figure}

Our results are summarized in Fig.~\ref{fig:PhaseDiagram}. We observe that when the magnetic field is increased there is a second order deconfinement transition, suggesting that indeed there is a form of Polyakov confinement at play. To our knowledge this is the first example in a holographic dual of a deconfinement transition that is not of first order, of Hawking--Page type. The deconfined phase after the transition is distinct from the one found by increasing the temperature, they are separated by a first order phase transition. Although the transition turns into a crossover for large enough magnetic fields, some properties still remain quantitatively different between the two deconfined phases.

The outline of the paper is as follows: in Section~\ref{sec:review} we summarize the main properties of the supergravity solution and the dual field theory in the absence of monopole magnetic field. In Section~\ref{sec:newsols} we introduce the monopole magnetic field and temperature and construct the supergravity solutions, finding both confining solutions and black branes dual to deconfined phases. In Section~\ref{sec:thermo} we study the thermodynamic properties of the solutions and describe the phase diagram. Finally, in Section~\ref{sec:discussion} we discuss the results for the phase diagram, future possible directions and speculate about the interpretation of the D0- and D2-branes as monopole operators in the field theory. The main text is complemented with several appendices containing some technical results and derivations. In Appendix \ref{sec:geometry} we collect some properties of the internal geometry. In Appendix~\ref{sec:truncation} we present a consistent truncation to four-dimensional supergravity. In Appendix~\ref{sec:groundstate} we collect previous results for the solutions in the absence of magnetic field and temperature. In Appendix~\ref{sec:D2pot} we compute the D2-brane configuration and action dual to a monopole-antimonopole pair. Finally, in Appendix~\ref{sec:numerics} we explain in detail the asymptotic expansions of the solutions and the numerical calculation.

\section{Confining ground state and monopole condensation}\label{sec:review}

The type IIA supergravity solution describing the supersymmetric ground state of the system was originally found in \cite{Cvetic:2001ma}. It takes the form of a stack of deformed fractional D2-branes corresponding to D4-branes wrapped on a two-cycle. The internal manifold is $\CP^3$, seen as an ${\rm S}^2$ fibration over ${\rm S}^4$, which is squashed with respect to the standard Fubini--Study metric (see Appendix \ref{sec:geometry} for details on the geometry). The ten-dimensional metric is supported by the NS three- and RR four-forms and it is regular in the IR, with a collapsing two-cycle and a non-collapsing four-cycle which are ultimately responsible for confinement in the gauge-theory side. For completeness, we give the exact gravitational background in Appendix~\ref{sec:groundstate}. We will refer to this solution as the ``ground state'' in the following. The geometric mechanism for confinement is thus similar to other regular geometries with collapsing cycles in the internal space, like Witten QCD \cite{Witten:1998zw} or Klebanov--Strassler \cite{Klebanov:2000hb}.

This solution is expected to be dual to a three-dimensional gauge theory preserving $\NN=1$ supersymmetry, a non-conformal and less supersymmetric cousin of the ABJ and ABJM theories \cite{Aharony:2008ug,Aharony:2008gk}, without Chern--Simons terms. In the absence of fractional branes, the dual was proposed in \cite{Loewy:2002hu} to be a quiver with $\UU(N)\times \UU(N)$ gauge group together with bifundamental matter - analogous to the Klebanov--Witten quiver in four dimensions \cite{Klebanov:1998hh} or ABJM in three dimensions \cite{Aharony:2008ug} - although the precise details are difficult to pinpoint, essentially due to lack of holomorphicity. The presence of $M$ fractional branes should shift the rank of one of the gauge groups to $\UU(N)\times\UU(N+M)$. 

Fractional branes also produce an unequal running of the gauge couplings. This results in an RG flow similar to the one described by the Klebanov--Strassler solution \cite{Klebanov:2000hb}. There is a cascade of three-dimensional Seiberg-like dualities - of the type introduced in \cite{Karch:1997ux,Aharony:1997gp} - that reduces the rank of the gauge groups when one progresses towards the IR, until one is completely depleted, ending in a $\UU(M)$ gauge theory and confinement.\footnote{It is also possible to modify the model by adding Chern--Simons terms for both gauge groups. The corresponding family of supergravity solutions was found in \cite{Cvetic:2001pga}. Several of its physical properties were analysed in \cite{Hashimoto:2010bq,Faedo:2017fbv,Jokela:2020wgs}, including the fact that the topological interactions spoil confinement - in the sense of a linear quark/antiquark potential - leaving behind merely a mass gap \cite{Faedo:2017fbv}. In this work we will be mainly interested in the physics of confinement, so we will not consider the additional complications due to this deformation. 
}

The type of cascade we just described was first identified in gravity duals of $\NN=3$ deformations of the ABJ theory \cite{Aharony:2009fc}. It exhibits some important differences with respect to the four-dimensional Klebanov--Strassler case, most notably that it involves a finite number of steps so that the ranks of the gauge groups remain finite in the UV. This might be related to the fact that the gauge couplings of both groups are asymptotically free in three dimensions, in contrast to the four-dimensional case where one of the couplings always increases in the UV. We give the details of the cascade for the confining supergravity solutions in Appendix~\ref{sec:cascade}. A consequence of this analysis is that the rank $N$ has to be an integer multiple of $M$ in order for the cascade to be well defined and the background to be regular. Thus, in the following one should think of $M$ as scaling like $N$ in the large-$N$ limit. 

\begin{figure}[t!]
	\begin{center}
		\begin{subfigure}{.47\textwidth}
			\includegraphics[width=\textwidth]{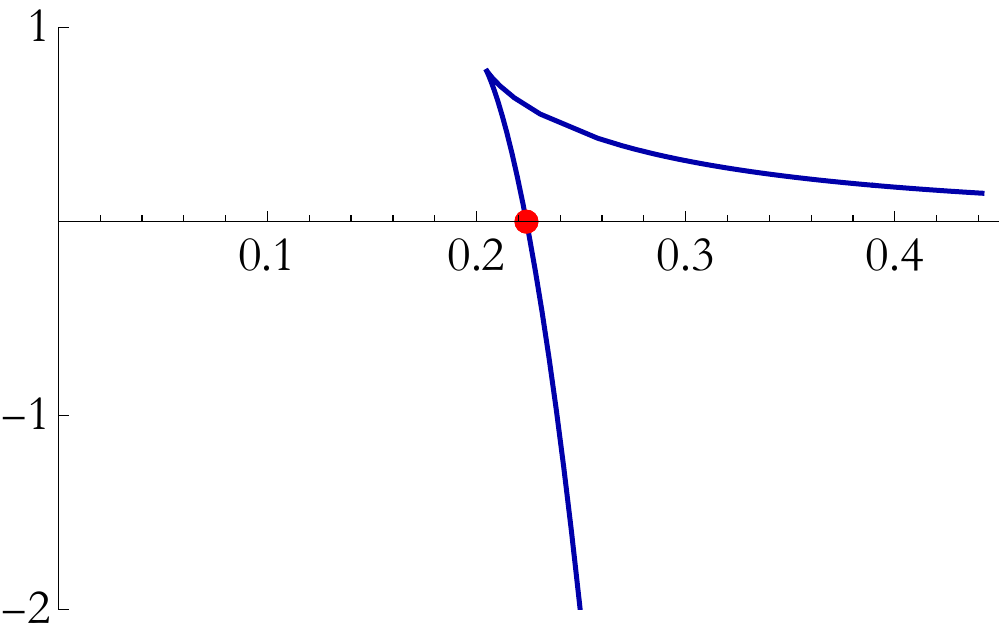} 
			\put(-205,135){ $10^ {3}G/(MN\LQCD^3)$}
			\put(-40,60){ $T/\LQCD$}
		\end{subfigure}\hfill
		\begin{subfigure}{.47\textwidth}
			\includegraphics[width=\textwidth]{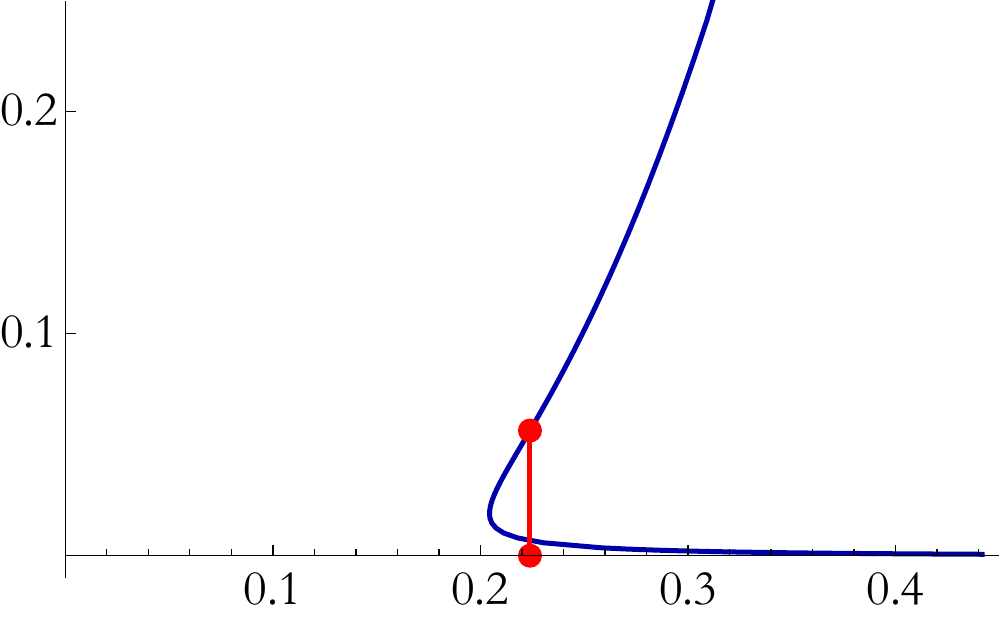} 
			\put(-205,135){ $S/(MN\LQCD^2)$}
			\put(-40,20){ $T/\LQCD$}
		\end{subfigure}
		\caption{\small {\bf Left}: difference in free energy between the deconfined plasma phase and the confining thermal state as a function of the temperature. When the curve is below the horizontal axis the plasma phase (black brane solutions) is dominant. There is a first order confining/deconfining phase transition when the curve crosses the axis. {\bf Right}: corresponding curve for the entropy density as a function of temperature. At the phase transition, the entropy jumps from some finite value to zero, as depicted by the solid red line. We are plotting the quantities in units of the IR scale \eqref{eq:lambdaQCD}.}\label{fig.B8confPhase_transition}
	\end{center}
\end{figure}

The homogeneous phases at finite temperature were found in \cite{Elander:2020rgv}. At low temperatures the dominant solution is a confining thermal state which is obtained by compactifying the Euclidean time of the ground state on a circle. On the other hand, at high temperatures the preferred state is a deconfined plasma phase captured by a black brane solution. These two phases are connected through a first-order confinement/deconfinement transition at a critical temperature
\be
T_{\text{\tiny c}}=0.2239\dots\,\,\lambda\left(\frac{M}{N}\right)^3\,, 
\ee
with $\lambda=\gs\ls^{-1}N$ the three-dimensional 't Hooft coupling, with dimensions of energy. At this temperature the free energy is continuous but the entropy jumps, as shown in Fig.~\ref{fig.B8confPhase_transition}. It is useful to define an IR energy scale associated to the deconfinement transition 
\be\label{eq:lambdaQCD}
\LQCD=\lambda\left(\frac{M}{N}\right)^3\,,
\ee
so that thermodynamic quantities will be measured in units of this scale. It can be seen that smallness of the ten-dimensional curvature in string units in the interior of the geometry requires $M\ll N$, so this scale is parametrically smaller than the scale $\lambda$.  This last determines the transition between weak coupling in the UV and strong coupling in the IR, and determines the region of the geometry where the dual gravity description is appropriate.

\subsection{Confinement of electric charges and screening of monopoles}

The supergravity solution gives some clues about the symmetries of the dual field theory. In the reduction to four-dimensional supergravity there are two gauge fields that couple to D0-branes and D2-branes wrapping a $\CP^1$ cycle
\begin{equation}
A_{\text{\tiny D0}}=C_1, \ \ A_{\text{\tiny D2}}=\int_{\CP^1} C_3\,,
\end{equation}
where $C_1$ and $C_3$ are the one- and three-form RR potentials. The D0 field is massless, but the D2 field is not (details can be found in Appendix~\ref{sec:truncation}), so there is only one conserved current in the dual field theory. Following the arguments in \cite{Bergman:2020ifi}, with Dirichlet boundary conditions for the gauge fields and the $B_2$ form, it follows that fundamental strings, D0-branes, and wrapped D2-branes can end at the boundary. On the other hand, D4-branes wrapping a $\CP^2$ and D6-branes wrapping a $\CP^3$ are not allowed to end at the boundary, so there are no local gauge-invariant dibaryon or baryon dual operators, as expected for the $\UU(N)\times \UU(N+M)$ theory. 

The interpretation of D0-branes and D2-branes ending at the boundary is as local magnetic monopole operators. In the ABJM theory, when the ranks of the groups are equal, a more precise map was proposed in  \cite{Bergman:2020ifi}. If we denote monopole operators as $\MM_{m_1,m_2}$, with $m_1$ and $m_2$ the Abelian magnetic flux in each of the gauge groups, the D0-brane corresponds to an operator with magnetic flux in the diagonal group, $\MM_{1,1}$, while the wrapped D2 corresponds to an operator with magnetic flux in the anti-diagonal $\MM_{1,-1}$. When the Chern--Simons level $k$ is non-zero, the $\MM_{1,-1}$ operator is in a $k$-symmetric representation of the gauge groups, so it is not gauge invariant. This maps in the bulk to the fact that there is a coupling in the D2-brane worldvolume
\begin{equation}
\int_{\text{D2}} A \wedge F_2\sim \left(\int_{\CP^1} F_2\right)\int A\sim k \int A\,, 
\end{equation}
where $A$ is the gauge field on the D2-brane. This induces a charge on the brane that has to be compensated by attaching $k$ fundamental strings. Since in the confining geometry we are considering there is no $F_2$ flux, the Chern--Simons level is zero in the dual theory, and the monopole operator dual to the wrapped D2-brane is gauge invariant.

Given this map between branes and operators, it is natural to interpret the massless $A_{\text{\tiny D0}}$ field as dual to the magnetic global symmetry with topological one-form current 
\begin{equation}
J=\operatorname{tr}_{_N}\,\star {\cal F}_{_N}+\operatorname{tr}_{_{N+M}}\,\star {\cal F}_{_{N+M}},
\end{equation}
with ${\cal F}$ the color field strengths of the $\UU(N)$ and $\UU(N+M)$ groups in obvious notation. We will denote this global symmetry as $\UM$. 

We will present some evidence in the following indicating that the solutions we are considering indeed describe a confining theory and that confinement is likely produced by monopole condensation in three dimensions {\it \`a la} Polyakov, at least for the Abelian component of the gauge group. The monopoles that condense are dual to the wrapped D2-branes, which in our setup are tensionless at the origin of the geometry and carry $n=N/M$ units of D0-brane charge at the boundary (see Appendix~\ref{sec:D2pot} for more details). The D0 charge appears due to the the coupling of the D2-brane to the background $B_2$ form
\begin{equation}
\int_{\text{D2}} C_1 \wedge B_2 \sim \left( \int_{\CP^1} B_2\right) \int C_1\,.
\end{equation}
At the asymptotic boundary of the geometry the integral over $\CP^1$ of $B_2$ gives precisely the coupling of $n$ D0-branes to $C_1$. At the origin of the geometry the integral of $B_2$ vanishes, so possibly the D0 charge is screened as one progresses towards the IR.

To show confinement, we will discuss the potential between electric and magnetic charges showing that while it is linearly increasing for the first, it is screened for the second. The quark-antiquark potential was computed in \cite{Faedo:2017fbv} introducing a rectangular time-like Wilson loop in the usual way. The gravity description is a fundamental string attached to the curve defining the Wilson line at the asymptotic boundary. It was found that the quark-antiquark potential increases linearly with the separation, an indication of confinement. This is possible when the string tension remains finite through all the geometry, as one can easily check it is the case.

\begin{figure}[t!]
\begin{center}
\includegraphics[width=0.5\textwidth]{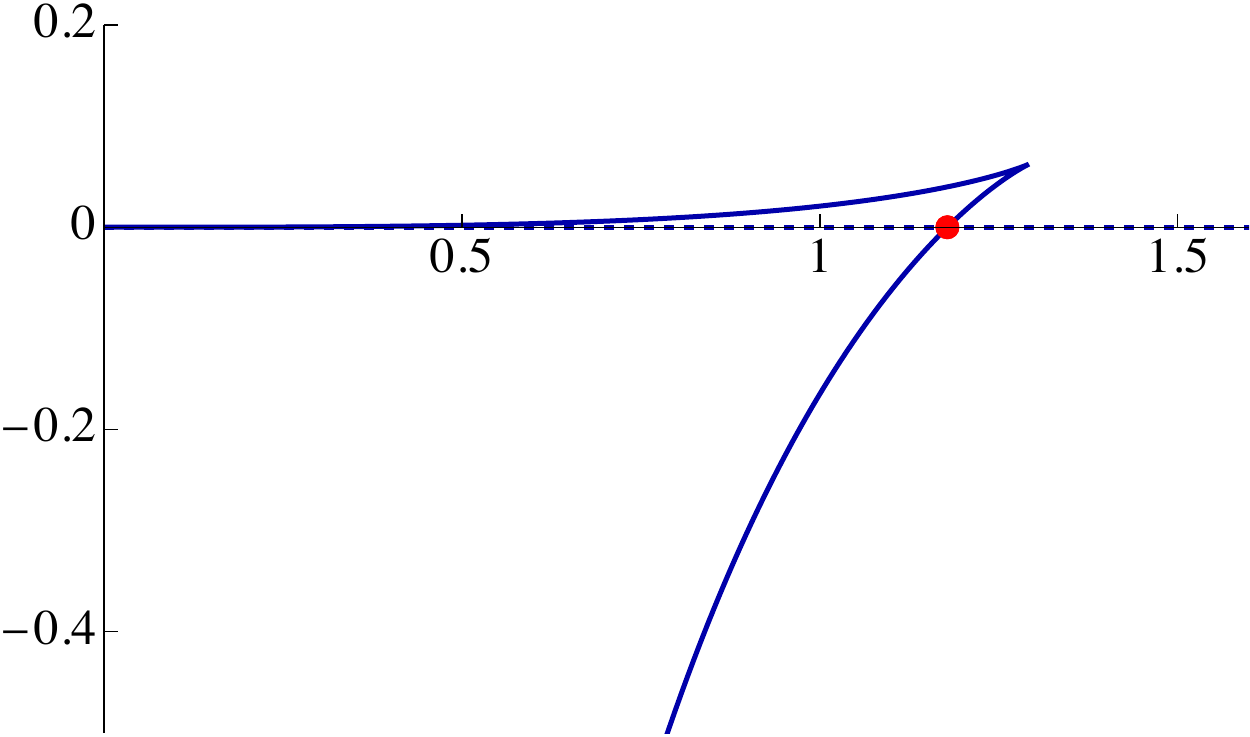} 
\put(-265,120){ $\Delta S_{\text{\tiny D2}}/M$}
\put(-30,62){ $L \LQCD$}
\caption{The action of a wrapped D2-brane, $\Delta S_{\text{\tiny D2}}=S_{\text{\tiny conn}}-S_{\text{\tiny discon}}$, as a function of length. The red dot stands for the transition between connected and disconnected configuration. Technical details of the calculation can be found in Appendix~\ref{sec:D2pot}.}\label{fig:D2pot}
\end{center}
\end{figure}

An indication that there is a monopole condensate is that wrapped D2-branes become tensionless at the origin, since the two-cycle they wrap collapses to zero size. If confinement was produced by monopole condensation as we propose, then we would expect that the monopole-antimonopole interaction would be screened. This can be computed in a similar way to the quark-antiquark potential by introducing a wrapped D2-brane extending between two points separated in a spatial direction at the asymptotic boundary. The action of this configuration, $S_{\text{\tiny D2}}$, gives a contribution to the spatial correlator of the monopole operators $\MM_{\text{\tiny D2}}$ dual to the wrapped D2-branes 
\begin{equation}
\vev{\MM_{\text{\tiny D2}}^\dagger(x) \MM_{\text{\tiny D2}}(0)} \sim e^{-S_{\text{\tiny D2}}}\,.
\end{equation}
Contrary to the fundamental string, there can be a disconnected configuration where spatially separated D2- and anti-D2-branes extend along the radial direction all the way to the origin, since their worldvolume ends smoothly where the two-cycle collapses to zero size. In Fig.~\ref{fig:D2pot} we show the difference in action between the connected and disconnected configurations, as a function of the asymptotic separation between the branes. The details of the calculation are collected in Appendix~\ref{sec:D2pot}. We can observe that for separations  $L \gtrsim 1.178\, \LQCD^{-1}$ the disconnected configuration dominates, after the difference in action vanishes. Moreover, for $L\gtrsim 1.292\,\LQCD^{-1}$, the connected configuration ceases to exists after meeting with an unstable branch.

In the following sections we will introduce a magnetic field for the monopoles dual to the D0-branes. Since the wrapped D2-brane carries non-zero D0 charge, this could affect the monopole condensate, possibly even completely depleting it for large enough magnetic fields, in which case we expect that the theory would no longer be confining. As we will see, the phase diagram we find seems to confirm this picture.

\section{Solutions at nonzero monopole magnetic field}\label{sec:newsols}

If monopole condensation is indeed behind confinement, as the arguments of the previous section suggest, we should expect a deconfinement transition when the cost of introducing a monopole becomes too large. We will discuss now how to do this within the supergravity description.

As we have argued, the dual of a monopole is a D2-brane wrapped on $\CP^1$, which close to the asymptotic boundary has an action
\begin{equation}
S_{\text{\tiny D2}}=M\left(\hat{S}_{\text{{\tiny DBI}}} -\frac{1}{\gs \ls} \frac{N}{M^2} \int C_1\right)\,,
\end{equation}
where we have extracted an explicit $M$ factor from the DBI action, in such a way that $\hat{S}_{\text{{\tiny DBI}}}$ is dimensionless and has no explicit dependence on parameters of the field theory like the 't Hooft coupling or the ranks of the groups (see Appendix~\ref{sec:D2pot} for more details). We can modify the action of the wrapped D2-brane by turning on the $C_1$ form, whose components along the field theory directions correspond, in the dual theory, to the conserved $\UM$ current and an external gauge field coupled to it
\begin{equation}\label{eq:physvec}
C_1=\gs \ls \frac{M^2}{N} \mathsf{A}_\mu \dd x^\mu\,,
\end{equation}
where the normalization is chosen to match the DBI part of the wrapped D2-brane action. We can thus modify the action of the monopole-antimonopole configuration by switching on the spatial components of $\mathsf{A}_\mu$. In particular, a $\UM$ magnetic field can be characterized in a gauge-invariant way and preserves rotational invariance, simplifying the analysis significantly. Furthermore, one might expect that a magnetic field would effectively gap charged degrees of freedom, which are the monopoles, potentially triggering a deconfinement transition. We will see in Sec.~\ref{sec:thermo} that this scenario is indeed realized for the solutions we will construct in this section.

\subsection{Ten--dimensional ansatz and equations}\label{sec:10D}

All the solutions we will discuss are completely regular in ten-dimensional type IIA supergravity. The internal geometry is a particular realization of $\CP^3$ whose characteristics are given in Appendix~\ref{sec:geometry}. In the following we use the notation stated there. The solutions we are interested in asymptote to a stack of coincident color D2-branes in the UV, like the ground state and thermal backgrounds discussed in Section~\ref{sec:review}. Inspired by this, we take the string-frame ansatz for the metric and dilaton as
\begin{eqnarray} \label{eq:ansatz_metric}
\dd s_{\rm st}^2 &=&h^{-\frac12}\left(-\mathsf{b}\ \dd t^2 + \dd x_1^2 + \dd x_ 2^2\right)+h^{\frac12} \left(\frac{\dd r^2}{\mathsf{b}}+e^{2f}\dd\Omega_4^2+e^{2g}\left[\left(E^1\right)^2+\left(E^2\right)^2\right] \right)\,,\nonumber\\[2mm]
e^\Phi&=&h^{\frac14} \, e^\Lambda \,,
\end{eqnarray}
with $f$, $g$, $h$, $\mathsf{b}$ and $\Lambda$ depending only on the radial coordinate $r$. The function $\mathsf{b}$ in \eqref{eq:ansatz_metric} breaks three-dimensional Lorentz invariance in the directions parallel to the color branes whenever it is non-constant. This is necessary both because we will look for black brane solutions and because charge or magnetic fields will be turned on. The internal space is topologically ${\rm S}^2\times {\rm S}^4$, with the forms $E^1$ and $E^2$ spanning the two-cycle, of volume $X_2=E^1\wedge E^2$. This can be combined with another two-form $J_2$ to produce a closed form $\dd\left(X_2-J_2\right)=0$. More details about the internal geometry can be found in Appendix \ref{sec:geometry}. 
 
Our conventions for type IIA supergravity are such that the forms satisfy the Bianchi identities
\begin{equation}\label{eq:IIABianchis}
\dd H_3=0\,,\,\qquad\qquad\quad \dd F_2=0\,,\qquad\qquad\quad \dd F_4=H_3\wedge F_2\,,
\end{equation}
as well as the string-frame equations of motion 
\begin{equation}\label{eq:IIAeoms}
\begin{aligned}
\dd*F_4+H_3\wedge F_4 &=  0\,,\\[2mm] 
\dd*F_2+H_3\wedge * F_4 &= 0\,,\\[2mm]
\dd\left(e^{-2\Phi}*H_3\right) -F_2\wedge*F_4-\frac12 F_4\wedge F_4&= 0\,.
\end{aligned}
\end{equation}
The Bianchi identities are solved by the ansatz for the forms
\begin{equation}
H_3 = \dd B_2\,,\qquad\qquad F_2 = \dd C_1+F_2^{\rm fl}\,,\qquad\qquad F_4 = \dd C_3+B_2\wedge F_2+F_4^{\rm fl}\,,
\end{equation}
where we have introduced the closed but non-exact terms
\begin{equation}{\label{eq:fluxes1}}
F_2^{\rm fl}=Q_k\,\left(X_2-J_2\right)\,,\qquad\qquad\qquad F_4^{\rm fl}=q_c\,\left(J_2\wedge J_2-X_2\wedge J_2\right)\,.
\end{equation}
in correspondence with the two- and four-cycle respectively. It is instructive to explain the ansatz for the forms in relation to the consistent truncation and the four-dimensional model of Appendix~\ref{sec:truncation}. First of all, from the two fluxes \eqref{eq:fluxes1} allowed by the geometry, we take the two-form to vanish, that is, $Q_k=0$. This is related to the fact that we do not want to include a Chern--Simons term in the dual gauge theory (see Eq.~\eqref{eq:gaugeparam} and the discussion below it). The four-form flux on the other hand must be non-vanishing, so $q_c\ne0$. This constant corresponds to the Page charge of D4-branes and it is therefore quantized. It is related to the shift in the UV rank of one of the gauge groups in the dual field theory $M$ as
\begin{equation}\label{eq:qc}
q_c = \frac{3\pi \ls^3 \gs}{4} \,M\,.
\end{equation}

The ansatz for the form potentials uses the left-invariant forms defined in Appendix~\ref{sec:geometry}. It reproduces the one used in the reduction to four dimensions, Eq.~\eqref{eq:formpot}, that is
\begin{equation}\label{eq:formansatz}
\begin{aligned}
B_2 &= b_2+b_X\,X_2+b_J\, J_2\,,\\[2mm]
C_1 &= a_1\,,\\[2mm]
C_3 &= a_3+\at\wedge X_2+\ah\wedge J_2+a_X \,X_3+a_J \, J_3\,,
\end{aligned}
\end{equation}
The functions $b_X$, $b_J$ and $a_J$ are scalars from the four-dimensional point of view and depend solely on the radial coordinate $r$ in \eqref{eq:ansatz_metric}. They are crucial to resolve the IR geometry of the ground state \cite{Faedo:2017fbv} and were non-trivial also in the thermal solutions in \cite{Elander:2020rgv}. The additional scalar $a_X$ is an axion - it appears in the equations always acted on by a derivative - giving mass to a vector (see Eq.~\eqref{eq:axions}) so it can be fixed to any constant value, in particular $a_X=0$. 

The rest of the terms are new with respect to \cite{Faedo:2017fbv,Elander:2020rgv}. There are three vectors (one-forms) that we parametrize as 
\begin{equation}
a_1 = a_t\left(r\right)\,\dd t+\gs\ls\,\frac{M^2}{N}\,\frac{\Bphys}{2}\left(x_1\dd x_2-x_2\dd x_1\right)\,,\qquad\quad\tilde{a}_1 = \tilde{a}_t\left(r\right)\,\dd t\,,\qquad\quad\hat{a}_1 = \hat{a}_t\left(r\right)\,\dd t\,,
\end{equation}
for some constant $\Bphys$. The prefactor has been chosen so that it corresponds to the physical magnetic field associated to the vector $\mathsf{A}$, defined in Eq.~\eqref{eq:physvec}. As reflected in \eqref{eq:eoma1b}, this is a massless vector and therefore associated to a U(1) symmetry. On the other hand, according to Eq.~\eqref{eq:axions}, the combinations $\tilde{a}_1-\hat{a}_1$ and $\tilde{a}_1+\hat{a}_1$ are Stueckelberg-coupled to axions and thus massive. They do not admit a magnetic field.  

We are then left with the three- and two-forms $a_3$ and $b_2$. The only non-trivial components allowed by the symmetries we want to preserve are
\begin{equation}
a_3=a_{t12}\left(r\right)\dd t\wedge\dd x_1\wedge\dd x_2\,,\qquad\qquad b_2=b_{12}\left(r\right)\dd x_1\wedge\dd x_2\,.
\end{equation}
Notice that the three-form lives in the external four-dimensional space and therefore it is non-dynamical. Indeed, it can be dualized to a constant $Q_c$ as explained in Eqs.~\eqref{eq:fieldstr}, \eqref{eq:f4der} and \eqref{eq:f4dual}. This is manifested in the relation 
\begin{equation}
a_{t12}'  = - \left[ \frac{e^{-4 f-2 g}}{h^2} \left[4a_J\left(b_X+b_J\right)+2q_c\left(b_X-b_J\right)+Q_c  \right] +\,  b_{12} \,a_t'   \right]\,,
\end{equation}
imposed by the equation of motion for $F_4$. Ultimately, the constant has to vanish due to regularity conditions in the IR of the geometry, as argued around Eq.~\eqref{eq:Qc0}. In the following we fix $Q_c=0$. 

Finally, $b_2$ can be dualized to an axion $a$, defined in Eq.~\eqref{eq:b2dual}, which gives mass to the combination of vectors $\tilde{a}_1-\hat{a}_1$. This means that the two-form does not contain independent degrees of freedom. Taking the axion to vanish, this is reflected in the relation
\begin{equation}
b_{12}' \ =\  -\frac{e^{-4 f-2 g+2 \Lambda }}{\mathsf{b} \,  h} \, \left[ \, 4 a_J\left(\tilde{a}_t+\hat{a}_t\right) +  2 q_c \left(\tilde{a}_t-\hat{a}_t\right)\right]\,,
\end{equation}
which is deduced from \eqref{eq:b2dual} evaluated in our ansatz. This can be used to eliminate $b_{12}$ (which always appears with a derivative) in the rest of the equations. The content of \eqref{eq:IIABianchis} and \eqref{eq:IIAeoms} is then the dynamics of three scalars and three vector potentials, two of them massive. 

The equation of motion for the dilaton is in our conventions 
\begin{eqnarray}
R+4\nabla_M\nabla^M \Phi - 4 \nabla^M\Phi \nabla_M \Phi-\frac{1}{12} H^2 = 0\,,
\end{eqnarray}
while Einstein's equations read
\begin{equation}
R_{MN} + 2 \nabla_M\nabla_N \Phi -\frac{1}{4} H_{MN}^2 = e^{2\Phi}\left[
\frac{1}{2} (F_2^2)_{MN} + \frac{1}{12}(F_4^2)_{MN} - \frac{1}{4}g_{MN}\left(
\frac{1}{2}F_2^2 +\frac{1}{24}F_4^2\right)\right]\,, 
\end{equation}
in a self-explanatory notation.  From this set, one gets second order differential equations for the functions in the metric and dilaton, together with a first order constraint. Overall, the system consists of 11 second order differential equations subject to a first order constraint for the set of 11 functions $\{ f,\, g,\,  \Lambda,\, h,\, \mathsf{b},\, b_J,\, b_X ,\, a_J ,\, a_t ,\, \tilde{a}_t,\, \hat{a}_t \}\,$.

\subsection{Solutions and expansions}\label{sec:numsols}

In the following we will describe the main steps we took to construct the solutions at nonzero monopole magnetic field and describe their properties. A complete analysis can be found in Appendix~\ref{sec:numerics}.

We will discuss two types of new backgrounds at finite charge/magnetic field. One type, the ``confining solutions'', ends smoothly when the two-cycle of the internal space collapses to zero size, similarly to the zero magnetic field ground state. The other are black brane solutions with a regular non-extremal horizon. Their thermodynamic properties will be explained in Sec.~\ref{sec:thermo}. These solutions were found numerically by means of a shooting method. In a few words, we specified the desired boundary conditions both at the asymptotic boundary of spacetime (the UV) and at the origin/horizon (the IR) by using series expansions in the radial coordinate near these two regions. Such series are written in terms of a set of undetermined coefficients, which are fixed by demanding continuity and differentiability of the numerical solution at some intermediate point in the bulk.

It is advantageous to work with a dimensionless radial coordinate $\xi$ such that
\begin{equation}\label{eq:coordinate}
\dd r = - \frac{\rho_0}{\xi^2\sqrt{1-\xi^4}}\dd\xi\,.
\end{equation} 
where $\rho_0$ is some constant with dimensions of length whose precise value will be determined presently. In this coordinate, the asymptotic boundary is at $\xi\to 0$ and the origin/horizon at $\xi\to 1$ for the confining solutions and at $\xi=\xi_h<1$ for the black brane solutions.

Near the boundary we impose that all our solutions have the same leading D2-brane asymptotics, coincident with the ground state, except for the control parameters, which are in this case the magnetic field and chemical potential. In this way we ensure, through the holographic dictionary, that the gauge theory we consider is not modified in the UV. This condition fixes the value of $\rho_0$ to\footnote{The arguments leading to this relation are analogous to those resulting in Eq.~\eqref{eq:rho0}, which is recovered for the ground state value of $b_0$.}
\begin{equation}\label{eq:ratiobrho}
	\rho_0 =  |b_0| \frac{\ls^ 2}{2}\,\lambda\,\frac{M^2}{N^ 2}\,,
\end{equation}
where $b_0$ is a dimensionless constant that is determined by imposing regularity of the confining solutions. At vanishing magnetic field its value is $b_0(\Bphys=0)=-3 K(-1)$, with $K(m)$ the complete elliptic integral of the first kind. It turns out that $b_0<0$ for all the solutions we have found. For black brane solutions the value of $b_0$ can be taken to be the same as the ground state solution.

The boundary expansions of the metric and dilaton take the form
\be\label{eq:boundexpmet}
\begin{split}
&e^{2f}  =   \frac{\rho_0^2}{2\xi^2} \left( 1 +\cdots  + f_5 \xi ^5   \right)+\cdots\,,  \qquad\qquad 	e^{2g} = \frac{\rho_0^2}{4\xi^2}+\cdots\,,  \\	
&\,\,\,\,\, h = \frac{128\, q_c^2 }{15\rho_0^ 6}|b_0|\,\xi^5+\cdots \,,\qquad\qquad \mathsf{b} =  1  + \mathsf{b}_5 \xi^5+\cdots \,, \qquad \qquad e^\Lambda= 1+\cdots\,,
\end{split}
\ee
where we have showed only the leading terms and the independent subleading coefficients appearing later in the expressions for thermodynamic quantities. Similarly, for the scalars the expansions are 
\be\label{eq:boundexpscal}
b_J =\frac{2q_c}{3\rho_0 }\, b_0+\cdots\,, \qquad\qquad b_X = -  \frac{2q_c}{3\rho_0 }\, b_0 +\cdots\,, \qquad\qquad a_J = \frac{q_c}{6} +\cdots\,.
\ee
Finally, the vector potentials are written as
\begin{equation}\label{eq:boundexpgauge}
a_t  = \frac{\rho_0^3}{q_c}  \, (v_0+v_1\xi)+\cdots,  \qquad	\hat{a}_t = \rho_0^2 \frac{2b_0v_1}{15}\xi+\cdots , \qquad	\tilde{a}_t = -\rho_0^2 \frac{2b_0v_1}{15}\xi+\cdots  \,. 
\end{equation}
It is convenient also to introduce the dimensionless magnetic field $\BET$ defined as 
\begin{equation}\label{eq:dimlessB}
\Bphys = \frac{1}{\gs\ls}\,\frac{N}{M^2}\,  \frac{\rho_0^5}{q_c^2}\, \BET=\frac{|b_0|^5}{18\pi^2}\LQCD^2 \BET\,.
\end{equation}
The dimensionful factors in \eqref{eq:boundexpmet}-\eqref{eq:dimlessB} have been chosen in such a way that the coefficients $f_5$, $\mathsf{b}_5$, $v_0$ and $v_1$ are dimensionless and moreover $\rho_0$ and $q_c$ drop from the equations of motion.

The boundary expansion of the warp factor determines the number $N$ of D2-branes as follows
\begin{equation}
	h \simeq \frac{16}{5}\, \frac{\QD}{r^ 5}\,,\qquad \qquad \QD = 3\pi^2 \ls^5 g_s N\,.
\end{equation}
Comparing with \eqref{eq:boundexpmet}, one can check, after changing coordinates according to \eqref{eq:coordinate}, that this is consistent with the identifications \eqref{eq:ratiobrho} and \eqref{eq:qc}. Among the coefficients that appear in the expansions, $f_5$ and $\mathsf{b}_5$ in \eqref{eq:boundexpmet} determine the expectation value of the energy-momentum tensor in the dual field theory, while $v_0$ and $v_1$ in \eqref{eq:boundexpgauge} are related to the value of the chemical potential and the charge density. To be more precise, the charge density is actually not only determined by $v_1$ but it also receives additional contributions depending on the magnetic field, as we will see in more detail later.

Let us now discuss the two different IR boundary conditions. These correspond to two different phases in the dual gauge theory.
\\[.5ex]

\paragraph{Magnetized confining phase} {\ }\\[1ex]

\begin{figure}[t!]
	\begin{center}
		\includegraphics[width=.65\textwidth]{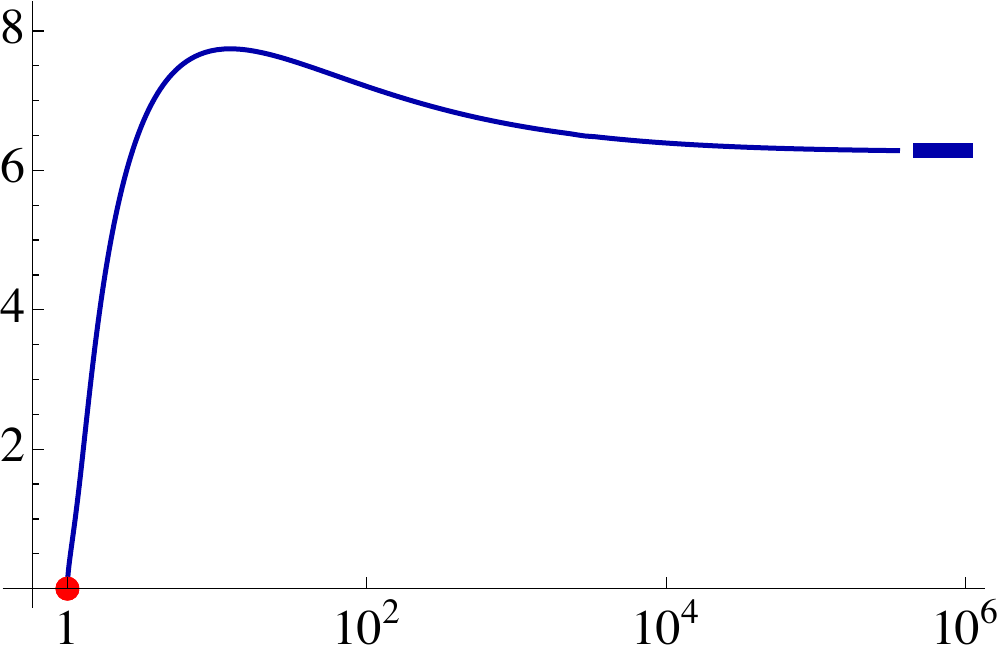} 
		\put(-280,190){$\Bphys/\LQCD^2$}
		\put(-30,30){$\icli$}
		\caption{\small Magnetic field, in units of the confinement scale \eqref{eq:lambdaQCD}, as a function of the IR value of the dilaton factor $e^\Lambda$. The ground state corresponds to $\cli = 1$ and $\Bphys = 0$, represented by the red dot. The magnetic field reaches a maximum value $\Bphys_{\text{\tiny max}} \approx 7.744\,\LQCD^2$ and saturates to $\Bphys_{\text{\tiny sat}} \approx2\pi\,\LQCD^2$ (thick segment) as $\cli\to0$.} \label{fig.bounded_magneticField}
	\end{center}
\end{figure}

One of our main results is that the confining phase is still present when the magnetic field is gradually turned on, but it disappears when the magnetic field is large enough. In the gravitational dual this means that there is a solution at finite magnetic field with similar IR boundary conditions as the ground state, meaning that the two-sphere collapses smoothly while ${\rm S}^4$ keeps a finite size (see Eq.~\eqref{eq:IRgeom}).  This implies that $e^{2g}\to 0$ when $\xi\to 1$, while $e^{2f}$, $h$ and the dilaton remain constant. 

We impose regularity on the solutions, which in this case also implies that the form potentials on the collapsing cycle - described by the form $X_2$ - as well as the associated radial flux, should vanish when $\xi\to 1$. Given the ansatz for the fluxes  \eqref{eq:formansatz}, this means that $b_X$, $\at$, and their field strengths should vanish in this limit. We also use the gauge freedom to set $a_t=0$ at $\xi=1$, although it can be shifted to an arbitrary value at the cost of shifting $v_0$ in \eqref{eq:boundexpgauge}. This freedom indicates that the solutions are actually independent of the chemical potential, which can be set to an arbitrary value. Notice as well that nothing prevents us from compactifying the (Euclidean) time direction, as was done with the ground state in \cite{Elander:2020rgv}, so the solutions we discuss in the following should also be thought of as having an arbitrary temperature. The details of the expansion verifying all these conditions can be found in Appendix~\ref{sec:conf_newsols}. 

When solving the equations through the shooting method, it turns out that the magnetic field itself is not the most convenient label for this family because for values of $\Bphys$ in certain interval there are two branches of solutions. An appropriate parameter is for instance $\cli$, the value at $\xi=1$ of the factor $e^\Lambda$ that enters in the dilaton ansatz \eqref{eq:ansatz_metric}. This parameter turns out to be restricted to the range $0<\cli\le1$. The values of the magnetic field in this interval are shown in Fig.~\ref{fig.bounded_magneticField}, where the horizontal axis represents $\icli $. The value $\cli=1$ corresponds to the ground state solution in the absence of magnetic field and is represented by a red dot on the figure. As $\cli$ is decreased the magnetic field grows up to a maximum value $\Bphys_{\text{\tiny max}} \approx 7.744\,\LQCD^2$, reached when $\cli \approx 0.0818$. Then it decreases, saturating at $\Bphys_{\text{\tiny sat}} \approx2\pi\,\LQCD^2$ in the limit $\cli\to0$. This gives rise to two branches of confining solutions for $\Bphys\in\left(\Bphys_{\text{\tiny sat}},\,\Bphys_{\text{\tiny max}}\right)$. Moreover, this points towards the existence of additional solutions that complete the phase diagram for values of the magnetic field beyond $\Bphys_{\text{\tiny max}}$. 

\begin{figure}[t!]
	\begin{center}
		\begin{subfigure}{.47\textwidth}
			\includegraphics[width=\textwidth]{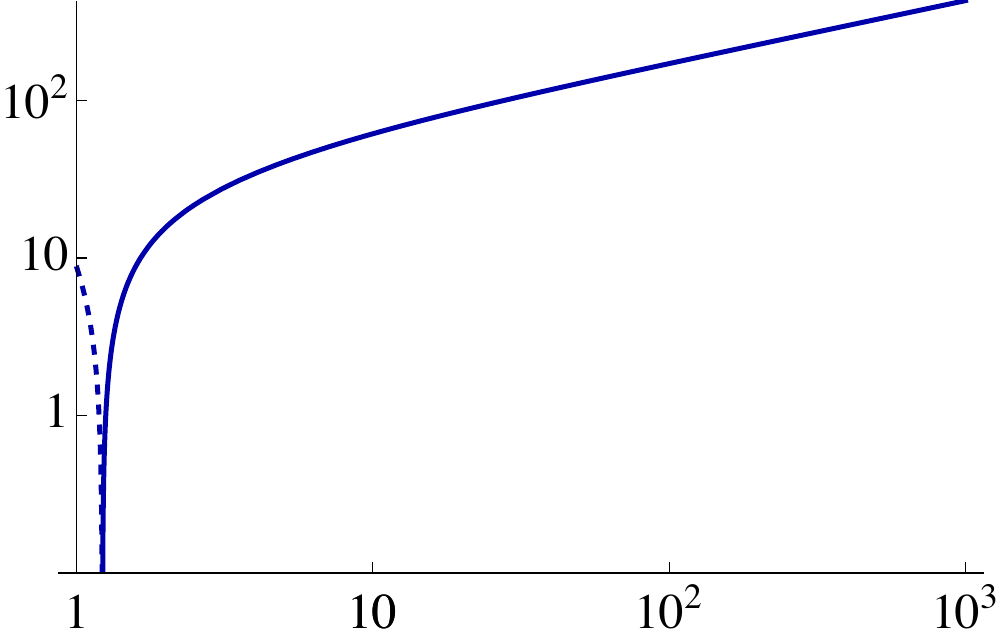} 
			\put(-180,130){$\frac{N}{M}\ls^2\cdot \mathrm{R}_{\text{\tiny10D}}$}
			\put(-30,25){${\icli}$}
		\end{subfigure}\hfill
		\begin{subfigure}{.47\textwidth}
			\includegraphics[width=\textwidth]{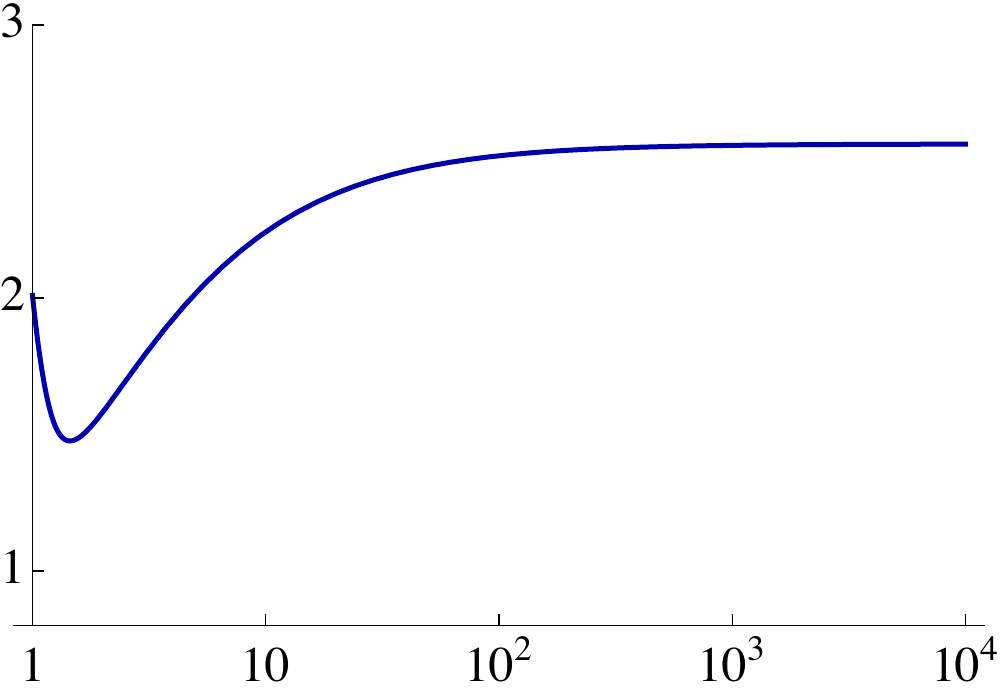} 
			\put(-185,130){$M^ {\frac{2}{3}} \ell_p^2 \cdot \mathrm{R}_{\text{\tiny{11D}}}$}
			\put(-30,25){${\icli}$}
		\end{subfigure}
		\caption{\small {\bf Left}: Absolute value of the ten-dimensional Ricci scalar evaluated at the origin of the geometry, in string units, as a function of the IR value of the dilaton factor $e^\Lambda$.  
		The solid curve stands for solutions with negative IR Ricci scalar, whereas for the dashed curve it is positive. {\bf Right}: Eleven-dimensional Ricci scalar evaluated at the origin of the geometry, in Planck units, in terms of the same parameter.}  
		\label{fig.curvatures}
	\end{center}
\end{figure}

Finally, it is worth stressing that all these confining solutions are regular at the origin, as the S$^2$ shrinks smoothly and the transverse topology becomes that of an $\R^3$ bundle over ${\rm S}^4$, exactly as in the ground state geometry \eqref{eq:IRgeom}. In particular, the curvature invariants evaluated at the IR are finite. As $\cli$ decreases and the magnetic field saturates to $\Bphys_{\text{\tiny sat}}$, some of these curvature invariants grow, as seen in the left panel of Fig.~\ref{fig.curvatures}. This could be signalling that an IR singularity is developing. However, all these solutions can be uplifted to eleven-dimensional supergravity. It turns out that in this limit the eleven-dimensional curvatures stay finite and saturate to a constant value (see right panel of Fig.~\ref{fig.curvatures}).

\begin{figure}[t!]
	\begin{center}
		\includegraphics[width=.65\textwidth]{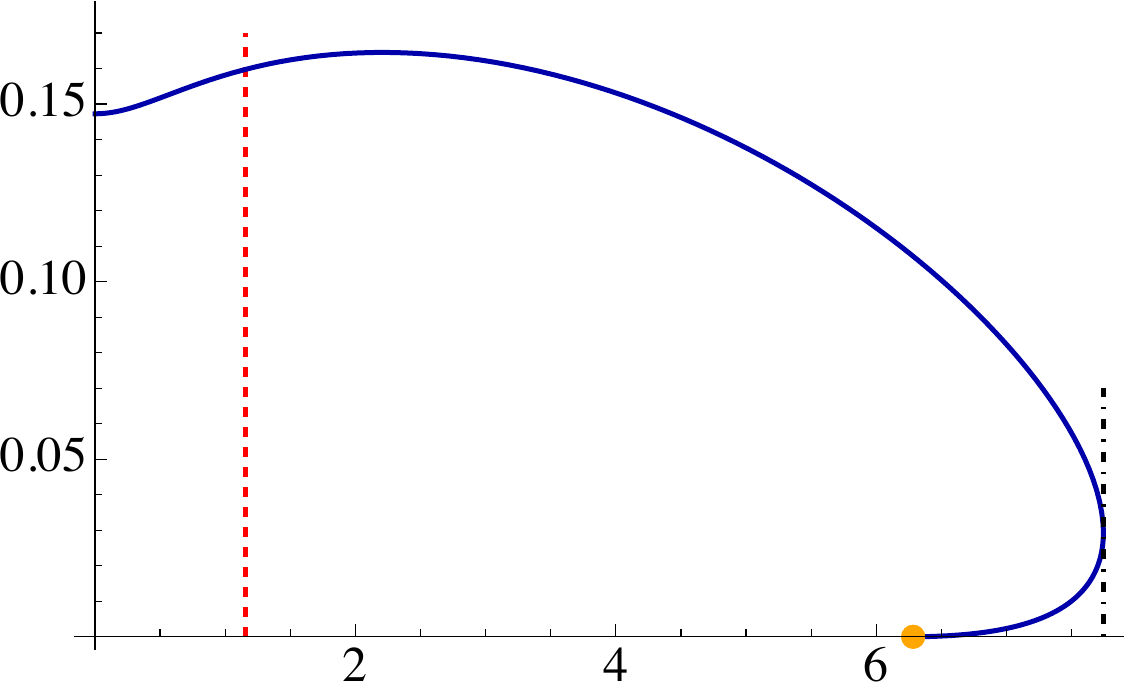} 
		\put(-285,180){$N^2\sigma/(M^2\LQCD^2)$}
		\put(-30,0){$\Bphys/\LQCD^2$}
		\caption{\small String tension in units of the confinement scale, computed from the quark/antiquark potential, as a function of the magnetic field. The red dashed line indicates the magnetic field at which the confining backgrounds cease to be dominant. The black, dot-dashed line corresponds to $\Bphys_{\text{\tiny max}}$ while the orange dot labels $\Bphys_{\text{\tiny sat}}$.} \label{fig.StringTension}
	\end{center}
\end{figure}

This hints that the family of backgrounds with these IR boundary conditions may be continuously connected to another branch of solutions, which would be singular from the ten-dimensional point of view but regular in eleven dimensions,\footnote{This is not an uncommon feature and happens for instance if one turns on a Chern--Simons interaction in the gauge theory dual, as detailed in \cite{Faedo:2017fbv}.} perhaps extending to values of the magnetic field above the maximum we encountered. We have checked that the M-theory circle shrinks as the limiting magnetic field is approached. The arguments in \cite{Faedo:2017fbv} then suggest that this putative branch of solutions may be gapped but non-confining. This picture is supported by the behavior of the string tension computed from the quark/antiquark potential, which vanishes as $\cli\to0$ and the limiting value of the magnetic field is approached. This can be seen in Fig.~\ref{fig.StringTension}.

In this work we focus on the phase diagram at nonzero temperature. As we will see in the next section the branch that reaches the limit $\Bphys\to\Bphys_{\text{\tiny sat}}$ is always thermodynamically disfavoured, so we leave the construction of this new branch at vanishing temperature for future studies. 

\paragraph{Magnetized plasma phase} {\ } \\[1em]
In order to complete the phase diagram at finite temperature we will construct black branes at non-zero magnetic field. These are dual to deconfined plasma states in the field theory side. The existence of a horizon is encoded in a simple zero of the blackening factor $\mathsf{b}$. Regularity for the rest of the functions imposes that they reach a finite value, with the additional condition that the time components of the vector potentials must vanish at the horizon (see Appendix~\ref{sec:plasma_newsols}). The leading terms in the expansion of the metric and dilaton are
\be\label{eq:exphormet}
\begin{split}
&e^{2f} =\rho_0^2\,  \cfh+\cdots\,,  \qquad	e^{2g} = \rho_0^2 \,\cgh+\cdots\,,  \qquad
h = \frac{128\, q_c^2 }{9\rho_0^ 6}\, \chh+\cdots \,,\\ &\mathsf{b}= \cbh (\xi-\xi_h)+\cdots\,,\qquad e^{\Lambda}= \clh+\cdots\,.
\end{split}
\ee
The values of the dimensionless coefficients $\cfh,\cgh,\chh,\cbh,\clh$ determine the value of physical properties such as the entropy and temperature. In this case we decided to fix the parameter $b_0$ - and accordingly $\rho_0$ through Eq.~\eqref{eq:ratiobrho} - to its ground-state value, $b_0=-3K(-1)$, as in \cite{Elander:2020rgv}. Then, in the shooting procedure all the coefficients of the IR and UV expansions are fixed except for three control parameters: $v_0$, $\xi_h$ and $\BET$ or, equivalently, the chemical potential, the temperature and the magnetic field. We have thus a three-parameter family of black branes. Their thermodynamic properties are investigated in the next section.

\section{Thermodynamics and the phase diagram}\label{sec:thermo}

In this section we discuss the main thermodynamic properties of the different solutions we have constructed. We uncover an interesting structure of phase transitions in the temperature and monopole magnetic field plane, focusing on the case of vanishing chemical potential. The map between parameters of the gravity solution and field theory quantities is collected in Table~\ref{tab:dictionary} of Appendix~\ref{sec:groundstate}.

\subsection{Thermodynamic quantities and relations}\label{sec:thermoexpressions}

The renormalized four-dimensional bulk action $\Iren$ describing the system is obtained in Appendix \ref{sec:holoren}. The free-energy density $G$ of a particular state is given in terms of its Euclidean on-shell value by
\begin{equation}\label{eq:freeEnergy}
G = -\frac{\Iren}{\beta V_2}\,,
\end{equation}
with $V_2$ the (infinite) volume in the spacial directions and $\beta$ the period of the compact Euclidean time. It is related to the temperature through $T=1/\beta$.

Ultimately, the final expression for the free-energy density \eqref{eq:freeEnergy} in terms of the parameters of the solution depends on the particular phase we are considering. For any confining solution, it gets no IR contribution (see Eq.~\eqref{eq:Iren} and comments below) and the final result is
\begin{equation} \label{eq.freeEnergyConf}
{G}_{\text{\tiny conf}} \ = \ \frac{\rho_0^5}{2\kappa_4^ 2} \, \left(-\frac{7 \mathsf{b}_5}{2}\, -  f_5\right)= NM\LQCD ^ 3 \frac{(-b_0^ 5)}{3\cdot 2^{11} \pi^ 4}\left(-\frac{7 \mathsf{b}_5}{2}\, -  f_5 \right)\,,
\end{equation}
where in the second equality we have used the relation \eqref{eq:ratiobrho} and substituted the four-dimensional Newton's constant \eqref{eq:4dkappa}. On the other hand, in the plasma phase of Sec.~\ref{sec:plasma_newsols}, there is an additional contribution from the horizon of the black brane 
\begin{equation} \label{eq.freeEnergyPlasma}\begin{aligned}
{G}_{\text{\tiny plas}}  
&= NM\LQCD ^ 3  \frac{(-b_0^ 5)}{3\cdot 2^{11} \pi^ 4}\left[-\frac{7 \mathsf{b}_5}{2} -  f_5 + \frac{16 \cbh \cfh^2 \cgh \xi_h^2 (1-\xi_h^4)^{\frac{1}{2}}}{{\clh }^2}+
\frac{64}{135} b_0^2 v_0 (20 \BET b_0+27 v_1)\right].\\[2mm]
\end{aligned}
\end{equation}
In this paper we work in the grand canonical ensemble: if there are different states at the same temperature, chemical potential and magnetic field, the preferred one will be that with the lowest free-energy density. In our conventions, $G=0$ corresponds to the free-energy density of the supersymmetric ground state, which has $T=0$ and vanishing magnetic field.

As usual, for black brane solutions the temperature and entropy density are quantities obtained from horizon data. The former is determined from the requirement that the solution has no conical singularity at the horizon. The latter is the Bekenstein--Hawking entropy, given by the area of the horizon. In terms of the expansion parameters they read
\begin{equation}
\begin{aligned}\label{eq:entropy_temperature}
S_{\text{\tiny plas}} &= \frac{\rho_0^3 q_c}{2\kappa^2_4} \, \frac{512 \sqrt{2} \pi  \cfh^2 \cgh \chh^{\frac{1}{2}}}{3 \clh^2} =NM\LQCD ^ 2\frac{(-b_0^3)}{2^{11}\pi^3} \, \frac{512 \sqrt{2} \pi  \cfh^2 \cgh \chh^{\frac{1}{2}}}{3 \clh^2}\,,\\[2mm]
T_{\text{\tiny plas}}&= - \frac{\rho_0^2}{q_c}\, \frac{3 \cbh \xi_h^2 (1-\xi_h^4)^{\frac{1}{2}}}{32 \sqrt{2} \pi \chh^{\frac{1}{2}}} = - \LQCD\frac{b_0^2}{3\pi } \,  \frac{3 \cbh \xi_h^2 (1-\xi_h^4)^{\frac{1}{2}}}{32 \sqrt{2} \pi \chh^{\frac{1}{2}}}\,. 
\end{aligned}
\end{equation}
Notice that the entropy of the plasma grows as $NM\sim N^2$ in the large-$N$ limit, as expected for a deconfined phase of a Yang--Mills theory. On the other hand, the entropy vanishes in the absence of a horizon, so for the confining phase $S_{\text{\tiny conf}} = 0$. Similarly the temperature $T_{\text{\tiny conf}}$ is arbitrary, since there is no condition fixing the period of the time coordinate.

Our ansatz admits solutions with an external chemical potential $\mu$. For the black brane solutions it is fixed by the asymptotic value of $\mathsf{A}_t$
\begin{equation} \label{eq:chemical potential}
\mu_{\text{\tiny plas}} =\frac{N}{M^2}\,\frac{1}{\gs\ls}\, \frac{\rho_0^3}{q_c}\,v_0=\LQCD\frac{(-b_0^3)}{6\pi}\,v_0\,,
\end{equation}
demanding simultaneously that this vector potential vanishes at the horizon. In contrast, we need not impose such condition on the confining solutions, so $\mu_{\text{\tiny conf}}$ is arbitrary in that case. In this paper we only study solutions with $\mu_{\text{\tiny plas}} =\mu_{\text{\tiny conf}} =0$. Nevertheless, we keep $v_0$ explicit in our formulas for completeness. We leave the problem of turning on the chemical potential for future work. 

Despite the chemical potential being vanishing, some of our solutions, in particular the black branes, will still be charged. The charge density can be computed in several ways, for instance as the radial canonical momentum of $\mathsf{A}_t$. Equivalently, as a constant of integration in the equation of the massless vector, which can be written as a total derivative (see Eq.~\eqref{eq:eoma1}). In terms of the parameters of the solutions it reads
\begin{equation}\label{eq:charge}
Q = -\frac{M^2}{N}\,\frac{\gs\ls q_c\rho_0^2}{2\kappa_4^2}\,\frac{64}{135}b_0^2 (20 \BET b_0 + 27 v_1)=-NM\LQCD ^ 2\frac{b_0^4}{2160\pi^3}(20 \BET b_0 + 27 v_1)\,.
\end{equation}
Notice that the charge is not simply the normalizable mode in the expansion of the massless vector, $v_1$, but it gets shifted by the magnetic field $\BET$. This correction, which comes from the topological interactions in the action \eqref{eq:vectactionQk0}, is a reflection of the fact that Maxwell and Page charges do not necessarily coincide in the presence of such terms. The charge density vanishes identically when evaluated on confining solutions, $Q_{\text{\tiny conf}} = 0$, but it is generically non-zero in the plasma phase.

Varying the action with respect to the boundary metric we get the energy density, pressure and spatial components of the energy momentum tensor \eqref{eq.EMtensor}. They read 
\begin{equation} \label{eq.holoTmunu}
\begin{aligned}
{E}  & =  \frac{\rho_0^5}{2\kappa^ 2_4} \left(-\frac{7 \mathsf{b}_5}{2}\, -  f_5 \right) = NM\LQCD ^ 3 \, \frac{(-b_0^ 5)}{3\cdot 2^{11} \pi^ 4}\left(-\frac{7 \mathsf{b}_5}{2}\, -  f_5 \right)  \,, \\[2mm] 
T^x_{\ x}   =  {P} - \Bphys\Mphys & =   \frac{\rho_0^5}{2\kappa^2_4} \left( -\frac{3 \mathsf{b}_5}{2}\, +  f_5\right) = NM\LQCD ^ 3 \, \frac{(-b_0^ 5)}{3\cdot 2^{11} \pi^ 4} \left( -\frac{3 \mathsf{b}_5}{2}\, +  f_5\right)\,, \\[2mm]
\end{aligned}
\end{equation}
were $\Mphys$ is the magnetization. These expressions are valid both for the confined and deconfined phases, since they are given just in terms of UV data. Note that the energy density $E$ obtained from the energy momentum tensor coincides with the free energy density in a confined phase $G_{\text{\tiny conf}}$, Eq.~\eqref{eq.freeEnergyConf}. Identifying the pressure as 
\begin{equation}\label{eq:pressure}
P=-G\,,
\end{equation}
this is nothing but the first law of thermodynamics, $E + P = TS + \mu Q = 0$, in the confining phase, as the entropy and charge densities vanish. Moreover, taking into account Eqs.~\eqref{eq.freeEnergyPlasma} to \eqref{eq:pressure}, one gets as a consistency check that the first law also holds in the plasma phase.

Finally, from equations \eqref{eq.holoTmunu}, \eqref{eq:pressure} and \eqref{eq:dimlessB}, the magnetization can be written as
\begin{equation}\label{eq.magnetization}
\begin{aligned}
\Mphys_{\text{\tiny conf}}  &= NM\LQCD \, \frac{3 \cdot  5\, }{2^ {10} \pi^2 }   \frac{\mathsf{b}_5}{\BET}\,,\\[2mm]
\Mphys_{\text{\tiny plas}}  &=NM\LQCD \, \frac{3 \cdot  5\, }{2^ {10} \pi^2 }   \left[\frac{\mathsf{b}_5}{\BET} - \frac{16}{675} \left(\frac{27 \left(5 \cbh \cfh^2 \cgh \xi_h^2 (1-\xi_h^4)^{\frac{1}{2}}+4 b_0^2 \clh^2v_0 v_1\right)}{\BET \clh^2}+80 b_0^3 v_0\right)\right]\,,
\end{aligned}
\end{equation}
for the confined and plasma phases respectively.

These thermodynamic quantities should obey various relations, such as
\begin{equation}\label{eq.thermoRelations}
\Mphys = -  \left.\frac{\dd G}{\dd \Bphys} \right|_{T,\mu}\,, \qquad
S = - \left.\frac{\dd G}{\dd T}  \right|_{\Bphys,\mu} \,, \qquad
\text{and} \qquad Q = - \left.\frac{\dd G}{\dd \mu} \right|_{T,\Bphys}\,.
\end{equation}
Given that we have independent expressions in terms of the coefficients for the left hand side of these equalities, we can use them as a consistency check and to test our numerics. We have indeed verified that the first two identities hold when evaluated on our solutions. Since we focused on the case $\mu = 0$, we do not have enough data to differentiate with respect to the chemical potential and examine the last identity for the charge density.

\subsection{Phase diagram}\label{sec:phases}

In this section we analyze the phase structure of the system in the $(\Bphys,T)$-plane, determining the preferred phase for each choice of magnetic field and temperature. The two types of solutions that compete are the deconfined plasma states and the magnetized confining ones, both discussed in Sec.~\ref{sec:newsols}. Their free-energy density is given in Eqs.~\eqref{eq.freeEnergyConf} and \eqref{eq.freeEnergyPlasma} respectively. In the grand-canonical ensemble, when several solutions exist at the same values of temperature and magnetic field, the one with lower free energy will be thermodynamically preferred.

A convenient way to visualize the different types of phase transitions that are present in the system is to examine how it evolves when the temperature is lowered while the magnetic field is held fixed. We identify two special values of the magnetic field,
\begin{equation}
\Bphys_{\text{\tiny triple}} \approx 1.152\,\LQCD^2 \,, \qquad\qquad\qquad \Bphys_{\text{\tiny critical}} \approx1.615\,\LQCD^2\,,
\end{equation}
where the qualitative behavior changes. Accordingly, they determine the following regions in the phase diagram of Fig.~\ref{fig:PhaseDiagram}: \textbf {Region A} comprises the interval between zero magnetic field and the line of second order phase transitions at $\Bphys_{\text{\tiny triple}}$. \textbf {Region B} is the interval that lies between the second order phase transitions and the critical value of the magnetic field, $\Bphys_{\text{\tiny critical}}$, where the line of first order transitions ends. \textbf{Region C} corresponds to values of the magnetic field larger than the critical one. All three regions are defined for any value of the temperature. Within each region, we find the following phases and transitions when the temperature is changed:

\begin{description}
	\item[{\textbf{Region A:}}] $\Bphys\in [0,\Bphys_{\text{\tiny triple}})$.
	The phase structure in this case is similar to that of vanishing magnetic field, discussed in Sec.~\ref{sec:review}. In this region, we still encounter confinement/deconfinement phase transitions, which now occur between the magnetized plasma and the confining backgrounds with non-vanishing magnetic field. The effect of $\Bphys$ in the free energy and entropy can be seen in Fig.~\ref{fig.CaseA_small_magnetic_field}. Qualitatively, it lifts the free energy of the two different states - the plasma and the confining one - but their overall shape is similar to the non-magnetized solutions shown in Fig.~\ref{fig.B8confPhase_transition}. 
	\begin{figure}[t!]
		\begin{center}
			\begin{subfigure}{.47\textwidth}
				\includegraphics[width=\textwidth]{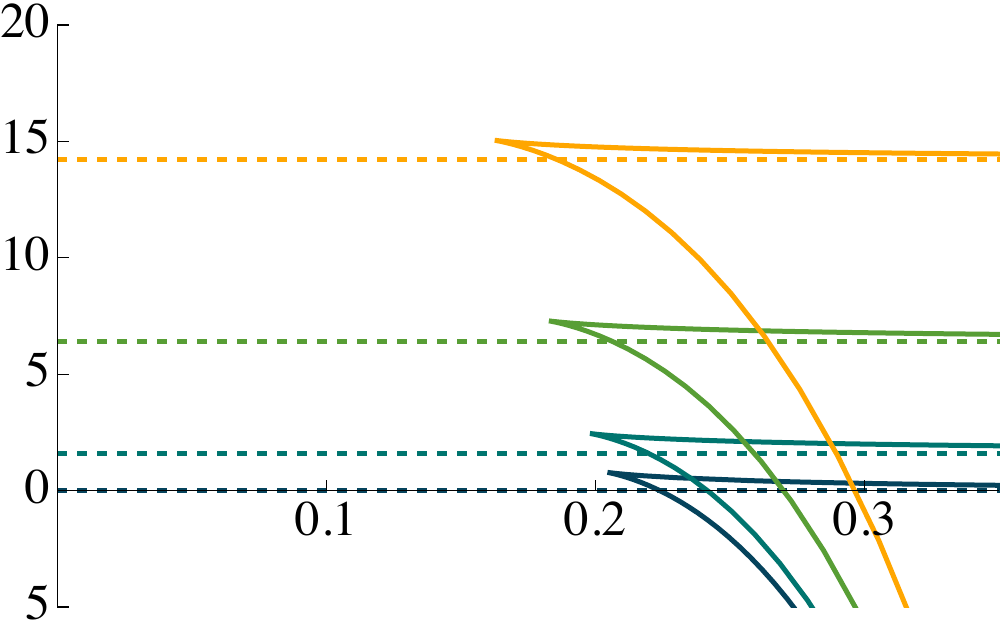} 
				\put(-200,140){ $10^3 \, G/(MN\LQCD^3)$}
				\put(-40,-10){ $T/\LQCD$}
			\end{subfigure}\hfill
			\begin{subfigure}{.47\textwidth}
				\includegraphics[width=\textwidth]{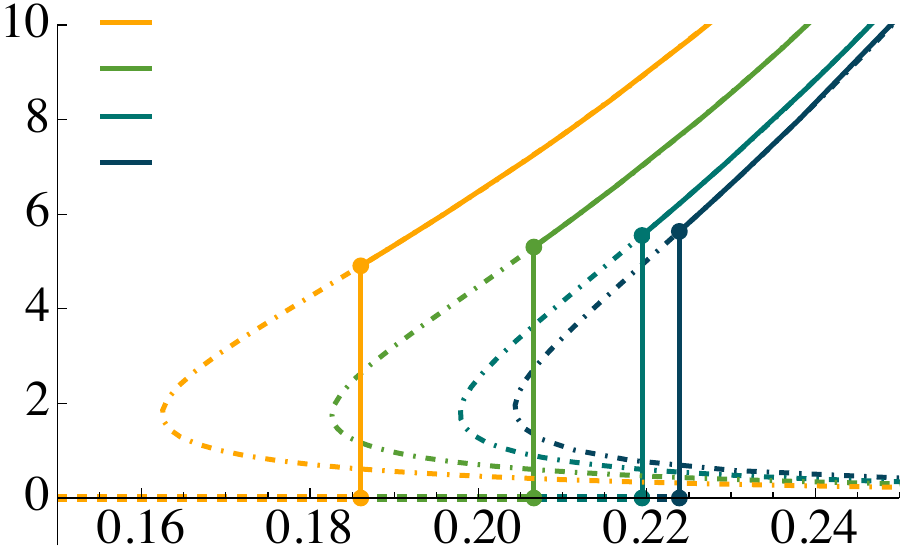} 
				\put(-164,117){\footnotesize $\Bphys = 0.794\,  \LQCD^2 $}
				\put(-164,106){\footnotesize $\Bphys = 0.530\,  \LQCD^2 $}
				\put(-164,95){\footnotesize $\Bphys = 0.264\,  \LQCD^2 $}
				\put(-164,84){\footnotesize $\Bphys =0\,  \LQCD^2 $}
				\put(-200,140){ $10^2\, S/(MN\LQCD^2)$}
				\put(-40,-10){ $T/\LQCD$}
			\end{subfigure}
			\caption{\small {\bf Left}: free-energy densities, as a function of temperature, for the magnetized plasma (solid curves) and magnetized confining phase (dashed lines) for different choices of magnetic field in the range $\Bphys\in [0,\Bphys_{\text{\tiny triple}})$, with $\Bphys$ increasing from bottom to top. The phase with the lowest free-energy density is preferred, so we encounter a confinement/deconfinement phase transition when the two curves cross. {\bf Right}: entropy density, as a function of temperature, for the same solutions. The dot-dashed curves indicate thermodynamically-disfavoured states of the system. The discontinuity at the phase transition indicates that it is first order. } \label{fig.CaseA_small_magnetic_field}
		\end{center}
	\end{figure}
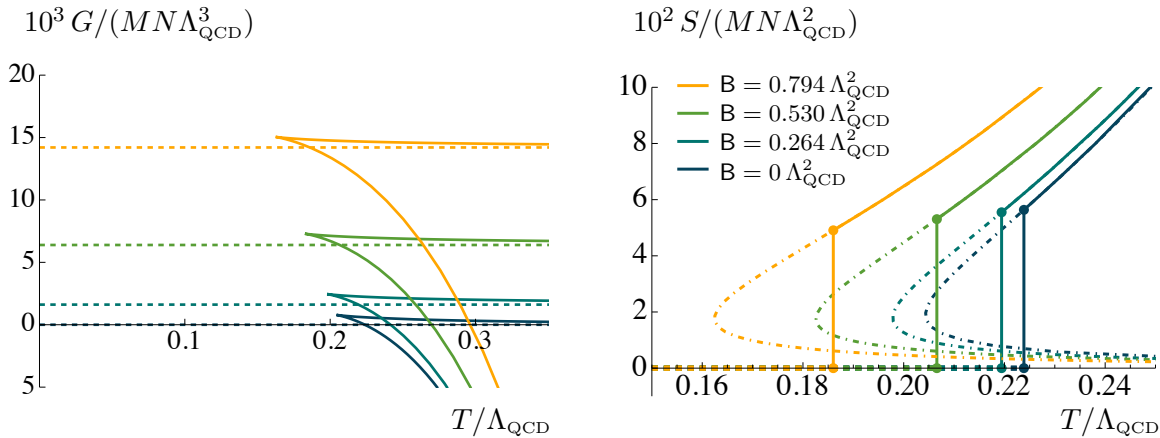
	
	Another important feature is that the phase transition is still first order, since quantities such as the entropy (see the right panel on Fig.~\ref{fig.CaseA_small_magnetic_field}), energy density or charge density are discontinuous across the phase transition.
	
	\item[{\textbf{Region B:}}] $\Bphys\in (\Bphys_{\text{\tiny triple}},\Bphys_{\text{\tiny critical}})$.
	When the magnetic field is raised above $\Bphys_{\text{\tiny triple}}$ there is an interesting effect. As shown in Fig.~\ref{fig.CaseB_triple}, a new stable branch of black brane solutions develops. The solutions on this branch have smaller free energy than the confining ones at the same value of the magnetic field. In a sense, we can think of the regular confining infrared as being ``covered by a horizon'' above this particular value of the magnetic field. From the field theory perspective, this signals the loss of confinement. Consequently, for values of the magnetic field in this region, there is still a first order phase transition, but now taking place between two deconfined phases.
	
This feature gives raise to a triple point in the $(\Bphys,T)$-plane, located in the intervals
	\begin{equation}
	\label{eq:triple_point}
	\Bphys_{\text{\tiny triple}} = \left(1.152 \pm 0.013\right)\LQCD^2\,,\qquad T_{\text{\tiny triple}} = \left(0.1456 \pm 0.0018\right)\LQCD\,.
	\end{equation}
At these particular values of temperature and magnetic field, three different phases coexist: the confining phase and two plasma phases. 
	
	\begin{figure}[t!]
		\begin{center}
			\begin{subfigure}{.47\textwidth}
				\includegraphics[width=\textwidth]{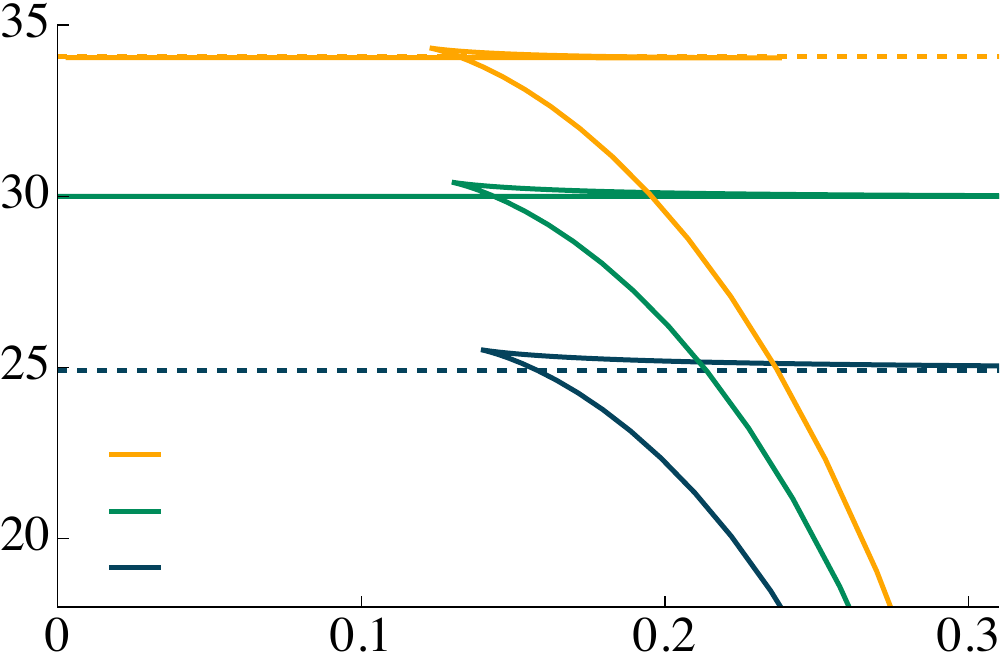} 
				\put(-200,145){ $10^3\, G/(MN\LQCD^3)$}
				\put(-35,-12){ $T/\LQCD$}
				\put(-164,41){\footnotesize $\Bphys = 1.24\,  \LQCD^2 $}
				\put(-164,29){\footnotesize $\Bphys = 1.17\,  \LQCD^2 $}
				\put(-164,17){\footnotesize $\Bphys = 1.06\,  \LQCD^2 $}
			\end{subfigure}\hfill
			\begin{subfigure}{.47\textwidth}
				\includegraphics[width=\textwidth]{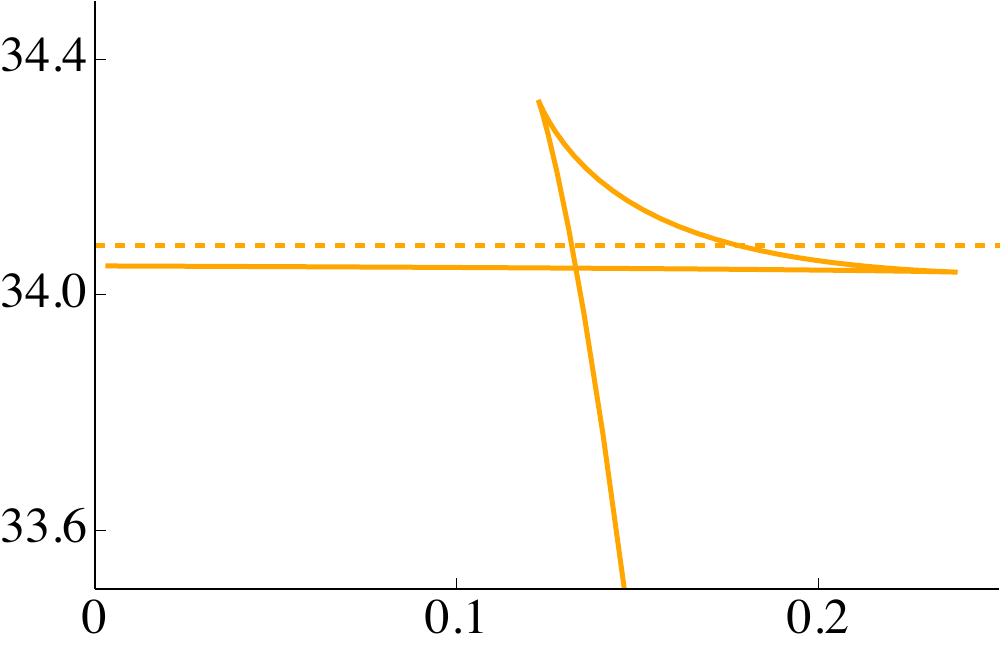} 
				\put(-200,145){ $10^3\, G/(MN\LQCD^3)$}
				\put(-35,-12){ $T/\LQCD$}
			\end{subfigure}
			\caption{\small {\bf Left}: free-energy density as a function of temperature for three different choices of magnetic field around $\Bphys_{\text{\tiny triple}}$, both for plasma (solid curves) and confining (dashed straight lines) states. The lowest curve has $\Bphys<\Bphys_{\text{\tiny triple}}$, the central one $\Bphys\approx\Bphys_{\text{\tiny triple}}$ and the upper one $\Bphys>\Bphys_{\text{\tiny triple}}$. {\bf Right}: zoom in version of the upper curve, where it can be seen that above $\Bphys_{\text{\tiny triple}}$ the confining phase is disfavored. There is a first order phase transition between deconfined phases when the two solid lines cross.} \label{fig.CaseB_triple}
		\end{center}
	\end{figure}
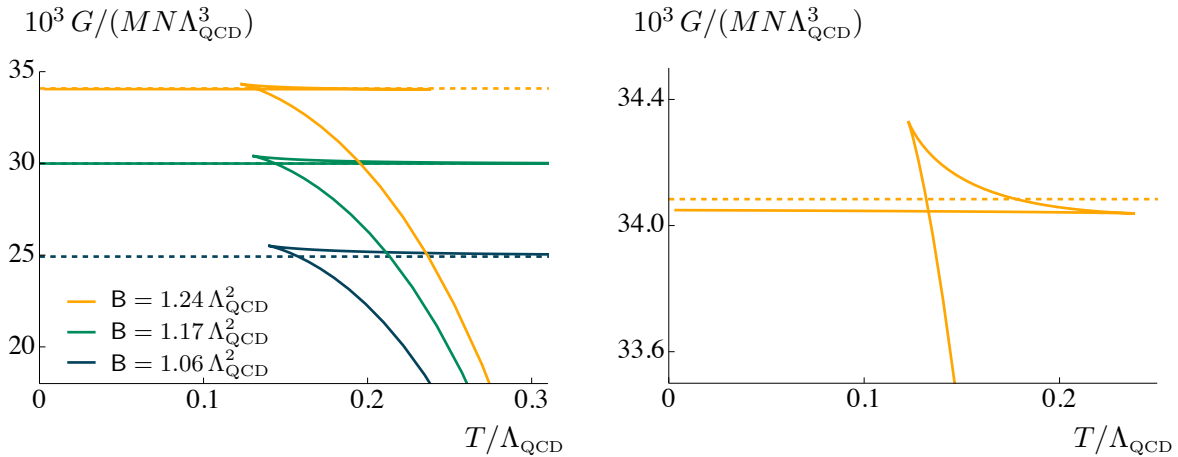

	\item[{\textbf{Region C:}} ] $\Bphys > \Bphys_{\text{\tiny critical}}$.
	The first order phase transitions between plasma phases that were identified in \textbf{Region B} cease to exist above the critical value of the magnetic field $\Bphys_{\text{\tiny critical}}$. The passage from \textbf{Region B} to \textbf{Region C} is reflected in Fig.~\ref{fig.CaseC_crossover}, where we see the loss of the swallowtail shape of the free energy characteristic of first order phase transitions. Therefore, for values of the magnetic field in this region, there are no phase transitions. Instead, thermodynamic quantities evolve smoothly between the low and high temperature behaviors.

	This means that there is a line of first order phase transitions between plasma states ending at a critical point, where a second order phase transition takes place. As for the triple point, we can locate this critical point in the intervals
	\begin{equation}
	\label{eq:critical_point}
	\Bphys_{\text{\tiny critical}} = \left(1.6145 \pm 0.0013\right)\LQCD^2\,,\quad T_{\text{\tiny critical}} =\left( 0.08801 \pm 0.00014\right)\LQCD\,.
	\end{equation}
	
		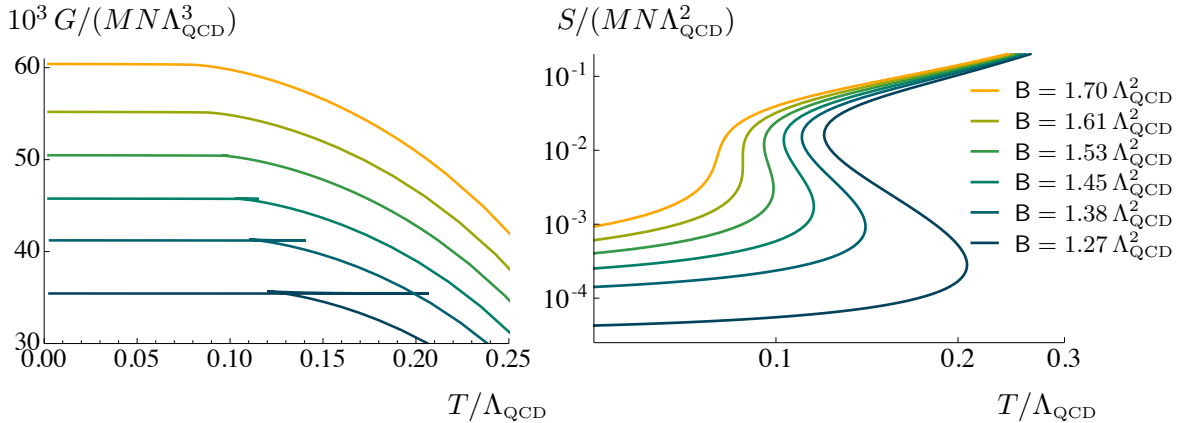
\begin{figure}[t!]
			\begin{subfigure}{.45\textwidth}
				\includegraphics[width=\textwidth]{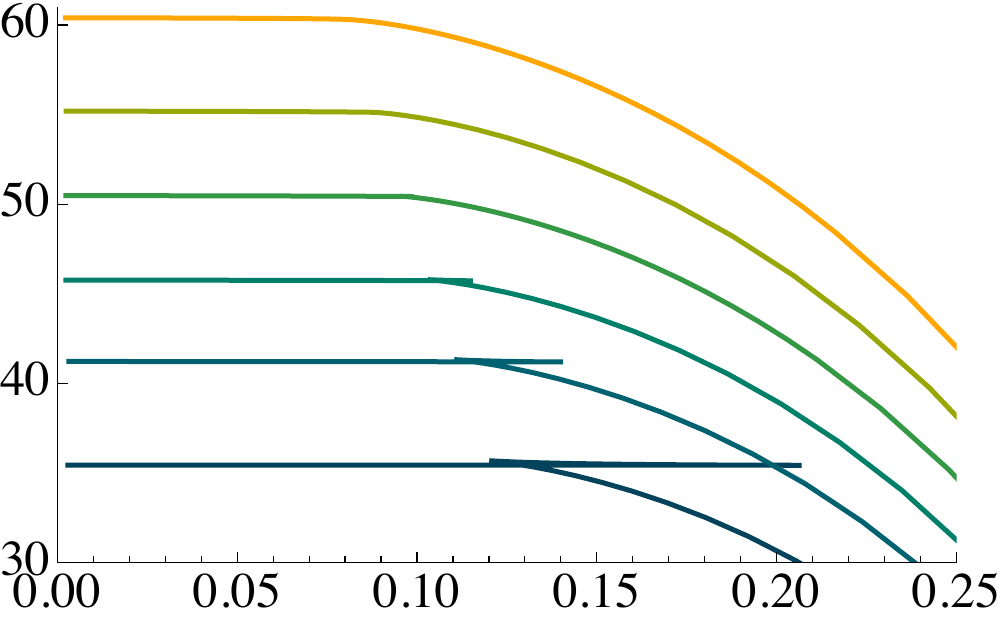} 
				\put(-200,130){ $10^3\, G/(MN\LQCD^3)$}
				\put(-35,-15){ $T/\LQCD$}
			\end{subfigure}
			\begin{subfigure}{.47\textwidth}
				\includegraphics[width=\textwidth]{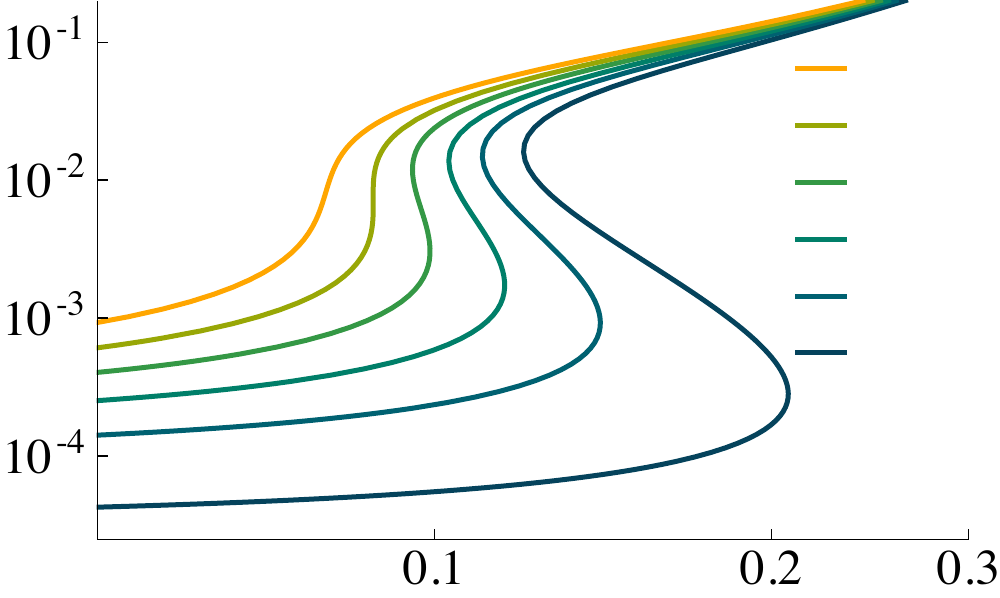} 
				\put(-200,130){ $S/(MN\LQCD^2)$}
				\put(-35,-15){ $T/\LQCD$}
				\put(-25,104){\footnotesize $\Bphys = 1.70\,  \LQCD^2 $}
				\put(-25,92.4){\footnotesize $\Bphys = 1.61\,  \LQCD^2 $}
				\put(-25,80.8){\footnotesize $\Bphys = 1.53\,  \LQCD^2 $}
				\put(-25,69.2){\footnotesize $\Bphys = 1.45\,  \LQCD^2 $}
				\put(-25,57.6){\footnotesize $\Bphys = 1.38\,  \LQCD^2 $}
				\put(-25,46){\footnotesize $\Bphys = 1.27\,  \LQCD^2 $}
			\end{subfigure}\hfill
			\caption{\small {\bf Left}: free-energy density as a function of temperature for different values of $\Bphys$ in the vicinity of the critical point, with the magnetic field increasing from bottom to top. The swallowtail is lost in the uppermost curves, which have $\Bphys\ge\Bphys_{\text{\tiny critical}}$. {\bf Right}: log-log plot of the entropy density as a function of temperature for the same values of the magnetic field.
			When the swallowtail shape disappears in the curves for the free energy, above $\Bphys_{\text{\tiny critical}}$, the phase transition becomes second order at the critical point and a smooth crossover for even larger values of $\Bphys$ .}  
				\label{fig.CaseC_crossover}
			\end{figure}

\end{description}

It remains to be seen how the transition between {\textbf{Region A} and {\textbf{Region B} proceeds as the magnetic field is increased for fixed temperatures below $T_{\text{\tiny triple}}$, since the low temperature phases in those regions are confined and deconfined respectively. Picking any value of the temperature $T<T_{\text{\tiny triple}}$, we can study how different thermodynamic quantities vary as the magnetic field is changed. As can be seen in Fig.~\ref{fig.charge_magnet_secondOrder}, the charge, magnetization and entropy are continuous across $\Bphys_{\text{\tiny triple}}$. This shows that the low-temperature confinement/deconfinement transition triggered by the magnetic field is second order. 

\begin{figure}[t!]
	\begin{center}
		\begin{subfigure}{.47\textwidth}
			\includegraphics[width=\textwidth]{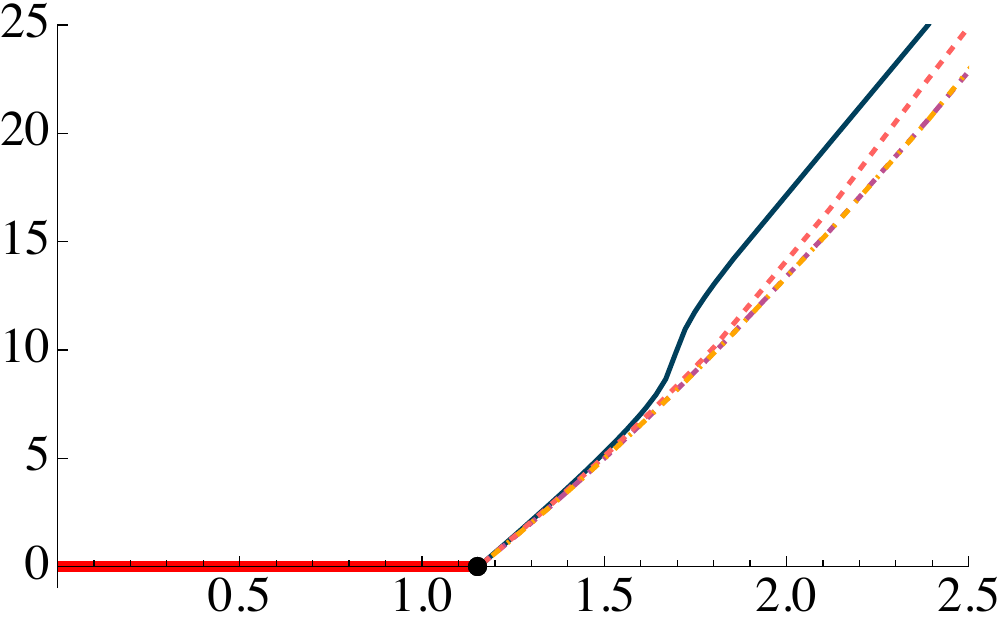} 
			\put(-190,130){ $10^3\, Q/(MN\LQCD^2)$}
			\put(-40,20){ $\Bphys/\LQCD^2$}
		\end{subfigure}\hfill
		\begin{subfigure}{.47\textwidth}
			\includegraphics[width=\textwidth]{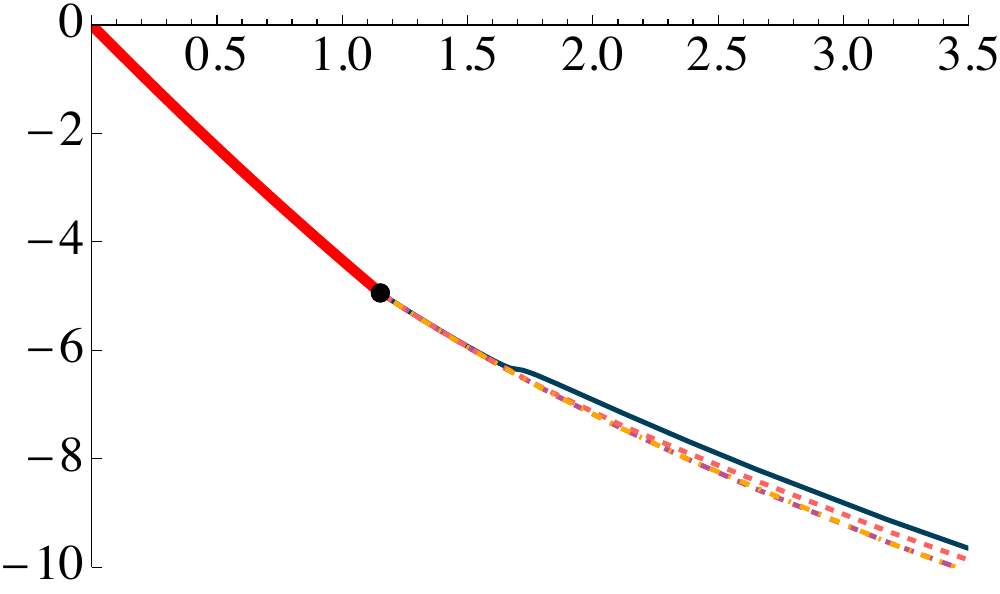} 
			\put(-200,130){ $10^2\, \Mphys/(MN\LQCD)$}
			\put(-40,80){ $\Bphys/\LQCD^2$}
		\end{subfigure}
	\begin{subfigure}{.65\textwidth}
		\includegraphics[width=\textwidth]{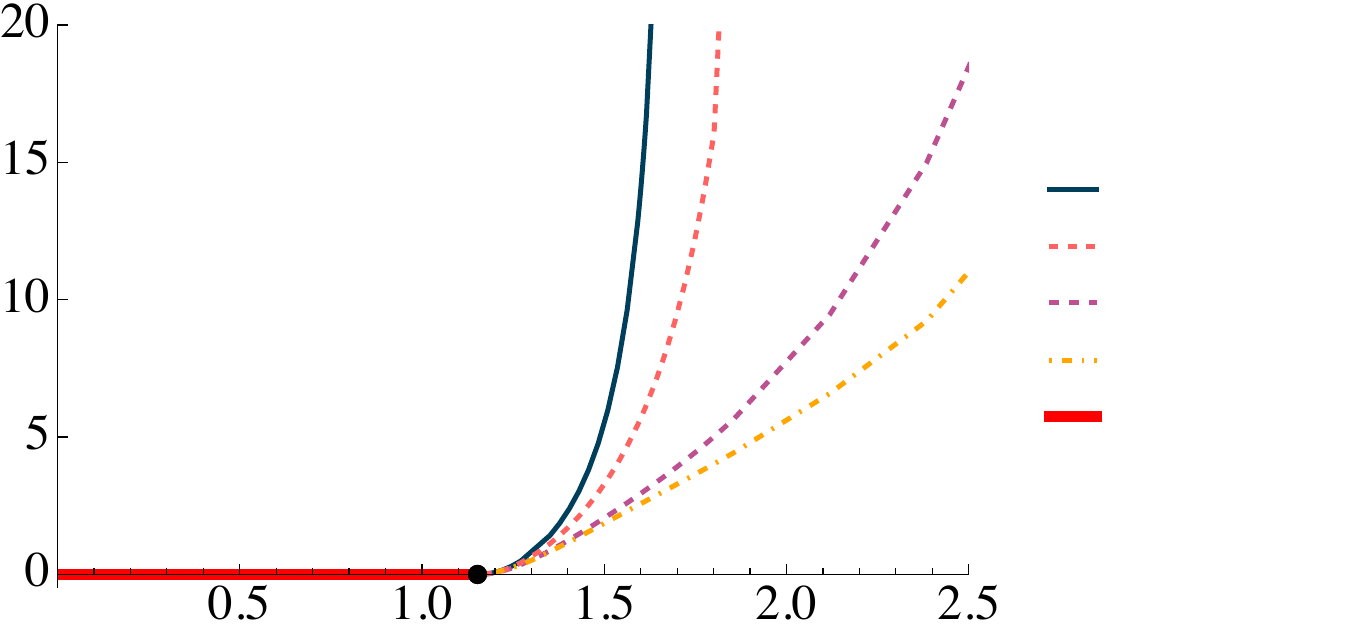} 
			\put(-180,170){}
		\put(-265,125){ $S/(MN\LQCD^2)$}
		\put(-115,20){ $\Bphys/\LQCD^2$}
		\put(-47,88){\footnotesize $T = 0.001\,  \LQCD^{-1} $}
		\put(-47,76.5){\footnotesize $T =  0.010\,  \LQCD^{-1}  $}
		\put(-47,65){\footnotesize $T =  0.050 \,  \LQCD^{-1} $}
		\put(-47,53.5){\footnotesize $T =  0.080 \,  \LQCD^{-1} $}
		\put(-47,42){\footnotesize Confining phase}
			\end{subfigure}
		\caption{\small Charge density (\textbf{left}), magnetization (\textbf{right}) and entropy density (\textbf{bottom}) as a function of the magnetic field in the vicinity of $\Bphys_{\text{\tiny triple}}$ (indicated by the black dot), for different choices of the temperature. In the confining phase, these quantities do not depend on the temperature, as indicated by the red, thick curves. The different curves, for any choice of the temperature, are continuous at $\Bphys_{\text{\tiny triple}}$, indicating that the transition is second order.} 
	\label{fig.charge_magnet_secondOrder}
	\end{center}
\end{figure}

\begin{figure}[t!]
	\begin{center}
		\scalebox{.90}{
		\includegraphics[width=.75\textwidth]{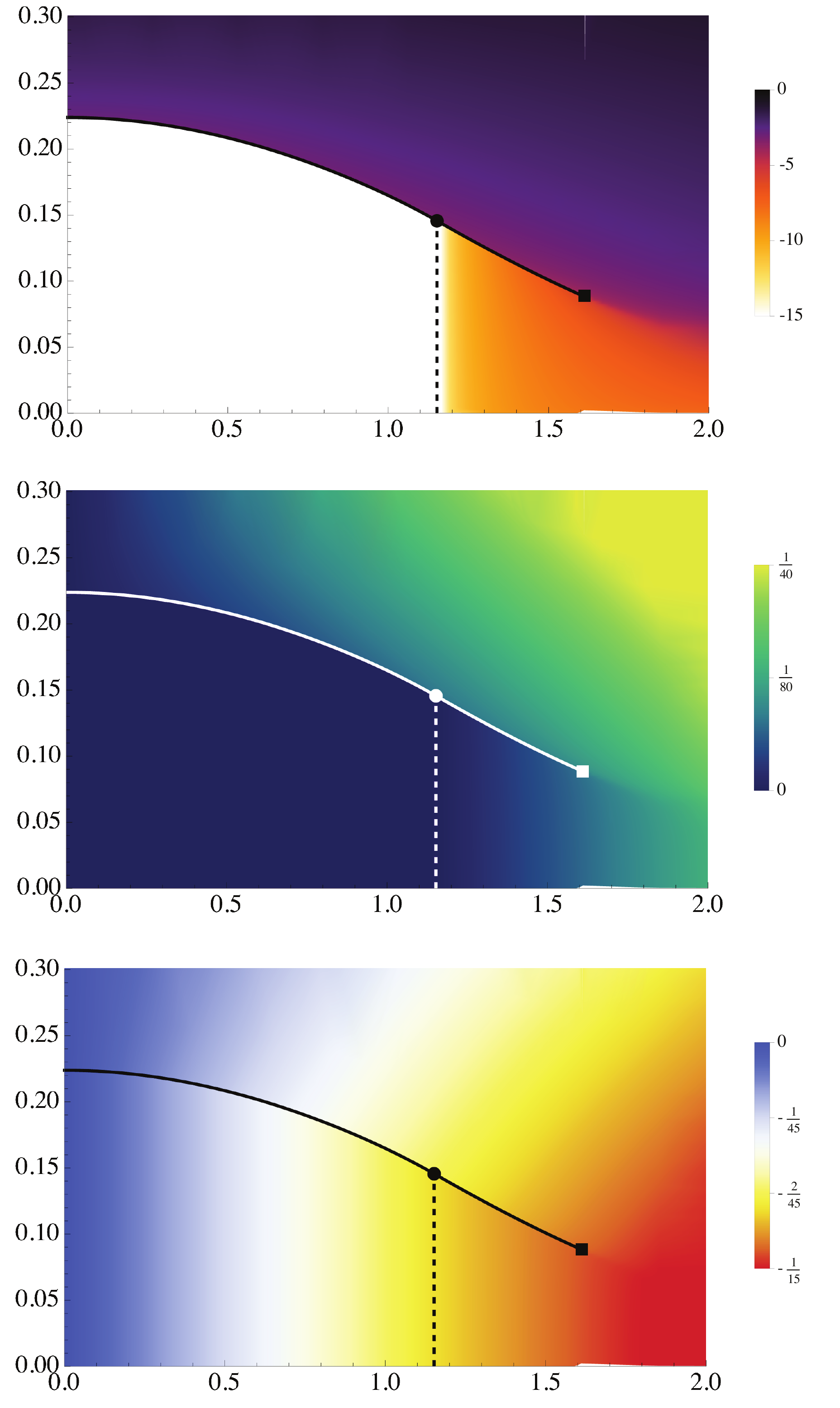} 
		\put(-360,530){$T/\LQCD$}
		\put(-40,380){$\Bphys/\LQCD^2$}
		\put(-35,525){$\log(S/(M\, N\, \LQCD^2))$}
		\put(-360,345){$T/\LQCD$}
		\put(-40,195){$\Bphys/\LQCD^2$}
		\put(-35,340){$Q/(M\, N\, \LQCD^2)$}
		\put(-360,160){$T/\LQCD$}
		\put(-40,10){$\Bphys/\LQCD^2$}
		\put(-35,155){$\Mphys/(M\, N\, \LQCD)$}
	}
		\caption{\small Density plots of the logarithm of the entropy density (\textbf{top}), charge density (\textbf{middle}) and magnetization (\textbf{bottom}) as a function of the external magnetic field and temperature. The solid curves correspond to the line of first order phase transitions between the plasma and the confining phase (when $\Bphys < \Bphys_{\text{\tiny triple}}$) and between different states of the plasma (when $\Bphys_{\text{\tiny triple}} < \Bphys < \Bphys_{\text{\tiny critical}}$). These curves end at a critical point, represented by a square. Finally, the dashed vertical lines correspond to second order confinement/deconfinement phase transitions between the confining  solution and the low-temperature plasma state. The triple point \eqref{eq:triple_point} is represented by a circle, where the three phases may coexist.} \label{fig.PhaseDiagramAll}
	\end{center}
\end{figure}

All this information about the behaviour of the system can be summarised by the phase diagram in the $(\Bphys,T)$-plane, as shown in Fig.~\ref{fig:PhaseDiagram} in the Introduction. There, the various regions and different types of phase transitions we have discussed are depicted, together with the triple and critical points given by equations \eqref{eq:triple_point} and \eqref{eq:critical_point} respectively. It is also instructive to show on this plane how different physical quantities vary as we change the temperature and the external magnetic field. This analysis can be found in Fig.~\ref{fig.PhaseDiagramAll}, where the density plots for the entropy density, charge density and magnetization are shown. These quantities change discontinuously across the line of first order phase transitions. Among them, the quantity that changes more abruptly is the entropy density. In contrast, both the magnetization and charge density are exactly zero for vanishing magnetic field, which means that their discontinuities become fainter as the magnetic field is switched off. Finally, it is interesting to note that both entropy and charge densities are constant (and zero) in the confining phase, whereas the absolute value of the magnetization grows with increasing magnetic field.

Notice that the confining phases that are thermodynamically preferred at low temperatures and small magnetic fields are continuously connected to the ground state with $\Bphys=0$. This is the branch of solutions to the left of the maximum magnetic field in Fig.~\ref{fig.bounded_magneticField}. The solutions to the right of the maximum are never realised since the plasma phases are always dominant for those values of $\Bphys$.

\section{Discussion}\label{sec:discussion}

The physical quantities plotted in Fig.~\ref{fig.PhaseDiagramAll} reveal several interesting aspects of the different phases. In the confined phase all quantities are temperature independent and the entropy vanishes, as expected at leading order in the large-$N$ expansion. As a consequence, the energy density and pressure satisfy the relation in vacuum $P=-E$, and the pressure is negative. Comparing the low and high temperature deconfined phases, one can appreciate that they remain mostly temperature independent in the first, but have a clear temperature dependence in the second. Also the entropy is much smaller in the low temperature phase than in the higher temperature one, even in the region where they are separated by a crossover, and the pressure and energy density are dominated by vacuum contributions $P\simeq -E$. This suggests that the low temperature phase retains some characteristics of the confined phase, perhaps indicating that a monopole condensate persists together with a plasma of monopoles in a normal phase, the last sourcing the non-zero monopole charge of the deconfined phase. Following this reasoning, in the high temperature phase one would expect a monopole condensate to be negligible or completely washed away.

Both deconfined phases have gravity duals that are black brane geometries. We thus expect them to behave as fluids at long wavelengths. Comparing the entropies, the low and high temperature phases resemble `liquid' and `gas' phases respectively. On the other hand, the viscosity of both phases is proportional to the entropy density, so that the viscosity is much larger in the high temperature phase, as can be appreciated in Fig.~\ref{fig.CaseC_crossover}. 

There are several directions in which the work presented here could be extended. First, it would be interesting to study in more detail the second order transition and in particular extract the critical exponents. At present we have not been able to do it because this requires finding solutions very close to the transition line, which is numerically very challenging. Finding solutions at very low temperatures is numerically demanding as well. We have been able to find some extremal solutions at nonzero magnetic field that we may present in a future work, but we have not determined yet whether they can be connected to the asymptotically D2 brane geometry, or if the right zero temperature solutions would be of a different kind. For instance, extending the unstable branch shown in Fig.~\ref{fig.bounded_magneticField} may require a different type of ansatz where the M-theory circle collapses to zero size at the origin.

There are many physical observables that can in principle be computed from the solutions we have obtained so far. Related to the phase transitions the monopole-antimonopole interaction and the Wilson line are obvious ones, but also the entanglement entropy can contain some useful information \cite{Klebanov:2007ws,Jokela:2020wgs}. The spectrum of fluctuations around the solutions can also provide some valuable information. It is for instance unclear whether the $\UM$ symmetry is spontaneously broken. If that were the case, there would be a gapless mode in the confining phase at zero temperature and magnetic field. If the mode exists it would be interesting to study how temperature and magnetic field affect its dispersion relation. In the black brane solutions we expect to have hydrodynamic modes whose dispersion relation would be interesting to obtain, as well as the value of other transport coefficients beyond the shear viscosity. In particular, since there are topological terms with field-dependent coefficients, it is possible that the Hall viscosity is non-zero through a mechanism analogous to the one described in \cite{Saremi:2011ab}.

Another interesting direction would be to study the phase diagram of a mirror dual \cite{Intriligator:1996ex,Hanany:1996ie},  in the particle-vortex dual sense \cite{Aharony:1997bx}. For the amount of supersymmetry we are considering, mirror duality of $\NN=1$ QED was studied in \cite{Gremm:1999su}, and there are generalizations for the Abelian theory of probe D2-branes on cones with special holonomy \cite{Gukov:2002es,Gukov:2002er}, and for non-Abelian theories with Chern--Simons terms \cite{Bashmakov:2021rci}, but there does not seem to be an extension to the non-Abelian case for vanishing Chern--Simons levels. In our case the expectation is that in the mirror dual the monopole magnetic field would translate into a baryon charge density \cite{future}, thus allowing to explore ``nuclear matter'' in a holographic setup without having to introduce additional flavor D-branes or instantons on those.

A different extension would be to introduce the monopole magnetic field in theories with a mass gap that are not confining \cite{Faedo:2017fbv}, by having a non-zero Chern--Simons level in the dual field theory. As we had discussed, in these theories the monopole operators dual to wrapped D2-branes are not gauge-invariant, and the massless field dual to the $\UM$ current corresponds to a different combination of vector fields in the reduction to four dimensional supergravity (see Appendix~\ref{sec:truncation}). Consequently, the type of charged objects that might be able to condense would probably be modified. Whether this translates to qualitative differences in the phase diagram is an interesting question that might be worth exploring.

Let us finally speculate about the field theory dual to the confining solutions. As we discussed previously, the geometry is likely dual to a quiver theory with $\UU(N)\times \UU(N+M)$ gauge group and bifundamental fields. The field theory contains gauge-invariant monopole operators dual to the D2-brane that carry $n=N/M$ units of D0-brane charge, with the D0-branes expected to be dual to symmetric monopole operators $\MM_{m,m}$ in the notation introduced in Sec.~\ref{sec:review}. In addition, in the reduction to four-dimensions described in Appendix \ref{sec:truncation} there is a topological term in \eqref{eq:topaction} of the form
\begin{equation}
\frac{32q_c}{\kappa_4^2}\int \,\left(\frac{\hat{a}_1-\tilde{a}_1}{2}\right)\wedge  \dd b_2 \,.
\end{equation}
The vector field is the combination that enters in $C_3$ as the coefficient of a closed two-form in the internal space. This term would be non-zero if we had for instance components of the vector fields along the $x_1$ direction and of the $b_2$ form along the $t,x_2$ directions and depending on the holographic radial coordinate. Setting $\hat{a}_1=-\tilde{a}_1$, the vector field and two-form couple respectively to a wrapped D2 brane and a fundamental string as follows
\begin{equation}
S_{\text{\tiny {D2}}}\supset \frac{1}{\pi \gs \ls^3} \int_{x_1} \tilde{a}_1,\ \ S_{\text{\tiny F1}}\supset \frac{1}{2\pi\ls^2} \int_{t x_2} b_2\,.
\end{equation}
The natural normalization of the fields is $\tilde{a}_1=\pi\gs \ls^3 \tilde{\mathsf{A}}_1$ and $b_2=2\pi \ls^2 \mathsf{b}_2$, in such a way that the vortex line associated to a D2-brane and the Wilson line associated to the string get multiplied by phases $e^{i\int \tilde{\mathsf{A}}_1}$ and $e^{i\int \mathsf{b}_2}$ respectively. This makes the topological term equal to
\begin{equation}
 \frac{M}{2\pi}\int\, \tilde{\mathsf{A}}_1 \wedge \dd\mathsf{b}_2\,. 
\end{equation}
Since we have Dirichlet boundary conditions of the two-form, following the arguments in \cite{Bergman:2020ifi}, there should be a $\mathds{Z}_M$ one-form symmetry in the dual field theory. 

A possible way to understand the monopole spectrum and the one-form symmetry is if the theory originates from an orbifold 
\begin{equation}
\begin{array}{c}
\UU(2N+M)\simeq \left[ \UU_{\text{\tiny diag}}(1)\times {\rm SU}(2N+M)\right]/\mathds{Z}_{N+M} \\ \Big\downarrow \\  \left[ \UU_{\text{\tiny diag}}(1)\times \UU(1)\times {\rm SU}(N)\times {\rm SU}(N+M)\right]/\mathds{Z}_{M}
\end{array}\,.
\end{equation}
We will denote the original $\UU(2N+M)$ theory as the `parent' and the theory obtained after orbifolding as the `daughter'. For $N=nM$, the fundamental representation with zero charge under the $\UU_{\text{\tiny diag}}(1)$ group splits into a $\left(N_{n+1},{(N+M)}_{-n}\right)$ representation, where the subindex denotes the charge under the non-diagonal $\UU(1)$. Then, a Wilson loop in the fundamental representation of the parent theory would produce line operators of the daughter theory that are charged under the non-diagonal $\UU(1)$, i.e. $W_{2N+M}\longrightarrow W_N W_{N+M}$ would have charge $1$ and would not be gauge invariant. A line operator neutral under the non-diagonal $\UU(1)$ could be obtained from the product of $2n+1$ fundamental Wilson lines in the parent theory 
\begin{equation}
W_{2N+M}^{2n+1}\longrightarrow W_N^n W_{N+M}^{n+1}\equiv {\mathsf W}\,.
\end{equation}
Using that $W_{2N+M}^{2N+M} \sim \mathds{1}$, $W_N^N\sim \mathds{1}$, $W_{N+M}^{N+M}\sim \mathds{1}$ - since in all cases they would be screened by fields in the adjoint representation - we obtain ${\cal W}^M\sim \mathds{1}$. This leads naturally to the $\mathds{Z}_M$ one-form symmetry.

Regarding the monopole spectrum, bifundamental fields would be in a $(N_{n+1},\overline{(N+M)}_{n})$ representation, thus having charge $2n+1$ under the non-diagonal $\UU(1)$. The phase that a particle in the bifundamental representation would pick when going around a monopole of charge $\MM_{m_1,m_2}$ is
\begin{equation}
\varphi_{\text{\tiny bif}}=2\pi \left(m_1 (n+1)+m_2 n\right)\,.
\end{equation}
For a symmetric monopole $m_1=m_2=\frac{1}{2n+1}$ is the minimal amount of magnetic charge allowed by Dirac quantization. The other minimal choice is $m_1=\frac{n+1}{2n+1}$, $m_2=-\frac{n}{2n+1}$. The symmetric monopole $\MM_{\frac{1}{2n+1},\frac{1}{2n+1}}$ would correspond to a D0-brane, while the natural interpretation of the D2-brane carrying $n$ D0-brane charge would be an operator $\MM_{\frac{1}{2n+1},\frac{1}{2n+1}}^n \MM_{\frac{n+1}{2n+1},-\frac{n}{2n+1}}$. This last would correspond in the parent theory to a monopole of unit magnetic charge.

\section*{Acknowledgments}\label{sec:acknowledgments}

We would like to thank Oren Bergman, Alexander Krikun, David Mateos and Ronnie Rodgers for useful discussions. Nordita is supported in part by NordForsk. We thank the PDC Center for High Performance Computing, KTH Royal Institute of Technology, Sweden, for providing access to the computing resources used in this research (project SNIC 2021/22-999). The work of A.F. and C.H is partially supported by the AEI and the MCIU through the Spanish grant PID2021-123021NB-I00 and by FICYT through the Asturian grant SV-PA-21-AYUD/2021/52177. A.F. is also supported by the ``Beatriz Galindo'' program, reference BEAGAL 18/00222.

\appendix

\section{Geometry of $\CP^3$}\label{sec:geometry}

In this paper we study solutions to type IIA supergravity whose internal geometry is $\CP^3$. For completeness, in this Appendix we give some details of this space and specify our conventions. For our purposes, the three dimensional complex projective space is seen as the coset Sp($2$)$/$U($2$), which is a two-sphere S$^2$ fibered over a four-sphere S$^4$. A suitable choice of coordinates goes as follows. Let $\omega^i$ be the set of left-invariant forms on the three-sphere, normalized so that $2\dd \omega^i=\epsilon_{ijk}\omega^j\wedge\omega^k$. A particular realization is for instance 
\begin{equation}
\begin{aligned}
\omega^1&=\cos\psi\,\dd\phi+\sin\psi\sin\phi\,\dd\chi\\[2mm]
\omega^2&=\sin\psi\,\dd\phi-\cos\psi\sin\phi\,\dd\chi\\[2mm]
\omega^3&=\dd\psi+\cos\phi\,\dd\chi
\end{aligned}
\end{equation}
with the ranges $0\le\phi\le\pi$, $0\le\chi\le2\pi$ and $0\le\psi\le4\pi$. Then, the metric of a unit radius four-sphere can be written as
\begin{equation}
\dd\Omega_4^2\,=\,\frac{4}{\left(1+\zeta^2\right)^2}\left[\dd \zeta^2+\frac{\zeta^2}{4}\omega^i\omega^i\right]\,,
\end{equation}
with $\zeta$ a non-compact coordinate in the range $0\le\zeta<\infty$. Taking $0\le\theta\le\pi$ and $0\le\varphi\le2\pi$ to be the angles on S$^2$, the non-trivial fibration is described by the vielbeins
\begin{eqnarray}
\label{introduced}
E^1&=&\dd \theta+\frac{\zeta^2}{1+\zeta^2}\left(\sin\varphi\,\omega^1-\cos\varphi\,\omega^2\right)\,,\nonumber\\[2mm]
E^2&=&\sin\theta\left(\dd\varphi-\frac{\zeta^2}{1+\zeta^2}\omega^3\right)+\frac{\zeta^2}{1+\zeta^2}\cos\theta\left(\cos\varphi\,\omega^1+\sin\varphi\,\omega^2\right)\,.
\end{eqnarray}
It is convenient to consider a rotated version of the vielbeins on the four-sphere
\begin{eqnarray}\label{eq:vielbeinsS4}
\mathcal{S}^1&=&\frac{\zeta}{1+\zeta^2}\left[\sin\varphi\,\omega^1-\cos\varphi\,\omega^2\right]\,,\nonumber\\
\mathcal{S}^2&=&\frac{\zeta}{1+\zeta^2}\left[\sin\theta\,\omega^3-\cos\theta\left(\cos\varphi\,\omega^1+\sin\varphi\,\omega^2\right)\right]\,,\nonumber\\
\mathcal{S}^3&=&\frac{\zeta}{1+\zeta^2}\left[\cos\theta\,\omega^3+\sin\theta\left(\cos\varphi\,\omega^1+\sin\varphi\,\omega^2\right)\right]\,,\nonumber\\
\mathcal{S}^4&=&\frac{2}{1+\zeta^2}\,\dd\zeta\,,
\end{eqnarray}
so that, despite the explicit dependence of \eqref{eq:vielbeinsS4} on the angles $\theta$ and $\varphi$, it is still verified that $\mathcal{S}^n\mathcal{S}^n=\dd \Omega_4^2$. Then, the metric on the coset reads
\be
\dd s^2\left(\CP^3\right)=\alpha^2\left[\left(E^1\right)^2+\left(E^2\right)^2\right]+\sum_{n=1}^4\mathcal{S}^n\mathcal{S}^n\,,
\ee
for some constant $\alpha$ controlling the squashing of the fiber over the base.\footnote{There are two Einstein points: $\alpha^2=1$, where the Fubini--Study metric is recovered, and $\alpha^2=1/2$ where the metric admits instead a nearly-K\"ahler structure.} It is also possible to write down the left-invariant forms on the coset in a compact manner using these vielbeins. This set contains the two-forms\footnote{The K\"ahler form associated to the Fubini--Study metric is in this language $X_2-J_2$.}
\begin{equation}
X_2\,=\,E^1\wedge E^2\,,\qquad\qquad\qquad J_2\,=\,\mathcal{S}^1\wedge\mathcal{S}^2+\mathcal{S}^3\wedge\mathcal{S}^4\,,
\end{equation}
as well as the three-forms 
\begin{eqnarray}
X_3&=&E^1\wedge\left(\mathcal{S}^1\wedge\mathcal{S}^3-\mathcal{S}^2\wedge\mathcal{S}^4\right)-E^2\wedge\left(\mathcal{S}^1\wedge\mathcal{S}^4+\mathcal{S}^2\wedge\mathcal{S}^3\right)\,,
\nonumber\\[2mm]
J_3&=&-E^1\wedge\left(\mathcal{S}^1\wedge\mathcal{S}^4+\mathcal{S}^2\wedge\mathcal{S}^3\right)-E^2\wedge\left(\mathcal{S}^1\wedge\mathcal{S}^3-\mathcal{S}^2\wedge\mathcal{S}^4\right)\,.
\end{eqnarray}
These are related by exterior differentiation as
\begin{equation}\label{eq:relationForms}
\dd X_2\,=\,\dd J_2\,=\,X_3\,,\qquad\qquad\qquad \dd J_3\,=\,2\left(X_2\wedge J_2+J_2\wedge J_2\right)\,.
\end{equation}
Higher-rank forms constructed by wedging of these will also be left-invariant. One finds the two four-forms $X_2\wedge J_2$ and $J_2\wedge J_2$, appearing in \eqref{eq:relationForms}, together with the volume form on $\CP^3$, namely $\Omega_6 =  (E_1 \wedge E_2)\wedge (\mathcal{S}^1\wedge\mathcal{S}^2\wedge\mathcal{S}^3\wedge\mathcal{S}^4)$. There are no adequate one- or five-forms. Moreover, the complete set closes under Hodge duality. 

The ansatz for the fluxes in the main text is given in terms of these left-invariant forms. This symmetry ensures that the ansatz is consistent, meaning that the angles drop out from the ten-dimensional equations and there is just dependence on the radial coordinate. Furthermore, there is a consistent truncation to four dimensions keeping just the invariant modes that we detail in Appendix \ref{sec:truncation}. 

A crucial characteristic of this manifold is that it has a two- and a four-cycle given by $\CP^1$ and $\CP^2$ respectively. The two-cycle is specified in our coordinates by $\theta$ and $\varphi$ at fixed coordinates on the four-sphere. We have thus the integrals 
\be\label{twocycle}
\int_{\CP^1}X_2=4\pi\,,\qquad\qquad\qquad\qquad\int_{\CP^1}J_2=0\,.
\ee 
On the other hand, the four-cycle is obtained by fixing $\theta=\varphi=\pi/2$, in which case the following integrals are obtained 
\be\label{fourcycle}
\int_{\CP^2}J_2\wedge J_2=\frac{16}{3}\pi^2\,,\qquad\qquad\qquad\qquad\int_{\CP^2}X_2\wedge J_2=-\frac{16}{3}\pi^2\,.
\ee 
Finally integrating over the entire volume we get
\be\label{sixcycle}
\int_{\CP^3}E^1\wedge E^2\wedge \mathcal{S}^1\wedge \mathcal{S}^2\wedge \mathcal{S}^3\wedge \mathcal{S}^4=\frac12\int_{\CP^3}X_2\wedge J_2\wedge J_2=\frac{32}{3}\pi^3\,.
\ee

\section{Truncation to four dimensions}\label{sec:truncation}

In this section we reduce type IIA supergravity on $\CP^3$ seen as the coset ${\rm Sp}(2)/{\rm U}(2)$. Consistency of the truncation will be ensured by left-invariance on the coset. The result is a four-dimensional $\mathcal{N}=2$ supergravity, first obtained in \cite{Cassani:2009ck}. Nevertheless, we will give the details to make contact with the variables used in \cite{Faedo:2017fbv,Elander:2020rgv}, keeping the vectors and forms that were discarded in those two references.  

The expansion of the ten-dimensional fields is performed using the set of left-invariant forms discussed in Appendix \ref{sec:geometry}. A particular realization in terms of coordinates is given there, which can be used for instance to compute the Hodge duals of these forms. 

We choose to solve the Bianchi identities \eqref{eq:IIABianchis} as 
\begin{equation}
H_3 = \dd B_2\,,\qquad\qquad F_2 = \dd C_1+F_2^{\rm fl}\,,\qquad\qquad F_4 = \dd C_3+B_2\wedge F_2+F_4^{\rm fl}\,,
\end{equation}
for some two-, one- and three-form potentials $B_2$, $C_1$ and $C_3$ respectively. The internal manifold is topologically ${\rm S}^2\times{\rm S}^4$, so it has non-trivial two- and four-cycles. This allows us to include the fluxes $F_2^{\rm fl}$ and $F_4^{\rm fl}$ for the two- and four-forms, which are closed but non-exact. In terms of the globally defined left-invariant forms they read
\begin{equation}{\label{eq:fluxes}}
F_4^{\rm fl}=q_c\,\left(J_2\wedge J_2-X_2\wedge J_2\right)\,,\qquad\qquad\qquad F_2^{\rm fl}=Q_k\,\left(X_2-J_2\right)\,,
\end{equation}
with $q_c$ and $Q_k$ some constants, related to gauge theory parameters through flux quantization, as will be detailed below. The manifold does not admit a three-form flux. The form-potentials are written in terms of the invariant forms as 
\begin{equation}\label{eq:formpot}
\begin{aligned}
B_2 &= b_2+b_X\,X_2+b_J\, J_2\,,\\[2mm]
C_1 &= a_1\,,\\[2mm]
C_3 &= a_3+\at\wedge X_2+\ah\wedge J_2+a_X \,X_3+a_J \, J_3\,,
\end{aligned}
\end{equation}
where $a_3$ is a three-form, $b_2$ is a two-form, $a_1$, $\at$ and $\ah$ are one-forms and the rest are scalars, all of them defined on the external manifold. This is the most general expansion compatible with the symmetries on the coset. Using these expansions together with the fluxes \eqref{eq:fluxes} the field strengths appearing in the equations of motion \eqref{eq:IIAeoms} are 
\begin{equation}
\begin{aligned}
H_3 &= \dd b_2+\dd b_X\wedge X_2+\dd b_J\wedge J_2+\left(b_J+b_X\right)X_3\,,\\[2mm]
F_2 &= \dd a_1+Q_k\,\left(X_2-J_2\right)\,,\\[2mm]
F_4 &= f_4+\tilde{f}_2\wedge X_2+\hat{f}_2\wedge J_2+Da_X\wedge X_3+\dd a_J\wedge J_3+f_0^X X_2\wedge J_2+f_0^JJ_2\wedge J_2\,,
\end{aligned}
\end{equation}
with the following covariant derivative and field strengths 
\begin{equation}\label{eq:fieldstr}
\begin{array}{rclcrcl}
Da_X&=&\dd a_X-\at-\ah\,,&\qquad\qquad\quad&f_4&=&\dd a_3+b_2\wedge\dd a_1\,,\\[3mm]
\tilde{f}_2&=&\dd \at+b_X\,\dd a_1+Q_k \,b_2\,,&\qquad\qquad\quad&\hat{f}_2&=&\dd \ah+b_J\,\dd a_1-Q_k\, b_2\,,\\[3mm]
f_0^X&=&2a_J+Q_k\left(b_J-b_X\right)-q_c\,,&\qquad\qquad\quad&f_0^J&=&2a_J-Q_k\,b_J+q_c\,.
\end{array}
\end{equation}
To complete the reduction ansatz we need to take the dilaton $\Phi$ to be purely external, while the string-frame metric reads\footnote{The relation between these scalars and the functions in the ten-dimensional ansatz \eqref{eq:ansatz_metric} is
\begin{equation}
e^\Phi=h^{1/4}e^\Lambda\,,\qquad\qquad e^{2U}=4h^{3/8}e^{2g-\Lambda/2}\,,\qquad\qquad e^{2V}=2h^{3/8}e^{2f-\Lambda/2}\,.
\end{equation}}
\begin{equation}\label{eq:ansatzmetric}
\dd s_{\rm st}^2\,=\,e^{\Phi/2}\left(e^{-2U-4V}\dd s_4^2+e^{2U}\,\frac14\left[\left(E^1\right)^2+\left(E^2\right)^2\right]+e^{2V}\,\frac12\dd\Omega_4^2\right)\,.
\end{equation}

Substituting the ansatz in the type IIA equations of motion and separating the components containing the different left-invariant forms we get a set of equations of motion for the fields in the reduction. From the equation for $F_4$ we get the following. The terms proportional to the internal volume-form give the condition\footnote{In this and the following equations the Hodge dual ``$*$'' is taken with respect to the four-dimensional metric $\dd s_4^2$ in \eqref{eq:ansatzmetric}.}
\begin{equation}\label{eq:f4der}
\begin{aligned}
0&=\dd\left(e^{6U+12V+\Phi/2}*f_4\right)+32f_0^X\dd b_J+32f^J_0\dd b_X+64\left(b_X+b_J\right)\dd a_J\\[2mm]
&=\dd\left[e^{6U+12V+\Phi/2}*f_4+16\left[4a_J\left(b_X+b_J\right)+2q_c\left(b_X-b_J\right)+Q_kb_J\left(b_J-2b_X\right)\right]\right]\,.
\end{aligned}
\end{equation} 
This can be immediately integrated to trade the four-form $f_4$ for a certain constant $Q_c$ such that 
\begin{equation}\label{eq:f4dual}
*f_4=-16e^{-6U-12V-\Phi/2}\left[Q_c+4a_J\left(b_X+b_J\right)+2q_c\left(b_X-b_J\right)+Q_kb_J\left(b_J-2b_X\right)\right]\,.
\end{equation}
 Next, the components proportional to $J_2\wedge J_2$ give the equation for $\at$, which reads 
\be\label{eq:eomat}
\dd\left(e^{-2U+4V+\Phi/2}*\tilde{f}_2\right)-4e^{-2U-4V+\Phi/2}*Da_X+2f_0^J\dd b_2+2\dd b_J\wedge\hat{f}_2=0\,.
\ee
Similarly, the components containing $X_2\wedge J_2$ yield the condition
\be\label{eq:eomah}
\dd\left(e^{2U+\Phi/2}*\hat{f}_2\right)-8e^{-2U-4V+\Phi/2}*Da_X+4f_0^X\dd b_2+4\dd b_J\wedge\tilde{f}_2+4\dd b_X\wedge\hat{f}_2=0\,.
\ee
From the terms proportional to $X_3$ one gets 
\be
\begin{aligned}
\dd\left(e^{-2U-4V+\Phi/2}*\dd a_J\right)-\dd b_2\wedge Da_X-\left(b_X+b_J\right)f_4&\\[2mm]
-2\left(f_0^Je^{-2U-12V+\Phi/2}+2f_0^Xe^{-6U-8V+\Phi/2}\right)*1&=0\,,
\end{aligned}
\ee
where $f_4$ is to be understood as given by (the dual of) \eqref{eq:f4dual}. Finally we have the equation for the axion $a_X$ coming from the components containing $J_3$, which reads
\be
\dd\left(e^{-2U-4V+\Phi/2}*Da_X\right)+\dd b_2\wedge \dd a_J=0\,.
\ee
Notice that this equation is not independent but given by the exterior derivative of either \eqref{eq:eomat} or \eqref{eq:eomah}.

The equation for $F_2$ gives a unique non-trivial condition, which plays the role of equation of motion for the vector $a_1$, and reads 
\be\label{eq:eoma1}
\begin{aligned}
\dd\left(e^{2U+4V+3\Phi/2}*\dd a_1\right)+e^{6U+12V+\Phi/2}\left(*f_4\right)\dd b_2+8e^{2U+\Phi/2}\dd b_J\wedge*\hat{f}_2&\\[2mm]
+16e^{-2U+4V+\Phi/2}\dd b_X\wedge*\tilde{f}_2+64\left(b_X+b_J\right)e^{-2U-4V+\Phi/2}*Da_X&=0\,,
\end{aligned}
\ee
where again $\left(*f_4\right)$ is given by \eqref{eq:f4dual}. 

Finally we have the equation for the NS three-form, whose component proportional to the internal volume form gives
\be\label{eq:eomb2}
\begin{aligned}
\dd\left(e^{4U+8V-\Phi}*\dd b_2\right)-e^{6U+12V+\Phi/2}\left(*f_4\right)\dd a_1-16Q_ke^{-2U+4V+\Phi/2}*\tilde{f}_2&\\[2mm]
+8Q_ke^{2U+\Phi/2}*\hat{f}_2-32f_0^J\,\tilde{f}_2-32f_0^X\,\hat{f}_2-64\,Da_X\wedge\dd a_J&=0\,.
\end{aligned}
\ee
Since the two-form turns out to be massive through a Stueckelberg coupling to a vector, we do not expect this equation to be independent of the rest. Indeed, its exterior derivative gives a combination of the vector equations of motion \eqref{eq:eomat} and \eqref{eq:eomah}. When the parameter $Q_k$ vanishes, it is convenient to dualize the two-form to an axion giving mass to a combination of the vectors, as will be detailed below.

The components containing $J_2\wedge J_2$ yield the equation for the scalar $b_X$
\be
\begin{aligned}
\dd\left(e^{-4U-\Phi}*\dd b_X\right)+8Q_ke^{-6U-8V+\Phi/2}f_0^X*1-2f_0^J\,f_4-\hat{f}_2\wedge\hat{f}_2&\\[2mm]
-4\left(b_X+b_J\right)e^{-4U-8V-\Phi}*1-e^{-2U+4V+\Phi/2}\dd a_1\wedge*\tilde{f}_2&=0\,,
\end{aligned}
\ee
while those with $X_2\wedge J_2$ correspond to $b_J$ and read
\be
\begin{aligned}
\dd\left(e^{-4V-\Phi}*\dd b_J\right)-16Q_ke^{-6U-8V+\Phi/2}f_0^X*1+8Q_ke^{-2U-12V+\Phi/2}f_0^J*1&\\[2mm]
-4f_0^X\,f_4-4\tilde{f}_2\wedge\hat{f}_2-8\left(b_X+b_J\right)e^{-4U-8V-\Phi}*1-e^{2U+\Phi/2}\dd a_1\wedge*\hat{f}_2&=0\,.
\end{aligned}
\ee

The ansatz in \cite{Faedo:2017fbv,Elander:2020rgv} is recovered by turning off all the vectors, the two-form and the scalar $a_X$, which solves their respective equations of motion and is therefore consistent. The remaining equations for $U$, $V$, $\Phi$ and the metric will not be presented in detail, since the only modifications with respect to \cite{Faedo:2017fbv,Elander:2020rgv} come from the kinetic terms of the vectors and two-form, that can be inferred from the previous equations. 

Defining the four-dimensional Newton's constant as 
\be\label{eq:4dkappa}
\frac{1}{2\kappa_4^2}=\frac{1}{2\kappa_{10}^2}\int_{\CP^3}\frac{1}{16}\,E^{12}\wedge \mathcal{S}^{1234}=\frac{4\pi^4}{3\left(2\pi\ell_s\right)^8 g_s^2}
\ee
the entire set of equations of motion can be obtained from the following action. First, we have the terms involving just the metric and scalars (in an obvious notation)
\begin{eqnarray}\label{eq:scalaction}
S_{\rm scal}&=&\frac{1}{2\kappa_4^2}\,\int\,\left[R*1-\frac12\left(\dd\Phi\right)^2-4\left(\dd U\right)^2-12\left(\dd V\right)^2-8\dd U\cdot\dd V-32e^{-2U-4V+\Phi/2}\left(\dd a_J\right)^2\right.\nonumber\\[2mm]
&&\quad\quad-4e^{-4V-\Phi}\left(\dd b_J\right)^2-8e^{-4U-\Phi}\left(\dd b_X\right)^2-32e^{-2U-4V+\Phi/2}\left(D a_X\right)^2-\mathcal{V}*1\bigg]\,,
\end{eqnarray}
with the potential
\begin{eqnarray}\label{eq:potential}
\mathcal{V}&=&128\,e^{- 6 U -12 V-\Phi/2} \left[Q_c + 4 a_J \left(b_J + b_X\right) +Q_kb_J\left(b_J-2b_X\right)+2 q_c \left(b_X - b_J\right)\right]^2 \nonumber\\[2mm]
&+&  32 \left(b_J + b_X\right)^2 e^{-4 U - 8 V - \Phi} + 64 \left[2 a_J +Q_k\left(b_J-b_X\right) -  q_c\right]^2 e^{-6 U - 8 V + \Phi/2} \nonumber\\[2mm]
&+& 32 \left(2 a_J -Q_kb_J + q_c\right)^2 e^{-2 U - 12 V + \Phi/2} +4Q_k^2 e^{-2 U - 8 V + 3\Phi/2 } \nonumber\\[2mm]
&+&8Q_k^2 e^{-6 U - 4 V + 3\Phi/2}- 24 e^{-2 U - 6 V} -  8 e^{-4 U - 4 V} + 2 e^{-8 V} \,. 
\end{eqnarray}
On the other hand there are the kinetic terms for the vectors and two-form
\begin{eqnarray}\label{eq:vectaction}
S_{\rm vec}&=&\frac{1}{2\kappa_4^2}\,\int\,\left[-\frac12e^{4U+8V-\Phi}\dd b_2\wedge*\dd b_2-\frac12e^{2U+4V+3\Phi/2}\dd a_1\wedge* \dd a_1\right.\nonumber\\[4mm]
&&\qquad\qquad-8e^{-2U+4V+\Phi/2}\tilde{f}_2\wedge*\tilde{f}_2-4e^{2U+\Phi/2}\hat{f}_2\wedge*\hat{f}_2\bigg]\,.
\end{eqnarray}
Finally one has the convoluted topological terms
\begin{eqnarray}\label{eq:topaction}
S_{\rm top}&=&\frac{1}{2\kappa_4^2}\,\int\,\bigg[\big[64\,a_JDa_X+32\,q_c\left(\ah-\at\right)+16\,Q_c\,a_1\big]\wedge\dd b_2-32\,q_cQ_k\,b_2\wedge b_2\nonumber\\[2mm]
&&\qquad\qquad-16\,b_X\,\hat{f}_2\wedge\hat{f}_2-32\,b_J\,\tilde{f}_2\wedge\hat{f}_2-16\,b_X\,b_J^2\,\dd a_1\wedge\dd a_1\nonumber\\[2mm]
&&\qquad\qquad+32\,b_X\,b_J\,\dd a_1\wedge\hat{f}_2+16\,b_J^2\,\dd a_1\wedge\tilde{f}_2\bigg]\,.
\end{eqnarray}

This reduced action is given in terms of the three constants associated to the fluxes, $Q_c$, $Q_k$ and $q_c$, which correspond to brane charges and are therefore quantized. Indeed, imposing the quantization conditions \cite{Hashimoto:2010bq}
\begin{equation}
\begin{aligned}\label{eq:Page}
k&=\frac{1}{2\kappa_{10}^2T_{{\rm D}6}}\int_{\CP^1}F_{2}\,,\\[4mm]
M-\frac{k}{2} &=\,\frac{1}{2\kappa_{10}^2T_{{\rm D}4}}\int_{\CP^2}\left(F_4-B_2\wedge F_2\right)\\[4mm]
N&=\,\frac{1}{2\kappa_{10}^2T_{{\rm D}2}}\int_{\CP^3}\left(-*F_4-B_2\wedge F_4+\frac12B_2\wedge B_2\wedge F_2\right)
\end{aligned}
\end{equation}
and using the integrals \eqref{twocycle}, \eqref{fourcycle} and \eqref{sixcycle}, one gets the relations
\begin{equation}\label{eq:gaugeparam}
Q_c\,=\,3\pi^2\ell_s^5 g_s\,N\,,\qquad\qquad q_c\,=\,\frac{3\pi\ell_s^3g_s}{4}\,\left(M-\frac{k}{2}\right)\,,\qquad\qquad Q_k\,=\,\frac{\ell_sg_s}{2}\,k\,,
\end{equation}
in the conventions 
\be
\frac{1}{2\kappa_{10}^2}=\frac{2\pi}{\left(2\pi\ell_s\right)^8 g_s^2}\,,\qquad\qquad\qquad T_{{\rm D}p}=\frac{1}{\left(2\pi\ell_s\right)^pg_s\ell_s}\,.
\ee
Notice that these correspond to Page charges, so the integers $N$, $M$ and $k$ count the number of different types of branes. However, Page charges change under large gauge transformations. This will be important later on when we discuss the cascade. 

The parameter $k$ corresponds to the Chern-Simons level in the gauge theory dual, which is vanishing in the solutions considered in this work. Moreover, regularity in the IR of the ground state forces $Q_c=0$, as explained around Eq.~\ref{eq:Qc0}, so in the remainder of this section - and in the bulk of the paper - we will fix $Q_c=Q_k=0$.

In this limit the action greatly simplifies. First, the equation of motion for the two-form can be written as the total derivative
\be
\dd\left[e^{4U+8V-\Phi}*\dd b_2+64a_J\,Da_X+32q_c\left(\ah-\at\right)\right]=0\,.
\ee
In this way it is possible to dualize the form into an axion $a$ defined as
\be\label{eq:b2dual}
*\dd b_2=16\,e^{-4U-8V+\Phi}\left[Da-4a_JDa_X\right]\,,
\ee
with the covariant derivatives
\be\label{eq:axions}
Da=\dd a+2q_c\left(\at-\ah\right)\,,\qquad\qquad\quad Da_X=\dd a_X-\left(\at+\ah\right)\,.
\ee
This immediately shows that the vectors $\at$ and $\ah$ are massive due to their Stueckelberg couplings to the axions $a_X$ and $a$. On the other hand, the remaining vector is massless and associated to a gauge symmetry in the bulk. Indeed, its equation of motion can be written as
\be\label{eq:eoma1b}
\begin{aligned}
&\dd\left[e^{2U+4V+3\Phi/2}*\dd a_1+16\,b_Xe^{-2U+4V+\Phi/2}*\tilde{f}_2+8\,b_Je^{2U+\Phi/2}*\hat{f}_2\right.\\[3mm]
&\left.\quad+32b_J^2b_X\dd a_1+16b_J^2\dd\at+32b_Xb_J\dd\ah\right]=0\,,
\end{aligned}
\ee
with the field strengths $\tilde{f}_2=\dd \at+b_X\,\dd a_1$ and $\hat{f}_2=\dd \ah+b_J\,\dd a_1$. This equation describes a massless vector, $a_1$, interacting with the massive ones through non-diagonal kinetic and Chern--Simons terms. Then, it is dual to the global $\UM$ symmetry we are interested in putting at finite density and/or magnetic field. 

The suitable action reproducing the correct equations of motion reads in this case
\begin{eqnarray}\label{eq:scalactionQk0}
S_{\rm scal}&=&\frac{1}{2\kappa_4^2}\,\int\,\left[R*1-\frac12\left(\dd\Phi\right)^2-4\left(\dd U\right)^2-12\left(\dd V\right)^2-8\dd U\cdot\dd V-4e^{-4V-\Phi}\left(\dd b_J\right)^2\right.\nonumber\\[2mm]
&&\qquad\qquad-8e^{-4U-\Phi}\left(\dd b_X\right)^2-32e^{-2U-4V+\Phi/2}\left(\dd a_J\right)^2-32e^{-2U-4V+\Phi/2}\left(D a_X\right)^2\nonumber\\[2mm]
&&\qquad\qquad-128e^{-4U-8V+\Phi}\left(Da-4a_JDa_X\right)^2-\mathcal{V}*1\bigg]
\end{eqnarray}
for the scalars and 
\begin{eqnarray}\label{eq:vectactionQk0}
S_{\rm vec}&=&\frac{1}{2\kappa_4^2}\,\int\,\left[-\frac12e^{2U+4V+3\Phi/2}\dd a_1\wedge *\dd a_1-8e^{-2U+4V+\Phi/2}\tilde{f}_2\wedge*\tilde{f}_2-4e^{2U+\Phi/2}\hat{f}_2\wedge*\hat{f}_2\right.\nonumber\\[4mm]
&&\qquad\qquad-16\,b_X\,\hat{f}_2\wedge\hat{f}_2-32\,b_J\,\tilde{f}_2\wedge\hat{f}_2-16\,b_X\,b_J^2\,\dd a_1\wedge\dd a_1+32\,b_X\,b_J\,\dd a_1\wedge\hat{f}_2\nonumber\\[4mm]
&&\qquad\qquad+16\,b_J^2\,\dd a_1\wedge\tilde{f}_2\bigg]
\end{eqnarray}
for the vectors. The potential $\mathcal{V}$ in \eqref{eq:scalactionQk0} is the one in \eqref{eq:potential} with $Q_c=Q_k=0$. 

\subsection{Holographic Renormalization}\label{sec:holoren}

In this appendix we perform the holographic renormalization of the action \eqref{eq:scalactionQk0} and \eqref{eq:vectactionQk0}. The action takes the general form
\begin{equation}
S=\frac{1}{2\kappa_4^2}\int\left(R*1-G_{AB}D\phi^A\wedge*D\phi^B-\mathcal{V}*1-H_{IJ}F^I\wedge*F^J+L_{IJ}F^I\wedge F^J\right)\,,
\end{equation}
with $F^I=\dd A^I$ and $D\phi^A=\dd\phi^A+K^A{}_IA^I$. We take for the eight scalars in our truncation the ordering $\phi^A=\left\{\Phi,U,V,b_J,b_X,a_J,a_X,a\right\}$ and for the vectors $A^I=\left\{a_1,\tilde{a}_1,\hat{a}_1\right\}$. With this ordering, the (scalar-dependent) matrices $H$ and $L$ are
\begin{equation}
\begin{aligned}
H&=
\begin{pmatrix}
\frac12e^{2U+4V+\frac{3\Phi}{2}}+4e^{2U+\frac{\Phi}{2}}b_J^2+8e^{-2U+4V+\frac{\Phi}{2}}b_X^2 &\quad\,\,8e^{-2U+4V+\frac{\Phi}{2}}b_X & \quad\,\,4e^{2U+\frac{\Phi}{2}}b_J \\
8e^{-2U+4V+\frac{\Phi}{2}}b_X & 8e^{-2U+4V+\frac{\Phi}{2}} &0\\
4e^{2U+\frac{\Phi}{2}}b_J & 0 & 4e^{2U+\frac{\Phi}{2}}
\end{pmatrix}\,,\\[2mm]
L&=
\begin{pmatrix}
-16b_J^2b_X &\quad\,\,-8b_J^2 & \quad\,\,-16b_Jb_X \\
-18b_J^2 & 0&-16b_J\\
-16b_Jb_X & -16b_J & -16b_X
\end{pmatrix}\,,
\end{aligned}
\end{equation}
while the non-vanishing components of the gauging are $K^7{}_2=K^7{}_3=-1$ together with $K^8{}_2=-K^8{}_3=2q_c$. Notice that the mass matrix for the vectors is then given by the product
$M_{IJ}=G_{AB}K^A{}_IK^B{}_J$.

In our conventions, this same action reads in components
\begin{equation}\label{eq:actcomponents}
S=\frac{1}{2\kappa_4^2}\int\sqrt{-g}\left(R-G_{AB}g^{\mu\nu}D_\mu\phi^AD_\nu\phi^B-\mathcal{V}-\frac12H_{IJ}g^{\mu\nu}g^{\rho\sigma}F_{\mu\rho}^IF_{\nu\sigma}^J\right)\dd^4x+S_{\text{\tiny{top}}}\,,
\end{equation}
with the topological term
\begin{equation}
S_{\text{\tiny{top}}}=\frac{1}{2\kappa_4^2}\int\frac14\tilde{\epsilon}^{\mu\nu\rho\sigma}L_{IJ}F_{\mu\nu}^IF_{\rho\sigma}^J\,.
\end{equation}
Here $\tilde{\epsilon}$ is the Levi--Civita symbol (not tensor) verifying
\begin{equation}
\dd x^\mu\wedge \dd x^\nu\wedge \dd x^\rho\wedge\dd x^\sigma=\tilde{\epsilon}^{\mu\nu\rho\sigma}\dd^4x\,.
\end{equation}
Maxwell's equations deriving from this action are (recall that $\tilde{\epsilon}^{\nu\mu\rho\sigma}\partial_\mu F_{\rho\sigma}^J=0$ because of the Bianchi identity)
\begin{equation}
\partial_\mu\left(\sqrt{-g}H_{IJ}g^{\mu\rho}g^{\nu\sigma}F_{\rho\sigma}^J\right)-\sqrt{-g}G_{AB}K^B{}_Ig^{\nu\mu}D_\mu\phi^A+\frac12\tilde{\epsilon}^{\nu\mu\rho\sigma}F_{\rho\sigma}^J\partial_\mu L_{IJ}=0\,,
\end{equation}
while Einstein's equations read
\begin{equation}
R_{\mu\nu}=G_{AB}D_\mu\phi^AD_\nu\phi^B+\frac12g_{\mu\nu}\mathcal{V}+H_{IJ}g^{\rho\sigma}F_{\mu\rho}^IF_{\nu\sigma}^J-\frac14H_{IJ}g_{\mu\nu}F_{\rho\sigma}^IF^{J\,\rho\sigma}\,.
\end{equation}

Our task is to write the on-shell action as a total derivative. Let us first manipulate the gravitating part of the action in \eqref{eq:actcomponents}. The trace of Einstein's equations gives
\begin{equation}
R-G_{AB}g^{\mu\nu}D_\mu\phi^AD_\nu\phi^B=2\mathcal{V}\,,
\end{equation}
so the on-shell action reduces to
\begin{equation}
I_{\text{\tiny{grav}}}=\frac{1}{2\kappa_4^2}\int\sqrt{-g}\left(\mathcal{V}-\frac12H_{IJ}g^{\mu\nu}g^{\rho\sigma}F_{\mu\rho}^IF_{\nu\sigma}^J\right)\dd^4x\,.
\end{equation}
Next, for a diagonal metric 
\begin{equation}
\dd s^2=g_{tt}\dd t^2+g_{xx}\dd x^2+g_{yy}\dd y^2+g_{rr}\dd r^2\,,
\end{equation}
with the metric components depending only on $r$, the mixed $tt$ component of Einstein's equations reads
\begin{equation}
R^t{}_t=g^{tt}G_{AB}K^A{}_IK^B{}_JA_t^IA_t^J+\frac12\mathcal{V}+H_{IJ}g^{tt}g^{\rho\sigma}F_{t\rho}^IF_{t\sigma}^J-\frac14H_{IJ}F_{\rho\sigma}^IF^{J\,\rho\sigma}\,,
\end{equation}
where we have imposed that $\phi^A=\phi^A\left(r\right)$. Using this we can get rid of the potential in the action, that results 
\begin{equation}
I_{\text{\tiny{grav}}}=\frac{1}{\kappa_4^2}\int\sqrt{-g}\left(R^t{}_t-g^{tt}G_{AB}K^A{}_IK^B{}_JA_t^IA_t^J-H_{IJ}g^{tt}g^{\rho\sigma}F_{t\rho}^IF_{t\sigma}^J\right)\dd^4x\,.
\end{equation}
Now we want to eliminate the mass term for the vectors. In order to do so, we use the temporal component of Maxwell's equations
\begin{equation}
\partial_r\left(\sqrt{-g}H_{IJ}g^{rr}g^{tt}F_{rt}\right)-\sqrt{-g}g^{tt}G_{AB}K^A{}_IK^B{}_JA_t^J+\tilde{\epsilon}^{trxy}F_{xy}^J\partial_r L_{IJ}=0\,,
\end{equation}
where we have imposed that in our ansatz the only non-vanishing components of the field strengths are $F_{rt}^I$ and $F_{xy}^I$. Substituting this in the mass term in the action and taking into account that $F_{rt}^I=\partial_rA_t^I$ we get
\begin{equation}
I_{\text{\tiny{grav}}}=\frac{1}{\kappa_4^2}\int\left[\sqrt{-g}R^t{}_t-\partial_r\left(\sqrt{-g}H_{IJ}A_t^IF^{J\,rt}\right)-\tilde{\epsilon}^{trxy}A^I_tF_{xy}^J\partial_r L_{IJ}\right]\dd^4x\,.
\end{equation}
The last term is not yet a total derivative, but we can combine it with the topological one 
\begin{equation}
I_{\text{\tiny top}}=\frac{1}{\kappa_4^2}\int \tilde{\epsilon}^{rtxy}L_{IJ}F_{rt}^IF_{xy}^J\,,
\end{equation}
so that the total on-shell action reads
\begin{equation}
I_{\text{\tiny{grav}}}+I_{\text{\tiny top}}=\frac{1}{\kappa_4^2}\int\left[\sqrt{-g}R^t{}_t-\partial_r\left(\sqrt{-g}H_{IJ}A_t^IF^{J\,rt}\right)-\partial_r\left(\tilde{\epsilon}^{trxy}L_{IJ}A^I_tF_{xy}^J \right)\right]\dd^4x\,.
\end{equation}
Finally, it is always verified for this type of metrics that 
\begin{equation}
R^t{}_t=\frac{1}{\sqrt{-g}}\partial_r\left(\sqrt{-\gamma}K^t{}_t\right)
\end{equation}
with $K^t{}_t$ the temporal component of the extrinsic curvature and $\gamma$ the boundary metric. In this way the complete on-shell action is a total derivative. 

This is generically UV-divergent and has to be regularized by introducing a UV cutoff $\Lambda_{\text{\tiny UV}}$. To renormalize we need to add the Gibbons--Hawking term 
\begin{equation}
I_{\text{\tiny GH}}=\frac{1}{\kappa_4^2}\int\sqrt{-\gamma}K\dd^3 x\,,
\end{equation}
with $K$ the trace of the extrinsic curvature. The appropriate counterterm to regularize the scalar contributions (we do not expect divergencies from the vectors) is
\begin{equation}
I_{\text{\tiny ct}}=-\frac{1}{2\kappa_4^2}\int\sqrt{-\gamma}4\mathcal{W}\dd^3 x
\end{equation}
with $\mathcal{W}$ the superpotential
\begin{eqnarray}\label{superpotential}
\mathcal{W}&=&e^{-4 V} + 2 e^{-2U -2 V} - 8 e^{-3 U - 6 V - \Phi/4} \left[2 a_J \left( b_J + b_X\right) +q_c \left(b_X - b_J\right) \right] \,.
\end{eqnarray}
In this way, the on-shell action plus these contributions is finite and we can remove the cutoff. The complete renormalized action is then
\begin{equation}\label{eq:Iren}
\begin{aligned}
I_{\text{\tiny ren}}&=-\frac{\beta V_2}{\kappa_4^2}\lim_{\Lambda_{\text{\tiny UV}}\to\infty}\left[\sqrt{-\gamma}\left(K^t{}_t-K+2\mathcal{W}\right)-\sqrt{-g}H_{IJ}A^I_tF^{J\,rt}-\tilde{\epsilon}^{trxy}L_{IJ}A^I_tF_{xy}^J\right]_{\Lambda_{\text{\tiny UV}}}\\[4mm]
&+\frac{\beta V_2}{\kappa_4^2}\left[\sqrt{-\gamma}K^t{}_t-\sqrt{-g}H_{IJ}A^I_tF^{J\,rt}-\tilde{\epsilon}^{trxy}L_{IJ}A^I_tF_{xy}^J\right]_{r_H}
\end{aligned}
\end{equation}
for solutions that have a horizon, such as the black branes of Sec.~\ref{sec:plasma_newsols}. For the solutions with confining IR conditions of Sec.~\ref{sec:conf_newsols} the only contribution comes from the boundary, so the second line is absent. 

From this renormalized action we can compute the energy momentum tensor of the dual gauge theory by varying with respect to the induced metric, evaluated at the boundary. It is related to the energy density, pressure, magnetic field and magnetization density as \cite{Hoyos:2019pyz} 
\begin{equation}\label{eq.EMtensor}
T^i_{\ j}  \ = \ - \frac{1}{\kappa_4} \lim_{\Lambda_{\text{\tiny UV}}\to\infty} \left[\sqrt{-\gamma}\, \left(\, K^i_{\ j} - \delta^i_{\ j}  (K-2\mathcal{W})  \, \right) \right]_{\Lambda_{\text{\tiny UV}}} \ = \  \text{diag}(-E,\, P \, - \, \Bphys\Mphys ,\, P\, - \, \Bphys\Mphys)\,,
\end{equation}
where $E$ is the energy density, $P$ is the pressure, $\Bphys$ is the magnetic field and $\Mphys$ is the magnetization.

\section{Ground state}\label{sec:groundstate}

The confining ground state solution is given by the D2-like string-frame metric 
\be
\dd s_{\rm st}^2\,=\,h^{-1/2}\dd x_{1,2}^2+h^{1/2}\dd s_7^2\,,
\ee
with the transverse metric 
\be
\dd s_7^2=\frac{\dd \rho^2}{\left(1-\frac{\rho_0^4}{\rho^4}\right)}+\frac14\,\rho^2\left(1-\frac{\rho_0^4}{\rho^4}\right)\left[\left(E^1\right)^2+\left(E^2\right)^2\right]+\frac12\,\rho^2\,\dd\Omega_4^2\,.
\ee
This space ends at $\rho=\rho_0$, where the metric in the radial coordinate $\rho_0\left(\rho-\rho_0\right)=\tilde{\rho}^2$ turns into 
\be\label{eq:IRgeom}
\dd s_7^2=\dd \tilde\rho^2+\tilde{\rho}^2\left[\left(E^1\right)^2+\left(E^2\right)^2\right]+\frac12\,\rho_0^2\,\dd\Omega_4^2\,.
\ee
which is regular.\footnote{It is topologically an $\mathds{R}^3$ bundle over ${\rm S}^4$.} 
Moreover, in terms of the dimensionless radial coordinate $z=\rho/\rho_0$ the warp factor is given by  
\begin{equation}
\label{given}
h=\frac{128\,q_c^2}{9\,\rho_0^6}\int_z^\infty\left[\frac{2-3\sigma^4}{\sigma^3\left(\sigma^4-1\right)^2}+\frac{\left(4-9\sigma^4+9\sigma^8\right)\mathcal{U}(\sigma)}{\sigma^4\left(\sigma^4-1\right)^{5/2}}+\frac{2\left(1-3\sigma^4\right)\mathcal{U}(\sigma)^2}{\sigma^5\left(\sigma^4-1\right)^3}\right]\dd\sigma\,.
\end{equation}
The (dimensionless) function $\mathcal{U}$ is defined as 
\begin{equation}
\mathcal{U}(z)\,=\,\int_1^z\left(\sigma^4-1\right)^{-1/2}\dd \sigma=K\left(-1\right)-F\left(\arccsc z|-1\right)=\frac{\sqrt{\pi}\,\Gamma\left(5/4\right)}{\Gamma\left(3/4\right)}-F\left(\arccsc z|-1\right)\,,
\end{equation}
with $K\left(m\right)$ the complete elliptic integral of the first kind and $F\left(\phi|m\right)$ the elliptic integral of the first kind. This warp factor is finite as $z\to1$ so the entire ten-dimensional metric is regular in this limit. On the other hand, it has the following leading behavior at the UV
\be\label{eq:hUV}
h=\frac{128\,q_c^2\,K\left(-1\right)}{5\,\rho_0^6}\,\frac{1}{z^5}\left[1+\mathcal{O}\left(\frac{1}{z}\right)\right]=\frac{128\,q_c^2\,K\left(-1\right)}{5\,\rho_0}\,\frac{1}{\rho^5}\left[1+\mathcal{O}\left(\frac{1}{\rho}\right)\right]\,.
\ee
Notice that the warp factor does not depend at all on the parameter $Q_c$ that, according to \eqref{eq:gaugeparam}, counts the number of (ordinary) D2-branes, so in particular one could force it to vanish. 

The general fluxes regularizing the solution are given by \eqref{eq:formpot} with $b_2=a_1=a_3=\tilde{a}_1=\hat{a}_1=a_X=0$ and 
\begin{equation}
\begin{aligned}
b_J&=\frac{Q_c}{4q_c}+\frac{2q_c}{3\rho_0}\left[\frac{z\sqrt{z^4-1}-\left(3z^4-1\right)\mathcal{U}(z)}{z^4-1}\right]\,,\\[2mm]
b_X&=-\frac{Q_c}{4q_c}-\frac{2q_c}{3\rho_0}\left[\frac{z\sqrt{z^4-1}-\left(3z^4-1\right)\mathcal{U}(z)}{z^4}\right]\,,\\[2mm]
a_J&=\frac{q_c}{6}+\frac{2q_c\,\mathcal{U}(z)}{3z\sqrt{z^4-1}}\,.
\end{aligned}
\end{equation}
Moreover, the two-form $F_2$ is vanishing, which in particular means that $Q_k=0$ and therefore there is no Chern--Simons term. 

The three-form flux is non-vanishing in the IR. In terms of the flat-space radial coordinate reads
\begin{equation}
H_{\text{\tiny IR}}=-\frac{4q_c}{3\rho_0^2}\,\dd\tilde{\rho}\wedge J_2\,,
\end{equation}
while the only remaining piece of the four-form is the flux on the four-cycle
\begin{equation}
F_4^{\text{\tiny IR}}=2q_c\,J_2\wedge J_2\,.
\end{equation}

Throughout the paper we will encounter several quantities defined in the gravity theory that translate to different field theory parameters. Although the map is explained at the relevant points in the discussion, in order to facilitate the reader the conversion between the two, we give a complete set of relations in Table \ref{tab:dictionary}.

\begin{table}[h!]
\begin{center}
\begin{tabular}{|c|c|c|}
\hline 
\begin{minipage}{0.25\textwidth} 
$$
2\kappa_{10}^2=(2\pi)^7 \gs^2 \ls^8
$$
\end{minipage}
& 
\begin{minipage}{0.25\textwidth} 
$$
2\kappa_4^2=3\cdot2^6 \pi^4 \gs^2 \ls^8
$$ 
\end{minipage}
&
\begin{minipage}{0.25\textwidth} 
$$
T_{\text{\tiny D$p$}}=\frac{1}{(2\pi \ls)^p \gs\ls}
$$
\end{minipage} 
 \\[2em]
\hline 
\begin{minipage}{0.25\textwidth} 
$$
q_c=\frac{3\pi}{4} \gs\ls^3 M
$$ 
\end{minipage}
&
\begin{minipage}{0.25\textwidth} 
 $$
 \QD = 3\pi^2 \gs \ls^5 N
 $$ 
\end{minipage}
 &
 \begin{minipage}{0.25\textwidth} 
 $$
 \rho_0=\frac{|b_0|}{2}\gs \ls\frac{M^2}{N}
 $$
\end{minipage}
 \\[2em]
\hline 
\begin{minipage}{0.25\textwidth} 
$$
b_0(\Bphys=0)=-3K(-1)
$$  
\end{minipage}
&
\begin{minipage}{0.25\textwidth} 
$$
\lambda=\gs\ls^{-1} N
$$ 
\end{minipage}
& 
\begin{minipage}{0.25\textwidth} 
$$
\LQCD=\lambda\left(\frac{M}{N}\right)^3
$$ 
\end{minipage}
\\[2em] \hline
\end{tabular}
\caption{\small Dictionary between gravity and field theory parameters. $N$ and $M$ determine the rank of the gauge groups, $\lambda$ is the 't Hooft coupling and $\LQCD$ the characteristic scale of confinement. On the gravity side $q_c$ is the $F_4$ flux in the internal space, $\QD$ is the coefficient of the warp factor when the geometry asymptotes to that of color D2-branes, $\rho_0$ determines the radial position where the two-cycle in the geometry collapses to zero size and $b_0$ is a numerical integration constant determined by the equations of motion and regularity. The remaining $\gs$ and $\ls$ are the string coupling and length, $\kappa_{10}^2$ and $\kappa_4^2$ are the gravitational constants in the ten- and four-dimensional theories respectively, and $T_{\text{\tiny D$p$}}$ is the D$p$-brane tension.}\label{tab:dictionary}
\end{center}
\end{table}

\subsection{The cascade}\label{sec:cascade}

In this section we discuss how the supergravity solution implements the gauge-theory cascade, which is similar to the cascade of $\NN=3$ deformations of ABJ \cite{Aharony:2009fc}. A related analysis, in the presence of Chern--Simons terms, can be found in \cite{Hashimoto:2010bq}. 

The two-form potential on the two-cycle - corresponding to the difference between the (inverse) gauge couplings of both gauge groups in the quiver - is 
\begin{equation}\label{eq:Bflux}
\frac{1}{\left(2\pi\ell_s\right)^2}\int_{\CP^1}B_2=\frac{b_X}{\pi\ell_s^2}=b_{\infty}\,\mathcal{B}\left(z\right)+\frac{Q_c}{4\pi\ell_s^2q_c}\left[\mathcal{B}(z)-1\right]\,,
\end{equation}
with
\begin{equation}
\mathcal{B}\left(z\right)=\frac{\left(3z^4-1\right)\mathcal{U}(z)-z\sqrt{z^4-1}}{3K(-1)z^4}\,.
\end{equation}
This of course runs with the energy, identified with the holographic radial coordinate, and interpolates smoothly between $\mathcal{B}\left(\infty\right)=1$ and $\mathcal{B}\left(1\right)=0$. Its UV and IR values are 
\begin{equation}
b_{\infty}=-\frac{Q_c}{4\pi q_c\ell_s^2}+\frac{2q_cK\left(-1\right)}{\pi\rho_0\ell_s^2}\,,\qquad\qquad\qquad b_{\text{\tiny IR}}=-\frac{Q_c}{4\pi q_c\ell_s^2}\,.
\end{equation}
This forces 
\be\label{eq:Qc0}
Q_c=0
\ee 
if we want this flux to vanish at the point where the two-sphere shrinks, as it should if the background is to be regular. Following Eq.~\eqref{eq:gaugeparam}, this seems to indicate that there are no (ordinary) D2-branes on the background. Nevertheless, as we have observed the warp factor does not depend on $Q_c$. Moreover, it has the asymptotic behavior of a D2-brane, which in our radial coordinate would be
\be
h=\frac{16\,Q_{\text{\tiny D2}}}{5\,\rho^5}\,,
\ee
with $Q_{\text{\tiny D2}}$ quantized as $Q_c$ in \eqref{eq:gaugeparam}. Comparing with \eqref{eq:hUV}, this suggests an identification of the parameter $\rho_0$, which has dimensions of length, as (see \cite{Herzog:2002ss} for an equivalent identification)
\be\label{eq:rho0}
\rho_0=\frac{8\,q_c^2\,K\left(-1\right)}{Q_{\text{\tiny D2}}}=\frac{3g_s\ell_sK\left(-1\right)}{2}\,\frac{M^2}{N}\,.
\ee
Using this, the difference between the UV and IR values of the two-form flux is exactly  
\begin{equation}\label{eq:binfty}
b_{\infty}-b_{\text{\tiny IR}}=b_{\infty}=\frac{2q_cK\left(-1\right)}{\pi\rho_0\ell_s^2}=\frac{N}{M}\,.
\end{equation}
It is also instructive to compute the Maxwell charges for the D2- and D4-branes 
\begin{equation}\label{eq:Maxwell}
\begin{aligned}
N^{\text{\tiny Max}}_4 &=\,\frac{1}{2\kappa_{10}^2T_{{\rm D}4}}\int_{\CP^2}F_4\,,\\[4mm]
N^{\text{\tiny Max}}_2&=\,\frac{1}{2\kappa_{10}^2T_{{\rm D}2}}\int_{\CP^3}\left(-*F_4\right)\,.
\end{aligned}
\end{equation}
The one for the D4 brane does not run and coincides with the Page charge. Taking already $Q_c=0$, the one for the D2-branes reads 
\be
N^{\text{\tiny Max}}_2=\frac{64\pi^3}{3g_s\left(2\pi\ell_s\right)^5}\left[2a_J\left(b_J+b_X\right)+q_c\left(b_X-b_J\right)\right]\,.
\ee
This charge vanishes in the IR, while it takes the UV value 
\be
N^{\text{\tiny Max}}_{\text{\tiny UV}}=\frac{16\sqrt{2}q_c^2\Gamma\left(5/4\right)^2}{3\rho_0g_s\ell_s^5\pi^{5/2}}=N\,,
\ee
where we have used in the last step the identification \eqref{eq:rho0}. This means that we have lost exactly $N$ branes from the UV to the IR, suggesting the cascade
\begin{equation}\label{eq:cascading}
{\rm U}(N)\times{\rm U}(N+M)\,\,\to\,\, {\rm U}(M)
\end{equation}
whose IR is confining. 

Each step of the cascade proceeds as follows. The quantized Page charges for ordinary D2- and D4-branes are not gauge invariant, since they change under a large gauge transformation for $B_2$. This is because we can add to the NS-form a closed but non-exact piece
\begin{equation}
B_2\to B_2+\mathcal{B}\left(X_2-J_2\right),
\end{equation}
with $\mathcal{B}$ constant, that does not change the supergravity solution but does alter the Page charges, as seen in \eqref{eq:Page}. Changing the flux \eqref{eq:Bflux} by one unit corresponds to $\mathcal{B}=\pi\ell_s^2$. The quantities that do not change under this transformation are the Maxwell charges \eqref{eq:Maxwell}. Indeed, this large gauge transformation can be seen as the shift in the Page charges 
\begin{equation}
b_{\infty}\to b_{\infty}-1\,,\qquad\qquad N\to N+M\,,\qquad\qquad M\to M\,,
\end{equation}
under which the Maxwell charges are invariant. 

Crucially, the correct gravitational description requires the two-form flux to be in the range $(0,1)$. If that is not the case, we can add or subtract one unit of flux to put it back to the correct range. Imagine that we follow the flow from the IR, where the gauge theory is expected to be purely U($M$). As we flow towards the UV, the two-form flux \eqref{eq:Bflux} (with $Q_c=0$), which was initially vanishing, grows. The moment it reaches 1, we perform a gauge transformation to put it back to zero, so the rank of the gauge groups in the dual field theory jumps $M$ units to 
\begin{equation}
 {\rm U}(M)\,\,\to\,\,{\rm U}(M)\times{\rm U}(2M)\,.
\end{equation}
The flux will then continue growing, and when it reaches unity again we perform once more a gauge transformation that shifts the rank $M$ units. This continues as long as the flux can grow above 1. Since it is bounded by $b_{\infty}$ in \eqref{eq:binfty},\footnote{This is in sharp contrast with the cascade on the conifold \cite{Klebanov:2000hb}, whose two-form flux is not bounded and grows indefinitely towards the UV, resulting in infinitely many duality steps. This might be a reflection of the fact that three-dimensional Yang--Mills theories are asymptotically free.} this can happen $N/M$ times, resulting in the UV gauge group stated in \eqref{eq:cascading}. Notice that if $N/M$ is not an integer, the last step of the cascade in the IR would take us to 
\begin{equation}
{\rm U}(p)\times{\rm U}(p+M)\,.
\end{equation}
There are now two possibilities. If $p\ll M$ the second gauge group is weakly ('t Hooft) coupled and its dynamics plays no role in our supergravity approximation, the background being equal to the one with $p=0$ to leading order in $M$. On the other hand, if $p\sim M$ the flux due to this additional D2-branes should be included, resulting in a singularity and the spoiling of confinement. See \cite{Klebanov:2000hb} for the analogous discussion in the four-dimensional case. 

\section{Calculation of the monopole-antimonopole action}\label{sec:D2pot}

The ten-dimensional string-frame metric and dilaton are given in \eqref{eq:ansatz_metric}. Consider a D2-brane wrapping the $\CP^1$ two-cycle and extended in the radial direction and on a curve in the $(x_1,x_2)$-plane, with an embedding profile $X_1=x(r)$, $X_2=y(r)$. The induced metric on the brane is
\begin{equation}
\dd s_{\text{\tiny D2}}^2 =\frac{h^{\frac12}}{\mathsf{b}} \left[1+\frac{\mathsf{b}}{h} \left(\left(x'\right)^2+\left(y'\right)^2\right)\right]\dd r^2+h^{\frac12} e^{2g}\left(\dd \theta^2+\sin^2\theta\,\dd\varphi^2\right) \,.
\end{equation}
The D2-brane also couples to the components of the $B_2$ field along the two-cycle, whose pullback reads 
\begin{equation}
P\left[B_2\right]=b_X P\left[X_2\right]=b_X \sin\theta \,\dd\theta \wedge \dd \varphi\,.
\end{equation}
It is possible to turn on a magnetic field on the brane along those same directions
\begin{equation}
2\pi \ls^2 F=\beta_2 P\left[X_2\right]=\beta_2 \sin\theta\, \dd\theta \wedge \dd \varphi\,.
\end{equation}
Finally, we need to take into account the pullback of the RR one-form $C_1$ on the D2-brane worldvolume
\begin{equation}
P\left[C_1\right]=\gs\ls\,\frac{M^2}{N}\,\frac{\Bphys}{2} \left(x y'-y x'\right) \dd r\,.
\end{equation}
In Euclidean signature, the Dirac--Born--Infeld (DBI) part of the D2-brane action is then
\begin{equation}\label{eq:SDBID2}
\begin{split}
S_{\text{\tiny DBI}}=&\frac{1}{(2\pi)^2 \gs \ls^3} \int_{\rm D2} \dd^3 \eta \, e^{-\Phi}\sqrt{g_{\text{\tiny D2}}+B_2+2\pi \ls^2 F}\\[2mm]
=&\frac{1}{\pi \gs \ls^3}\int \dd r\, \left( \frac{h}{\mathsf{b}}\right)^{\frac12}\, e^{2g-\Lambda}\sqrt{\left(1+\frac{\mathsf{b}}{h} \left[\left(x'\right)^2+\left(y'\right)^2\right]\right)\left(1+\frac{1}{4}h^{-1}e^{-4g}\left(b_X+\beta_2\right)^2\right)}\,.
\end{split}
\end{equation}
On the other hand, the Wess--Zumino (WZ) terms are
\begin{equation}
S_{\text{\tiny WZ}}=-\frac{1}{(2\pi)^2\gs \ls^3}\int_{\rm D2} C_1\wedge \left(B_2+2\pi \ls^2 F\right)=-\frac{M^2}{N}\frac{\Bphys}{2\pi \ls^2}\int \dd r\, \left(xy'-yx'\right)\left(b_X+\beta_2\right)\,.
\end{equation}
When the magnetic field is vanishing, $\Bphys=0$, the configuration that becomes tensionless in the IR has $\beta_2=0$. At the boundary, the D2-brane carries an amount of D0 charge that is determined by the WZ coupling
\begin{equation}
S_{\text{\tiny WZ}}=-\frac{1}{(2\pi)^2\gs \ls^3}\left(\lim_{r\to\infty} \int_{\CP^1} B_2\right) \int C_1 =-\frac{1}{\gs\ls} \frac{N}{M} \int C_1\,.
\end{equation}
Therefore this wrapped D2 corresponds to a bound state of a D2-brane and $N/M$ D0-branes.

\subsection{Vanishing magnetic field}

In the absence of magnetic field, $\Bphys=0$, the WZ term vanishes, and the embedding can lie along one of the directions in the $(x_1,x_2)$-plane. For definiteness, we will take $X_2=y=0$ and $X_1= x(r)$.
From \eqref{eq:SDBID2} we can derive the equation of motion for the embedding, using that $x$ is a cyclic variable 
\begin{equation}
\left(\frac{\mathsf{b}}{h}\right)^{\frac12} e^{2g-\Lambda}\sqrt{1+\frac{1}{4}h^{-1}e^{-4g}\left(b_X+\beta_2\right)^2} \,\frac{x'}{\sqrt{1+\frac{\mathsf{b}}{h} (x')^2}}=p_x\,,
\end{equation}
with $p_x$ an integration constant. Solving for $x'$ we get
\begin{equation}\label{eq:solD2}
x'=\left(\frac{h}{\mathsf{b}}\right)^{\frac12}\frac{p_x}{\sqrt{A^2 -p_x^2}}\,, \qquad\qquad A=e^{2 g-\Lambda}\sqrt{1+\frac{1}{4}h^{-1}e^{-4g}\left(b_X+\beta_2\right)^2}\,.
\end{equation}
For a connected configuration, it is possible to relate the integration constant with the lowest point in the radial direction reached by the embedding, $r_*$, as
\begin{equation}
p_x=A(r_*)\equiv A_*\,.
\end{equation} 
Therefore, the separation between the monopole and anti-monopole at the boundary is
\begin{equation}
L=2\, A_*\int_{r_*}^\infty dr\, \frac{h^{\frac12}\mathsf{b}^{-\frac12}}{\sqrt{A^2 -A_*^2}}\,.
\end{equation}
In terms of the dimensionless functions and radial coordinate defined in \eqref{eq:dimlessFUNC}, \eqref{eq:dimlessSCAL} and \eqref{eq:coordinate} this reads 
\begin{equation}
L=\ell_{\text{\tiny D2}}\, {\cal A}_*\int_0^{\xi_*} \frac{d\xi}{\xi^2\sqrt{1-\xi^4}}\, \frac{\mathbf{h}^{\frac12}\mathsf{b}^{-\frac12}}{\sqrt{{\cal A}^2 -{\cal A}_*^2}}\,,
\end{equation}
where we have defined
\begin{equation}
{\cal A}=e^{2\GG-\Lambda}\sqrt{1+\frac{1}{128}\mathbf{h}^{-1} e^{-4\GG}(\BX-\bar{\beta}_2)^2}\,,\qquad\qquad \beta_2=\frac{2 q_c}{3\rho_0} \bar{\beta}_2\,.
\end{equation}
The characteristic length scale is
\begin{equation}
\ell_{\text{\tiny D2}}=2\rho_0\left(\frac{128q_c^2}{9\rho_0^6}\right)^{\frac12}=\frac{16\sqrt{2}\,\pi}{|b_0|^2}\,\LQCD^{-1}\,.
\end{equation}
Introducing \eqref{eq:solD2} in the action \eqref{eq:SDBID2}, one finds
\begin{equation}
S_{\text{\tiny D2}}=\frac{1}{\pi g_s \ell_s^3}\int \dd r\,  \frac{h^{\frac12}\,\mathsf{b}^{-\frac12}\, A^2}{\sqrt{A^2 -A_*^2}}\,.
\end{equation}
The disconnected configuration corresponds to the case where the branes reach the bottom of the geometry, $r_*=r_0$, and therefore $A_*=p_x=0$. Then, the difference in the action between the connected and disconnected configurations is
\begin{equation}
\begin{split}
\Delta S_{\text{\tiny  D2}}&=\frac{2}{\pi g_s \ell_s^3}\left[\int_{r_*}^\infty \dd r\, \left(\frac{h^{\frac12}\,\mathsf{b}^{-\frac12}\, A^2}{\sqrt{A^2 -A_*^2}}-h^{\frac12}\,\mathsf{b}^{-\frac12}\, A\right)-\int_{r_0}^{r_*} \dd r\,h^{\frac12}\,\mathsf{b}^{-\frac12}\, A\right]\\[2mm]
&={\cal N}_{\text{\tiny D2}}\left[\int_0^{\xi_*} \frac{\dd\xi}{\xi^2\sqrt{1-\xi^4}}\, \left(\frac{\mathbf{h}^{\frac12}\,\mathsf{b}^{-\frac12}\, {\cal A}^2}{\sqrt{{\cal A}^2 -{\cal A}_*^2}}-\mathbf{h}^{\frac12}\,\mathsf{b}^{-\frac12}\, {\cal A}\right)-\int_{\xi_*}^1\frac{\dd\xi}{\xi^2\sqrt{1-\xi^4}}\,\mathbf{h}^{\frac12}\,\mathsf{b}^{-\frac12}\, {\cal A}\right]\,,
\end{split}
\end{equation}
where 
\begin{equation}
{\cal N}_{\text{\tiny D2}}=4\sqrt{2} \,M.
\end{equation}
The dominant configuration will be the one with least action, so the point where $\Delta S_{\text{\tiny D2}}=0$ determines the screening length for the monopoles. In addition, there is maximal possible separation $L$ for the connected configuration. 

\section{Numerics}\label{sec:numerics}

In this Appendix we give some details of the numerical strategy, including the boundary conditions (expansions) imposed both at the UV and the two possible IRs. These expansion are available to use upon request. 

In solving the equations numerically, it will be convenient to work with dimensionless quantities by rescaling the fields. For the functions in the metric we take
\be\label{eq:dimlessFUNC}
e^{2f}  = \rho_0^2\,  e^{2\FF},  \qquad\qquad	e^{2g} = \rho_0^2 \,e^{2\GG},  \qquad\qquad	h = \frac{128\, q_c^2 }{9\rho_0^ 6}\, \mathbf{h}\,,
\ee
with $\rho_0$ some constant with dimensions of length\footnote{The notation is not accidental, as it coincides with the parameter $\rho_0$ for the ground state solution in Appendix~\ref{sec:groundstate}. Notice that, as seen in \eqref{eq:qc}, the parameter $q_c$ has dimensions of length cubed.}. Similarly, the scalars are redefined to 
\be\label{eq:dimlessSCAL}
b_J = \frac{2q_c}{3\rho_0 }\, \BJ, \qquad\qquad b_X = -  \frac{2q_c}{3\rho_0 }\, \BX \,, \qquad\qquad a_J = -\frac{q_c}{2} - q_c\, \AJ\,,
\ee
while the vector potentials and magnetic field are written as
\begin{equation}\label{eq:dimlessVECS}
a_t  = \frac{\rho_0^3}{q_c}  \, \AO,  \qquad	\hat{a}_t = \rho_0^2\,  \AH, \qquad	\tilde{a}_t = \rho_0^2\,  \AT , \qquad \Bphys =\frac{1}{\gs\ls}\,\frac{N}{M^2}\,  \frac{\rho_0^5}{q_c^2}\, \BET\,.
\end{equation}
The functions $e^\Lambda$ and $\mathsf{b}$ are already dimensionless and do not need rescaling. Written in terms of the dimensionless radial coordinate $\xi$ defined in \eqref{eq:coordinate} and the rescaled functions in \eqref{eq:dimlessFUNC}, \eqref{eq:dimlessSCAL} and \eqref{eq:dimlessVECS}, the dimensionful quantities $\rho_0$ and $q_c$ factor out and one ends up with a completely dimensionless set of equations. 

\subsection{UV expansion}\label{sec:UVexpan}

The expansion of the dimensionless functions compatible with the desired D2-brane UV asymptotics takes the following form\footnote{The logarithms appear at a relatively high order; for instance, in the metric function $e^{2\FF}$, the first non-vanishing coefficients in $\log\xi$ and $\log^2\xi$ are $\cfUV{1}{9}$ and $\cfUV{2}{18}$ respectively.  Moreover, all of them vanish if the magnetic field is set to zero.}
\begin{equation}
\begin{aligned}
e^{2\FF} &= \left(\frac{1}{2\xi^2}\right)\left( \ 1 \ + \  \sum_{k=1}^{\infty}\,  \sum_{l=1}^{\infty}\   \cfUV{l}{k}\,  \xi^ k\, \log^l\xi \ \right)\,, \\[2mm]
e^{2\GG} &= \left(\frac{1}{4\xi^2}\right)\left( \ 1 \ + \  \sum_{k=1}^{\infty}\,  \sum_{l=0}^{\infty}\   \cgUV{l}{k}\,  \xi^ k\, \log^l\xi \ \right)\,, \\[2mm]
e^{2\Lambda} &=  1 \ + \  \sum_{k=1}^{\infty}\,  \sum_{l=0}^{\infty}\   \clUV{l}{k}\,  \xi^ k\, \log^l\xi \,, \qquad
\mathbf{b} =  1 \ + \  \sum_{k=5}^{\infty}\,  \sum_{l=0}^{\infty}\   \cbUV{l}{k}\,  \xi^ k\, \log^l\xi  \,, \\[2mm]
\mathbf{h}& =  \xi^5\, \sum_{k=0}^{\infty}\,  \sum_{l=0}^{\infty}\   \chUV{l}{k}\,  \xi^ k\, \log^l\xi  \,,\qquad\,\AAA_J =  \sum_{k=0}^{\infty}\,  \sum_{l=0}^{\infty}\   \caJUV{l}{k}\,  \xi^ k\, \log^l\xi  \,, \\[2mm]
\BB_J &=   \sum_{k=0}^{\infty}\,  \sum_{l=0}^{\infty}\   \cbJUV{l}{k}\,  \xi^ k\, \log^l\xi  \,, \qquad
\,\,\BB_X =  \sum_{k=0}^{\infty}\,  \sum_{l=0}^{\infty}\   \cbXUV{l}{k}\,  \xi^ k\, \log^l\xi  \,, \\[2mm]
\AO &=   \sum_{k=0}^{\infty}\,  \sum_{l=0}^{\infty}\   \caoneUV{l}{k}\,  \xi^ k\, \log^l\xi  \,, \qquad 
\,\,\AH =   \sum_{k=1}^{\infty}\,  \sum_{l=0}^{\infty}\   \cahUV{l}{k}\,  \xi^ k\, \log^l\xi  \,, \\[2mm]
\AT &= \sum_{k=1}^{\infty}\,  \sum_{l=0}^{\infty}\   \catUV{l}{k}\,  \xi^ k\, \log^l\xi  \,. \\[2mm]
\end{aligned}
\end{equation}
Some parameters in these series are left undetermined by the equations, while the rest are given in terms of them. We chose the free coefficients in the previous expansions to be
\noindent 
\begin{center}
	\begin{tabular}{cccccc}
		\hline
		\multicolumn{1}{|c|}{$\cfUV{0}{4}$}  & \multicolumn{1}{c|}{$\cfUV{0}{5}$} & \multicolumn{1}{c|}{$\cfUV{0}{10}$}   & \multicolumn{1}{c|}{$\cbJUV{0}{0}$}   & \multicolumn{1}{c|}{$\cbJUV{0}{4}$}   & \multicolumn{1}{c|}{$\cbJUV{0}{6}$}   \\ \hline
		\multicolumn{1}{|c|}{$f_4$}          & \multicolumn{1}{c|}{$f_5$}         & \multicolumn{1}{c|}{$f_{10}$}         & \multicolumn{1}{c|}{$b_0$}            & \multicolumn{1}{c|}{$b_4$}            & \multicolumn{1}{c|}{$b_6$}            \\ \hline
		\multicolumn{1}{l}{}                 & \multicolumn{1}{l}{}               & \multicolumn{1}{l}{}                  & \multicolumn{1}{l}{}                  & \multicolumn{1}{l}{}                  & \multicolumn{1}{l}{}                  \\ \hline
		\multicolumn{1}{|c|}{$\cbJUV{0}{9}$} & \multicolumn{1}{c|}{$\cbUV{0}{5}$} & \multicolumn{1}{c|}{$\caoneUV{0}{0}$} & \multicolumn{1}{c|}{$\caoneUV{0}{1}$} & \multicolumn{1}{c|}{$\catUV{0}{4}$} & \multicolumn{1}{c|}{$\cahUV{0}{5}$} \\ \hline
		\multicolumn{1}{|c|}{$b_9$}          & \multicolumn{1}{c|}{$\mathbf{b}_5$}         & \multicolumn{1}{c|}{$v_0$}            & \multicolumn{1}{c|}{$v_1$}            & \multicolumn{1}{c|}{$v_4$}            & \multicolumn{1}{c|}{$v_5$}            \\ \hline
	\end{tabular}
\end{center}
\vspace{2mm}
We rename them as in the second row for the sake of clarity. This is the notation used in the bulk of the paper. In terms of these independent coefficients the expansions read 
\begin{equation}
\begin{aligned}\label{eq:UVexpansion}
e^{2\FF} &= \frac{1}{2\xi^2} \left( 1 + f_4 \xi ^4 + f_5 \xi ^5 + \cdots + f_{10} \xi ^{10} +\OO\left(\xi^{11}\right)  \right)\,, \quad\quad e^{2\GG} = \frac{1}{4\xi^2}\left( 1  +\OO\left(\xi^4\right) \right)\,,\\[2mm]
e^{2\Lambda} & =  1  +\OO\left(\xi^5\right) \,, \qquad\qquad\quad \mathbf{h} =  -\frac{3b_0}{5} \xi^5 +\OO\left(\xi^6\right)  \,, \qquad\quad\quad\mathsf{b} =  1  + \mathsf{b}_5 \xi^5  +\OO\left(\xi^6\right)  \,, \\[2mm]
  \BB_J&= b_0 +\cdots + { b_4} {\xi^4} +\cdots + { b_6} {\xi^6} + \cdots + { b_9} {\xi^9} +\OO\left(\xi^{10}\right)   \,, \\[2mm]
\BB_X &= b_0 +\OO\left(\xi\right)\,, \qquad\quad\quad \,\,\AAA_J =  -\frac{2}{3}  + \OO\left(\xi\right) \,,\qquad\qquad\quad\,\,\, \AO =   v_0 + v_1 \xi + \OO\left(\xi^2\right)\,, \\[2mm] 
\AH &= \frac{2b_0v_1}{15}\xi +\cdots + v_5\xi^5 + \OO\left(\xi^6\right) \,,\qquad\qquad \AT = -\frac{2b_0v_1}{15}\xi +\cdots + v_4\xi^4 + \OO\left(\xi^5\right)  \,. \\[2mm]
\end{aligned}
\end{equation}

 The parameter $\cfUV{0}{1}$ is also free. However, it is related to a shift in the original radial coordinate $r$, 
\begin{equation}
r \, \mapsto r+a \,,
\end{equation}
and we are free to set it to zero, $\cfUV{0}{1}=0$. Some modes, such as $f_{10}$ and $b_9$, appear at a relatively high order. Thus, we reached a high order in the expansion so that they could be resolved in our numerical procedure. Namely, we solved up to  the term $\cfUV{0}{20}\xi^{20}$ in the $e^{2\FF}$ function and the corresponding terms in the rest of the functions which are solved at the same order. Below, we show the coefficients that appear in the renormalized action \eqref{eq:Iren}, and therefore are needed to obtain the thermodynamic quantities, in terms of the undetermined ones.  

\vspace{5mm}
\textbf{Coefficients appearing in $e^{2\FF}$:}
\noindent\begin{fleqn}
	\begin{equation}
	\begin{array}{*4{>{\displaystyle}l}}
	\cfUV{0}{1} =\cfUV{0}{2} =\cfUV{0}{3} =0 \,.
		\end{array}
	\end{equation}
\end{fleqn}

\textbf{Coefficients appearing in $e^{2\GG}$:}
\noindent\begin{fleqn}
\begin{equation}
\begin{array}{*3{>{\displaystyle}l}}
\cgUV{0}{1} =\cgUV{0}{2} =\cgUV{0}{3} =0 \,,\qquad\qquad\qquad & \cgUV{0}{4} =  -2 f_4-1 \,,\qquad\qquad\qquad  & \cgUV{0}{5} = f_5\,. \\[2mm]
\end{array}
\end{equation}
\end{fleqn}

\textbf{Coefficients appearing in $e^{\Lambda}$:}
\noindent\begin{fleqn}
\begin{equation}
\begin{array}{*2{>{\displaystyle}l}}
\clUV{0}{1} =\clUV{0}{2} =\clUV{0}{3}=\clUV{0}{4} =0 \,,\qquad\qquad\qquad &\clUV{0}{5} =  f_5 \,.  
\end{array}
\end{equation}
\end{fleqn}

\textbf{Coefficients appearing in $\BB_J$:}
\noindent\begin{fleqn}
	\begin{equation}
	\begin{array}{*4{>{\displaystyle}l}}
	\cbJUV{0}{1} = 4 \,,\qquad\qquad\qquad  & \cbJUV{0}{2} = 	\cbJUV{0}{3} = 0 \,,\qquad\qquad  \qquad
	& \cbJUV{0}{5} = \frac{2}{5} \left(16 f_4+7\right) \,. \\[4mm]
	\end{array}
	\end{equation}
\end{fleqn}

\textbf{Coefficients appearing in $\BB_X$:}
\noindent\begin{fleqn}
	\begin{equation}
	\begin{array}{*4{>{\displaystyle}l}}
	\cbXUV{0}{0} = b_0 \,,\qquad\qquad\qquad  & \cbXUV{0}{1} = 4 \,,\qquad\qquad\qquad  & \cbXUV{0}{2} = \cbXUV{0}{3} = 0 \,,\\[4mm]
	 \cbXUV{0}{4} = -\frac{b_4}{2} \,,\qquad \qquad\qquad&\cbXUV{0}{5} = -\frac{2}{5} \left(14 f_4+3\right)\,.&\qquad\qquad 
	\end{array}
	\end{equation}
\end{fleqn}

\textbf{Coefficients appearing in $\AAA_J$:}
\noindent\begin{fleqn}
	\begin{equation}
	\begin{array}{*4{>{\displaystyle}l}}
	\caJUV{0}{0} = - \frac{2}{3} \,,\qquad\qquad\qquad  & \caJUV{0}{1} = 0 \,.
		\end{array}
	\end{equation}
\end{fleqn}

\textbf{Coefficients appearing in $\mathbf{h}$:}
\noindent\begin{fleqn}
	\begin{equation}
	\begin{array}{*3{>{\displaystyle}l}}
	\chUV{0}{0} = -\frac{3 b_0}{5} \,,\qquad\qquad\qquad  & \chUV{0}{1} = -2 \,,\qquad\qquad\qquad  &\chUV{0}{2} = 0  \,,\qquad\qquad \\[4mm]
	\chUV{0}{3} = 0 \,, \qquad\qquad\qquad &\chUV{0}{4} = -\frac{b_0}{2}  \,,\qquad\qquad\qquad  &\chUV{0}{5} = \frac{1}{50} \left(45 b_0 f_5+24 f_4-92\right) \,, \\[2mm] 
	\end{array}
	\end{equation}
\end{fleqn}

\textbf{Coefficients appearing in} $\AH$\textbf{:}
\noindent\begin{fleqn}
	\begin{equation}
	\begin{array}{*2{>{\displaystyle}l}}
	\cahUV{0}{1} = \frac{2 }{15}b_0 v_1 \,,
		\end{array}
	\end{equation}
\end{fleqn}

\textbf{Coefficients appearing in} $\AT$\textbf{:}
\noindent\begin{fleqn}
	\begin{equation}
	\begin{array}{*2{>{\displaystyle}l}}
	\catUV{0}{1} =-\frac{2}{15} b_0 v_1 \,,
		\end{array}
	\end{equation}
\end{fleqn}

Imposing that the different backgrounds have the same leading asymptotic behaviour - except for $\AO$, whose leading term $v_0$ is related to the chemical potential - forces a relation between the parameter $b_0$ and the length scale $\rho_0$. This is most easily seen in the scalars $b_J$ and $b_X$. Their leading terms in the UV are
\be
b_J= \frac{2q_c}{3\rho_0 }\, \BJ=\frac{2q_c}{3}\,\frac{b_0}{\rho_0}+\cdots\,,\qquad\qquad\qquad b_X= -\frac{2q_c}{3\rho_0 }\, \BX=-\frac{2q_c}{3}\,\frac{b_0}{\rho_0}+\cdots\,,
\ee
so the ratio $b_0/\rho_0$ must be held fixed in comparing different solutions. To fix its value, we demand that the warp factor $h$ in the metric has the asymptotic behaviour of a D2-brane, which in our conventions is 
\begin{equation}
	h = \frac{16}{5}\, \frac{\QD}{r^ 5}+\cdots=\frac{128 \,q_c^2 }{15\rho_0^ 6} \left|b_0\right| \xi^5+\cdots\,.
\end{equation}
Using that the relation between radial coordinates is $r=\rho_0/\xi$ in the asymptotic UV, one gets 
\begin{equation}
	\frac{\rho_0}{b_0} = - \frac{8}{3}\frac{q_c^2}{\QD} = -  \frac{\ls^ 2}{2}\,\lambda\,\frac{M^2}{N^ 2}\,,
\end{equation}
where in the last step we assumed that the D2-brane charge $\QD = 3\pi^2 \ls^5 g_s N$ is quantized in the usual manner (see Eq.~\eqref{eq:gaugeparam}). This argument is analogous to that leading to \eqref{eq:rho0} in Appendix~\ref{sec:cascade} for the ground state by selecting the appropriate value $b_0=-3K\left(-1\right)$, with $K(m)$ the complete elliptic integral of the first kind. 

Let us now discuss the two different IR boundary conditions. These correspond to two different phases in the dual gauge theory.

\subsection{Magnetized confining phase}\label{sec:conf_newsols}

This is a background at finite magnetic field sharing the IR boundary conditions with the ground state. In particular, the two-sphere collapses in a smooth manner while ${\rm S}^4$ remains at finite size (see Eq.~\eqref{eq:IRgeom}). Moreover, curvature invariants remain finite everywhere.

To be precise, in the radial coordinate \eqref{eq:coordinate}, with the IR located at $\xi\to1$, the equations admit the following perturbative solution
\begin{equation}\label{eq:confiningIR}
\begin{aligned}
e^{2\FF} &= \cfi + \OO(1-\xi)^1\,,\qquad e^{2\GG} = \frac{1-\xi}{\cbi} + \OO(1-\xi)^2\,,\qquad e^{\Lambda} = \cli + \OO(1-\xi)^1\,,\\[2mm]
\mathsf{b} &= \cbi + \OO(1-\xi)^1\,,\qquad  \mathbf{h} = \chir + \OO(1-\xi)^1\,, \qquad \BB_J = \cbji (1-\xi)^{\frac{1}{2}} + \OO(1-\xi)^\frac{3}{2},\\[2mm]
\BB_X &= \cbxi (1-\xi)^{\frac{3}{2}} + \OO(1-\xi)^\frac{5}{2}\,, \qquad  \AAA_J = -1  + \caji(1-\xi) + \OO(1-\xi)^2\,,\\[2mm]
\AO &= \caone (1-\xi) + \OO(1-\xi)^2\,,\qquad \AH = \cah (1-\xi)^{\frac{1}{2}} + \OO(1-\xi)^{\frac{3}{2}}\,,\qquad \AT = \OO(1-\xi)^{\frac{3}{2}}\,.\\[2mm]
\end{aligned}
\end{equation}
The equations admit constant terms for the vector potentials. However, the ones for $\AH$ and $\AT$ must be set to zero in order to prevent a non-vanishing flux on the collapsing two-sphere in the IR\footnote{The equations of motion set $\AH=-\AT$ in the IR. However, they leave undetermined the component $C_3\supset\AT\dd t\wedge X_2$. Since $X_2$ is a collapsing cycle, regularity forces this term to vanish in the IR.}. Additionally, we fix the constant parameter of the massless vector $\AO$ to zero, since one can always use gauge invariance to shift its value and we are keeping the analogous parameter $v_0$ in the UV. The constant parameters we made explicit are the ones that we chose to leave undetermined by the equations of motion.   

As we mentioned, this IR boundary condition gives rise to a regular ten-dimensional metric. Indeed, solving the change of coordinates \eqref{eq:coordinate} perturbatively in the IR, one finds that 
\begin{equation}
 1-\xi  = \frac{1}{\rho_0^2}\left(r-r_0\right)^2-\frac{11}{6\rho_0^4} \left(r-r_0\right)^4+\frac{71}{18\rho_0^6} \left(r-r_0\right)^6+\OO\left(r-r_0\right)^{8}\,.
\end{equation}
Substituting in the metric ansatz
\begin{equation}\begin{aligned}
\dd s^2 &\approx \frac{3\rho_0^3}{8 q_c \sqrt{2}}\chir^{-\frac{1}{2}}\left(-\cbi \dd t^2 + \dd x_1^2+ \dd x_2^2\right) + 
 \frac{8 q_c \sqrt{2}}{3\rho_0^3}\frac{\chir^{\frac{1}{2}}}{\cbi}\left(\dd  r^2 +  (r-r_0)^2\left[\left(E^1\right)^2+\left(E^2\right)^2\right]  \right) \\[2mm]
  &	\quad + \frac{8 q_c \sqrt{2}}{3\rho_0} \chir^{\frac{1}{2}} \cfi \dd \Omega_4^2\,,
\end{aligned}
\end{equation}
where we are keeping only the leading order in $(r-r_0)$ for every metric component. From this we see that the metric is regular, as the S$^2$ shrinks smoothly and the transverse topology becomes that of an $\R^3$ bundle over ${\rm S}^4$, exactly as in the ground state geometry \eqref{eq:IRgeom}. 

The UV and IR asymptotic expansions leave us with undetermined parameters that have to be found via numerical shooting methods. For each choice of magnetic field, the set of parameters we shoot for is 
\begin{equation}
	\{f_4, \,
	f_5, \,
	f_{10}, \,
	b_4, \,
	b_6, \,
	b_9; \,
	\mathsf{b}_5;\,
	b_0; \,
	v_0; \,
	v_1, \,
	v_4 , \,
	v_5; \,
	\cfi, \,	
	\cbi, \,
	\cli, \,
	\cbji, \,
	\cbxi, \,
	\caji, \,
	\chir, \,
	\caone ,\,
	\cah \} \,.
\end{equation}

The first 6 are UV parameters associated to normalizable modes in the scalar sector - comprising six non-axionic scalars, as seen in \eqref{eq:scalactionQk0} - and would correspond to vacuum expectation values on the field theory dual. The parameter $\mathsf{b}_5$ is a metric deformation allowed by the breaking of three-dimensional Lorentz invariance, in this case due to the magnetic field/chemical potential. The constant $b_0$ appears in the leading term of the scalars $b_J$ and $b_X$, as well as the warp factor $h$, and is also determined numerically. Since the ratio in \eqref{eq:ratiobrho} must be kept fixed, this forces us to change $\rho_0$ accordingly for each solution. 

The parameter $v_0$ is the leading term in the expansion of the massless vector and therefore controls the chemical potential associated to the $\UM$ symmetry. Its value will be determined by the shooting procedure, but, since the gauge invariance of the equations allows us to shift $a_t$ by an arbitrary constant, we can construct confining solutions with an arbitrary chemical potential. Equivalently, one could have retained the undetermined constant in the IR expansion of the vector potential, since the boundary conditions allow it, and determine it numerically for any fixed value of the chemical potential. This is in contrast with the black hole case, in which regularity forces the vector potential to vanish at the non-extremal horizon.    

The parameter $v_1$ is the subleading term in the expansion of the massless vector and as such it will be related to the charge density, while $v_4$ and $v_5$ are the normalizable modes in the massive vector potentials and thus correspond to vacuum expectation values for their dual operators. Finally, the remaining 9 constants are simply IR parameters to be found numerically.  

Once the magnetic field is fixed, there are in total 21 parameters to be determined by the numerical procedure. This matches the number of degrees of freedom of our system of equations (see Sec.~\ref{sec:10D}), which consist of 11 second order differential equations subject to a first order constraint. This means that, for every value of the temperature and chemical potential, there is a one-parameter family of solutions labelled by the magnetic field $\BET$. 

\subsection{Magnetized plasma phase}\label{sec:plasma_newsols}

These deconfined states correspond in the gravity dual to non-extremal black branes at non-vanishing magnetic field. In order to have a horizon the blackening factor $\mathsf{b}$ must have a simple zero. On the other hand, both the scalars and the rest of the metric functions reach a finite value at the horizon. Denoting the position of the horizon as $\xi=\xi_h$ in the dimensionless radial coordinate introduced in equation~\eqref{eq:coordinate}, this implies
\begin{equation}\label{Horizon_expansions_finite}
\begin{array}{rclrclrcl}
\AAA_J&=& \cajh+\OO(\xi-\xi_h) \,,&
\quad\BB_J&=& \cbjh +\OO(\xi-\xi_h)  \ ,&
\BB_X&=& \cbxh+\OO(\xi-\xi_h)   \ ,\\[2mm]
e^{2\FF} &=&  \cfh  +\OO(\xi-\xi_h)  \ ,&
\quad e^{2\GG} &=&  \cgh + \OO(\xi-\xi_h)  \ , &
 e^\Lambda&=& \clh + \OO(\xi-\xi_h) \ , \\[2mm]
\mathbf{h}& =& \chh +\OO(\xi-\xi_h)  \ ,&\quad\mathsf{b}&=&  \cbh (\xi-\xi_h)+\OO(\xi-\xi_h)^2&&&
\end{array}
\end{equation}
Regularity at the horizon implies additionally that the time component of the vectors must vanish, so they verify an expansion of the form 
\begin{equation}\label{Horizon_expansions_zero}
\begin{array}{rclrcl}
\AO&=&  \caoneh (\xi-\xi_h)+\OO(\xi-\xi_h)^2 \  , \qquad& \qquad\AH&=&  \cahh (\xi-\xi_h)+\OO(\xi-\xi_h)^2 \ ,\\[2mm]
\qquad\AT&=&  \cath (\xi-\xi_h)+\OO(\xi-\xi_h)^2\,,\\[2mm]
\end{array}
\end{equation}
All the subleading coefficients are fixed in terms of these eleven leading-order parameters. In this way, the constants that are determined by the numerical procedure are
\begin{equation}
	\{f_4, \,
	f_5, \,
	f_{10}, \,
	b_4, \,
	b_6, \,
	b_9; \,
	\mathsf{b}_5;\,
	v_1, \,
	v_4 , \,
	v_5; \,	
	\cfh, \,
	\cgh, \,	
	\cbh, \,	
	\clh, \,	
	\cbjh,\,
	\cbxh, \,
	\cajh, \,
	\chh, \,
\caoneh ,\,
\cahh ,\, 
\cath  
	\} \,.
\end{equation}
The first ten are UV parameters and have the same meaning as in the previous section. For these solutions we fix the parameter $b_0$ to its ground-state value, $b_0=-3K(-1)$, as in \cite{Elander:2020rgv}. The remaining eleven are horizon parameters that control physical properties such as the entropy and temperature. The total number matches again the degrees of freedom of our system of equations. This leaves three unfixed control parameters: $v_0$, $\xi_h$ and $\BET$ or, equivalently, the chemical potential, the temperature and the magnetic field, describing a three-parameter family of black branes.

\bibliographystyle{JHEP}
\bibliography{magnetic_ref}

\end{document}